\documentclass[10pt,reqno,a4paper]{amsart}
\usepackage[british]{babel}
\usepackage{amssymb,amstext,amsthm,eucal,bbm,mathrsfs,amscd,bbold,pifont}
\usepackage[margin=1in]{geometry}
\usepackage{array}
\usepackage[svgnames,table]{xcolor}
\usepackage[T1]{fontenc}
\usepackage[utf8]{inputenc}
\usepackage{enumitem}
\usepackage[small]{eulervm}
\usepackage{microtype}
\usepackage{booktabs} 
\usepackage{caption,cite}
\usepackage{tikz,pgfplots}
\usetikzlibrary{calc,arrows,shapes,cd}
\newsavebox{\uuunit}
\sbox{\uuunit}
    {\setlength{\unitlength}{0.825em}
     \begin{picture}(0.6,0.7)
        \thinlines
        \put(0,0){\line(1,0){0.5}}
        \put(0.15,0){\line(0,1){0.7}}
        \put(0.35,0){\line(0,1){0.8}}
       \multiput(0.3,0.8)(-0.04,-0.02){12}{\rule{0.5pt}{0.5pt}}
     \end {picture}}
\newcommand {\unity}{\mathord{\!\usebox{\uuunit}}}

\usepackage[unicode,pagebackref=true]{hyperref}
\usepackage{accents}
\usepackage{slashed}
\hypersetup{%
  pdftitle   = {A non-lorentzian primer},
  pdfkeywords = {non-lorentzian, carroll, newton-cartan, galilean, kinematical, spacetime},
  pdfauthor  = {Eric Bergshoeff, José Figueroa-O'Farrill, Joaquim Gomis},
  pdfcreator = {\LaTeX\ with package \flqq hyperref\frqq},
  linkcolor=NavyBlue,
  citecolor=ForestGreen,
  urlcolor=OrangeRed,
  anchorcolor=OrangeRed,
  colorlinks=true
}
\renewcommand*{\backref}[1]{}
\renewcommand*{\backrefalt}[4]{%
  \ifcase #1%
  \or [Page~#2.]%
  \else [Pages~#2.]%
  \fi%
}
\usepackage{multicol}
\theoremstyle{definition}
\newtheorem{definition}{Definition}

\theoremstyle{plain}

\newtheorem*{mainthm}{Theorem}
\definecolor{gris}{rgb}{0.5,0.5,0.5}
\definecolor{darkgreen}{rgb}{0,0.5,0}

\usepackage{mathtools}
\newcommand{\choice}[2]{\substack{\mathrlap{#1}\\ \mathrlap{#2}}}
\allowdisplaybreaks[0]
\numberwithin{equation}{section}
\usepackage{graphicx}
\newcommand{\be}{\nopagebreak[3]\begin{equation}}
\newcommand{\ee}{\end{equation}}
\newcommand{\ba}{\nopagebreak[3]\begin{eqnarray}}
\newcommand{\ea}{\end{eqnarray}}

\newcommand{\bal}{\nopagebreak[3]\begin{aligned}}
\newcommand{\eal}{\end{aligned}}

\newcommand{\bseq}{\nopagebreak[3]\begin{subequations}}
\newcommand{\eseq}{\end{subequations}\noindent}
\newcommand{\X}{\mathbb{X}}

\newcommand{\Ad}{\operatorname{Ad}}
\newcommand{\Ann}{\operatorname{Ann}}
\newcommand{\Sch}{\operatorname{Sch}}
\newcommand{\Diff}{\operatorname{Diff}}
\newcommand{\ad}{\operatorname{ad}}
\newcommand{\Aff}{\operatorname{Aff}}
\newcommand{\GL}{\operatorname{GL}}
\newcommand{\SL}{\operatorname{SL}}
\newcommand{\PSL}{\operatorname{PSL}}
\newcommand{\SO}{\operatorname{SO}}
\newcommand{\Ort}{\operatorname{O}}
\newcommand{\Af}{\mathbb{A}}
\newcommand{\LL}{\mathbb{L}}
\newcommand{\RR}{\mathbb{R}}
\newcommand{\ZZ}{\mathbb{Z}}
\newcommand{\bx}{\boldsymbol{x}}
\newcommand{\by}{\boldsymbol{y}}
\newcommand{\bv}{\boldsymbol{v}}
\newcommand{\bzero}{\boldsymbol{0}}
\newcommand{\e}{\boldsymbol{e}}
\renewcommand{\sl}{\mathfrak{sl}}
\renewcommand{\a}{\mathfrak{a}}
\renewcommand{\k}{\mathfrak{k}}
\newcommand{\gl}{\mathfrak{gl}}
\newcommand{\iso}{\mathfrak{iso}}
\newcommand{\fsim}{\mathfrak{sim}}
\newcommand{\co}{\mathfrak{co}}
\newcommand{\so}{\mathfrak{so}}
\newcommand{\g}{\mathfrak{g}}
\newcommand{\car}{\mathfrak{c}}
\newcommand{\s}{\mathfrak{s}}
\newcommand{\h}{\mathfrak{h}}
\newcommand{\n}{\mathfrak{n}}
\newcommand{\m}{\mathfrak{m}}
\renewcommand{\r}{\mathfrak{r}}
\newcommand{\eH}{\mathsf{H}}
\newcommand{\eE}{\mathscr{E}}
\newcommand{\eJ}{\mathscr{J}}
\newcommand{\eO}{\mathscr{O}}
\newcommand{\eX}{\mathscr{X}}
\newcommand{\eL}{\mathscr{L}}
\newcommand{\dS}{\mathsf{dS}}
\renewcommand{\d}{\partial}
\newcommand{\dvol}{d\text{vol}}
\newcommand{\B}{\boldsymbol{B}}
\renewcommand{\P}{\boldsymbol{P}}
\renewcommand{\L}{\boldsymbol{L}}
\newcommand{\Ggr}{\mathscr{G}}
\newcommand{\Hgr}{\mathscr{H}}

\begin{document}

\title{A non-lorentzian primer}
\author[Bergshoeff]{Eric A.~Bergshoeff}
\author[Figueroa-O'Farrill]{José M.~Figueroa-O'Farrill}
\author[Gomis]{Joaquim Gomis}
\address[EAB]{Van Swinderen Institute, University of Groningen, Nijenborgh 4, 9747 AG Groningen, The Netherlands}
\email{\href{mailto: e.a.bergshoeff@rug.nl}{e.a.bergshoeff[at]rug.nl}}
\address[JMF]{Maxwell Institute and School of Mathematics, The University
  of Edinburgh, James Clerk Maxwell Building, Peter Guthrie Tait Road,
  Edinburgh EH9 3FD, Scotland, United Kingdom}
\email{\href{mailto:j.m.figueroa@ed.ac.uk}{j.m.figueroa[at]ed.ac.uk}}
\address[JG]{Departament de Física Cuàntica i Astrofísica and Institut
  de Ciències del Cosmos, Universitat de Barcelona, Martí i Franquès
  1, E-08028 Barcelona, Spain}
\email{\href{mailto:gomis@ecm.ub.es}{gomis[at]ecm.ub.es}}
\begin{abstract}
  We review both the kinematics and dynamics of non-lorentzian theories
  and their associated geometries. First, we introduce non-lorentzian
  kinematical spacetimes and their symmetry algebras.  Next, we construct
  actions describing the particle dynamics in some of these kinematical
  spaces using the method of nonlinear realisations.  We explain the
  relation with the coadjoint orbit method.  We continue discussing
  three types of non-lorentzian gravity theories: Galilei gravity,
  Newton-Cartan gravity and Carroll gravity.  Introducing matter, we
  discuss electric and magnetic non-lorentzian field theories for
  three different spins: spin-0, spin-1/2 and spin-1, as limits of
  relativistic theories.
\end{abstract}
\thanks{EMPG-22-08}
\maketitle
\tableofcontents

\newpage

\section{Introduction}
\label{sec:introduction}

The recent interest in non-lorentzian theories and their associated
geometries is, among other things, due to the following developments:

\begin{enumerate}[label=(\roman*)]
\item \emph{non-relativistic holography}, which has applications in condensed
  matter physics \cite{sachdev2011quantum, zaanen2015holographic}.
  For example, it has been shown, in the context of Lifshitz holography
  \cite{Christensen:2013lma}, that interesting non-relativistic
  geometries appear at the boundary;

\item \emph{flat space holography}, see for example
  \cite{barnich2010symmetries, Bagchi:2010zz, bagchi2012bms} (general)
  and \cite{bondi1962gravitational,sachs1962gravitational} (BMS
  symmetries) and soft theorems as in \cite{weinberg1965infrared,
    strominger2018lectures};

\item \emph{carrollian physics}, see, for example,
  \cite{levy1965nouvelle, sen1966analogue}, which allows us to
  understand the symmetries of null hypersurfaces, such as black-hole
  horizons \cite{Donnay:2019jiz} and boundaries of asymptotically flat
  spacetimes \cite{Duval:2014uva};

\item \emph{non-relativistic string theories} \cite{Gomis:2000bd,
    Danielsson:2000gi},  carrollian string theories
  \cite{Cardona:2016ytk} and tensionless string theories
  \cite{Isberg:1993av,Bagchi:2015nca,Bagchi:2020fpr} as corners of the
  moduli space of solvable string theories (see the review
  \cite{Oling:2022fft}).  Conjecturally, non-relativistic string
  theory has its own holography probing a different class of boundary
  field theories;

\item \emph{post-Newtonian corrections} \cite{dautcourt1990newtonian,
    Dautcourt:1996pm, VandenBleeken:2017rij, Hansen:2019svu,
    Hansen:2020wqw} in the experimental and theoretical investigations
  of gravitational waves \cite{Planck:2018vyg};

\item \emph{fractons} \cite{Nandkishore:2018sel} \cite{Pretko:2020cko}
  which are condensed matter configurations with restricted mobility
  which display infrared/ultraviolet mixing with subsystem symmetries
  (see the review \cite{Grosvenor:2021hkn}).
\end{enumerate}

In this review we introduce some of the basic concepts and tools to
study these theories. We first introduce kinematical Lie algebras and
their associated homogeneous spacetimes. Some of these Lie algebras
arise as contractions of the isometry algebras of (anti) de~Sitter
spacetimes, following the pioneering work of Bacry and Lévy-Leblond
\cite{Bacry:1968zf}, but by far not all of them are obtained in this
way. We restrict ourselves to kinematical Lie algebras which preserve
space isotropy and hence the kinematical spacetimes we consider are
also spatially isotropic. They are adequate to describe particle
dynamics. In particular this means that we are not considering
so-called $p$-brane kinematical Lie algebras and their associated
spacetimes in which to describe non-lorentzian $p$-brane actions. We
refer the interested reader to \cite{Brugues:2004an, Gomis:2005pg,
  Brugues:2006yd, Barducci:2018wuj}.

We present a classification of (spatially isotropic) kinematical
and aristotelian\footnote{These are kinematical Lie algebras without
  boosts.} Lie algebras in generic dimension
\cite{Figueroa-OFarrill:2017ycu,Figueroa-OFarrill:2017tcy}.  Generic
means that they exist in all dimensions.  There are additional
kinematical Lie algebras in two, three and four spacetime dimensions,
to which we refer the reader to the classic work of Bacry and Nuyts
\cite{MR857383} (reviewed in \cite{Figueroa-OFarrill:2017ycu}) for
dimension $3+1$, \cite{Andrzejewski:2018gmz} for dimension $2+1$ and
the classic Bianchi classification of three-dimensional Lie algebras
\cite{Bianchi,MR1900159} for $1+1$.  After a brief review of
homogeneous geometry and the infinitesimal description of homogeneous
spaces in terms of Klein pairs, we present the classification of
spatially isotropic homogeneous spacetimes of kinematical Lie groups.
Again we list those which exist in generic dimension, which here it
means we are omitting some $1+1$ and $2+1$ dimensional spacetimes,
which can be found in
\cite{Figueroa-OFarrill:2018ilb,Figueroa-OFarrill:2019sex}.

In the study of particle dynamics on homogeneous kinematical
spacetimes, one meets homogeneous spaces of the kinematical groups
other that the actual spacetimes: namely, coadjoint orbits and their
associated evolution spaces.  We review the rôle played by these
homogeneous spaces in the construction of lagrangians describing
particle dynamics on the homogeneous spacetimes.

We review the method of nonlinear realisations and coadjoint orbits in
the construction of particle lagrangians and apply it in several
examples, among them the well-known relativistic massive and
massless particles. In the non-relativistic case we construct the
harmonic oscillator as a nonlinear realisation of a centrally extended
Newton--Hooke group.  We also consider the massless galilean particle
introduced by Souriau \cite{MR1066693,Souriau}.

In the case of Caroll and due to its causal structure, we consider a
massive timelike and tachyonic particles. Using the conformal algebra
in one dimension we derive the action of conformal mechanics of
de Alfaro, Fubini and Furlan \cite{deAlfaro:1976vlx} and the Schwarzian
action \cite{Kitaev:2017awl,Maldacena:2016hyu,Stanford:2017thb}.

The analogues of contractions for Lie algebras in dynamical systems
are limits of actions, such as non-relativistic, carrollian and flat
limits. The actions constructed using the nonlinear realisation method
are also obtained as nonrelativistic limits of the relativistic
actions. In general these limits produce terms that are divergent:
unwanted terms that can be eliminated by a suitable coupling of a
relativistic dynamical system to a gauge field in the case of a
particle or a $B$-field in the case of a string \cite{Gomis:2000bd}
\cite{Gomis:2005pg}. In some of these cases the divergent terms are
total derivatives. One can also eliminate divergences by a
redefinition of the parameters appearing in the first term of the
expansion \cite{Batlle:2016iel, Bergshoeff:2017btm}.

As for the case of non-lorentzian particles, we continue discussing
different aspects of non-lorentzian gravity theories. We first review
how general relativity can be described by a gauging procedure applied
to the Poincaré algebra. Next, we extend the discussion by applying
the same gauging procedure to the following non-lorentzian algebras:
Galilei, Bargmann and Carroll. These gaugings lead to Galilei gravity,
Newton-Cartan gravity\,\footnote{Newton-Cartan gravity in the context
  of non-relativistic holography was studied in \cite{Bagchi:2009my}.}
and Carroll gravity, respectively. We show how these same gravity
theories can be obtained by taking particular (galilean, Bargmann and
Carroll) limits of general relativity.  For recent work on electric
and magnetic theories of gravity see
\cite{Henneaux:2021yzg,Hansen:2021fxi,Perez:2021abf,Perez:2022jpr}.

Besides taking non-relativistic limits there are two more ways to
obtain non-relativistic theories that we do not explore any further in
this review.  First, instead of taking the Inönü--Wigner contraction
of a Lie algebra, one may also consider a Lie algebra expansion
\cite{Hatsuda:2001pp,Boulanger:2002bt,deAzcarraga:2002xi,Izaurieta:2006zz}
where the number of generators corresponding to the nonrelativistic
symmetries is increased. Second, one may obtain a non-relativistic
theory by the null reduction of a relativistic theory in one spatial
dimension higher, see, e.g.,
\cite{Gauntlett:1990xq,Julia:1994bs}. This null reduction is based
upon the fact that the Bargmann algebra embeds into the
Poincaré algebra in one spatial dimension higher.

Having discussed non-lorentzian gravity, we continue introducing
matter and discussing non-lorentzian field theories. We will do this
for a complex and real massive spin-0 particle, a massive spin-1/2
particle and a massless spin-1 particle. In particular, for spin-0, we
will discuss the Galilei, Bargmann and Carroll limits while for
spin-1/2 and spin-1 we will only discuss the Bargmann limit.

\section{Motivation}
\label{sec:motivation}

Let us motivate our discussion of kinematical symmetries and their
spacetimes by contrasting two classical models of the universe: the
Galilei spacetime of newtonian mechanics and Minkowski spacetime of
special relativity.  As we will see, both spacetimes are described by
a four-dimensional affine space, homogeneous under the action of a
kinematical Lie group; that is, a transformation group consisting of
rotations, boosts and translations in both space and time.  We will
contrast the invariant structures of the two spacetimes: a clock and a
ruler in the Galilei spacetime and a proper distance in Minkowski
spacetime.  The latter defines a lorentzian metric and the former, as
we will see, a (weak) Newton--Cartan structure.  We will also contrast
their Lie algebras of symmetries: the finite-dimensional Lie algebra
of isometries in Minkowski spacetime and the infinite-dimensional
Coriolis algebra in the Galilei spacetime.

\subsection{Affine space}
\label{sec:affine-space}

Let $\Af^4$ denote the four-dimensional affine space.  It is modelled
on the vector space $\RR^4$ in the sense that given any two points
$a,b \in \Af^4$ there exists a unique translation $v \in
\RR^4$ such that $b = a + v$.  We often refer to $v$ as $b-a$ and
identify translations with differences of points.  We will use an
explicit model for $\Af^4$ as the affine hyperplane in $\RR^5$
consisting of points $(x^1,x^2,x^3,x^4,x^5=1) \in \RR^5$, but we
should emphasise that the fifth dimension is an auxiliary construct
and has no physical meaning.  One cannot add points in $\Af^4$ (their
last entry would not equal $1$), but one can add differences, since
those lie in the hyperplane $x^5=0$.   In this model, the group
$\Aff(4,\RR)$ of affine transformations of $\Af^4$ is the subgroup of
$\GL(5,\RR)$ which preserves the hyperplane $x^5=1$.  It consists of
matrices of the form
\begin{equation}\label{eq:affine-trans}
  \begin{pmatrix}
    L & v \\ 0 & 1
  \end{pmatrix}
\end{equation}
where $v \in \RR^4$ and $L \in \GL(4,\RR)$.  We will see that the
relativity groups of the Galilei and Minkowski spacetimes are
subgroups of the affine group containing all the translations $v \in
\RR^4$ but with a restricted subgroup of linear transformations
consisting of rotations and boosts.

It follows from matrix multiplication that the affine group is the
semidirect product $\GL(4,\RR) \ltimes \RR^4$, with $\GL(4,\RR)$
acting on $\RR^4$ by matrix multiplication.  Multiplying $(x,1) \in
\RR^5$ by the matrix in equation~\eqref{eq:affine-trans}
gives $(Lx + v, 1)$, which is the effect of an affine transformation.
Both the Galilei and Minkowski spacetimes are described by $\Af^4$,
only that their invariant structures differ.  Points in $\Af^4$ are
called \textbf{(spacetime) events}.

\subsection{Galilei spacetime}
\label{sec:galilei-spacetime}

The following description of Galilei spacetime is essentially due to
Weyl \cite{MR988402}.

Galilei spacetime is defined by $\Af^4$ together with two invariant
notions:
\begin{itemize}
\item a \textbf{clock} $\tau : \RR^4 \to \RR$, sending $b-a \mapsto
  \tau(b-a)$ and measuring the time interval between two events $a,b
  \in \Af^4$.   If $a = (x, 1)$ and $b = (y, 1)$, then $\tau(b-a) =
  y^4 - x^4$.  Two events $a,b \in \Af^4$ are said to be
  \textbf{simultaneous} if $\tau(b-a) = 0$.  In other words,
  simultaneous events are related by translations in the
  kernel of $\tau$.  If we fix an event $a$, the set of events
  simultaneous to $a$ defines a three-dimensional affine subspace
  \begin{equation}
    a + \ker \tau = \left\{a + v ~\middle | ~\tau(v) = 0\right\}
  \end{equation}
  of $\Af^4$.  As the notation suggests, it is a coset of the subgroup
  $\ker \tau$ of the translation group $\RR^4$.  The quotient $\Af^4/\ker
  \tau$ is an affine line $\Af^1$, so that the clock gives a fibration $\pi: \Af^4 \to \Af^1$ whose fibre
  at $\pi(a)$ consists of all those events simultaneous to $a$, which
  constitute an affine hypersurface $\Af^3_a$ of $\Af^4$.  This is
  illustrated in Figure~\ref{fig:clock-fibration}.

\item a \textbf{ruler} $\lambda : \ker \tau \to \RR$, sending
  $b-a \mapsto \lambda(b-a)$ and measuring the euclidean distance
  between simultaneous events.  Explicitly, if $a = (\bx, x^4, 1)$ and $b = (\by, y^4,1)$ with
  $\bx,\by \in \RR^3$ and $x^4=y^4$ are simultaneous events, then
  $\lambda(b-a) = \|\by - \bx\| = \sqrt{(\by - \bx)\cdot (\by -
    \bx)}$, which is the euclidean distance between $\bx$ and $\by$.
\end{itemize}

\begin{figure}[h!]
  \centering
    \begin{tikzpicture}[x=1.0cm,y=1.0cm,scale=0.6]
      \coordinate [label=above right:{$\Af^4$}] (a4) at (4,1);
      \coordinate [label=right:{$\Af^1$}] (a1) at (4,-1);
      \coordinate [label=below:{$\pi(a)$}] (pa) at (-3,-1);
      \coordinate [label=below:{$\pi(b)$}] (pb) at (2,-1);
      \coordinate [label=left:{$a$}] (a) at (-3,3);
      \coordinate [label=right:{$b$}] (b) at (2,6);
      \coordinate [label=right:{$\Af^3_a$}] (a3) at (-3,7);
      \draw [black!50!white, thin] (-4,1) -- (4,1) -- (4,9) -- (-4,9) -- cycle;
      \draw [black!50!white, thin] (-4,-1) -- (4,-1);
      \draw [red, thick] (-3,1) -- (-3,9);
      \draw [red, thick] (2,1) -- (2,9);
      \draw[->,shorten >=2mm,shorten <=2mm,black,thick] (a)--(b) node[midway,sloped,above]{$b-a$};
      \draw[->,shorten >=1mm,shorten <=1mm,black,thick] (pa)--(pb) node[midway,above]{$\tau(b-a)$};
      \begin{scope}[transform canvas={xshift=0.7em}]
        \draw [->, shorten >=3mm,thin, black] (a4) -- node[above right] {$\pi$} (a1);
      \end{scope}
      \foreach \point in {pa,pb}
      \fill [red] (\point) circle (3pt);
      \foreach \point in {a,b}
      \fill [blue] (\point) circle (3pt);
  \end{tikzpicture}
  \caption{The clock fibration $\pi : \Af^4 \to \Af^1$}
  \label{fig:clock-fibration}
\end{figure}

The kinematical group of Galilei spacetime is called the
\textbf{Galilei group} and it consists of those affine
transformations of $\Af^4$ which preserve the clock and the ruler.  It
embeds in $\GL(5,\RR)$ as those matrices of the form
\begin{equation}
  \label{eq:gal-in-gl5}
  \begin{pmatrix}
    R & \bv & \boldsymbol{a} \\
    0 & 1 & s \\
    0 & 0 & 1
  \end{pmatrix},
\end{equation}
where $R \in \Ort(3)$, $\boldsymbol{a},\bv \in \RR^3$ and $s \in
\RR$.   This matrix is of the form~\eqref{eq:affine-trans}, but
where the general linear transformation $L$ is of the form
$\begin{pmatrix} R & \bv \\ 0 & 1\end{pmatrix}$.

The action of the matrix in equation~\eqref{eq:gal-in-gl5} on an event
$(\bx, t, 1)$ gives the event $(R\bx + t \bv + \boldsymbol{a}, t +
s,1)$ which we interpret as the composition of an orthogonal
transformation $\bx \mapsto R \bx$, a \textbf{Galilei boost} $\bx
\mapsto \bx + t \bv$, a spatial translation $\bx \mapsto \bx +
\boldsymbol{a}$ and a temporal translation $t \mapsto t + s$:
\begin{equation}
  \label{eq:gal-decomposition}
  \begin{pmatrix}
    R & \bv & \boldsymbol{a} \\
    0 & 1 & s \\
    0 & 0 & 1
  \end{pmatrix} =
  \begin{pmatrix}
    I & 0 & 0 \\
    0 & 1 & s\\
    0 & 0 & 1
  \end{pmatrix}
  \begin{pmatrix}
    I & 0 & \boldsymbol{a} \\
    0 & 1 & 0\\
    0 & 0 & 1
  \end{pmatrix}
  \begin{pmatrix}
    I & \bv & 0 \\
    0 & 1 & 0\\
    0 & 0 & 1
  \end{pmatrix}
  \begin{pmatrix}
    R & 0 & 0 \\
    0 & 1 & 0\\
    0 & 0 & 1
  \end{pmatrix}.
\end{equation}

Its Lie algebra is the \textbf{Galilei algebra}, which is isomorphic
to the subalgebra of $\gl(5,\RR)$ consisting of matrices of the form
\begin{equation}
  \label{eq:gal-algebra-gl5}
  \begin{pmatrix}
    A & \bv & \boldsymbol{a}\\
    0 & 0 & s\\
    0 & 0 & 0
  \end{pmatrix},
\end{equation}
where $A \in \so(3)$, $\bv,\boldsymbol{a} \in \RR^3$ and $s \in \RR$.
We may introduce a basis $L_{ab} = - L_{ba}, B_a, P_a, H$ by
\begin{equation}
  \label{eq:gal-basis}
  \begin{pmatrix}
    A & \bv & \boldsymbol{a}\\
    0 & 0 & s\\
    0 & 0 & 0
  \end{pmatrix} = \tfrac12 A^{ab} L_{ab} + v^a B_a + a^a P_a + s H.
\end{equation}
We can easily work out the Lie brackets of the Galilei algebra in
this basis.  The nonzero brackets are given by
\begin{equation}
  \label{eq:gal-algebra-brackets}
  \begin{split}
    [L_{ab},L_{cd}] &= \delta_{bc} L_{ad} - \delta_{ac} L_{bd} -  \delta_{bd} L_{ac} + \delta_{bd} L_{ac} \\
    [L_{ab}, B_b] &= \delta_{bc} B_a - \delta_{ac} B_b\\
    [L_{ab}, P_b] &= \delta_{bc} P_a - \delta_{ac} P_b\\
    [B_a, H] &= P_a.
  \end{split}
\end{equation}
This shows that $L_{ab}$ span an $\so(3)$ subalgebra, relative to
which $B_a,P_a$ transform according to the three-dimensional vector
representation (which is also the adjoint representation in this
dimension) and $H$ transforms as the one-dimensional scalar
representation.  We shall see that all kinematical Lie algebras (with
spatial isotropy) share these properties, which are strong enough to
allow for their classification.

\subsection{Minkowski spacetime}
\label{sec:minkowski-spacetime}

Minkowski spacetime is also described by $\Af^4$, but the invariant
notion is now that of a \textbf{proper distance} $\Delta : \RR^4 \to
\RR$, sending $b-a \mapsto \Delta(b-a)$, where if $a = (x,1)$ and $b =
(y,1)$,
\begin{equation}
  \label{eq:proper-distance}
  \Delta(b-a) = (y-x)^T \eta (y-x),
\end{equation}
where
\begin{equation}
  \label{eq:minkowski-IP}
  \eta =
  \begin{pmatrix}
    -1 & 0 & 0 & 0 \\
    \phantom{-}0 & 1 & 0 & 0\\
    \phantom{-}0 & 0 & 1 & 0\\
    \phantom{-}0 & 0 & 0 & 1
  \end{pmatrix}.
\end{equation}
We no longer have a separate clock and ruler, or as Minkowski himself
put it \cite{zbMATH02638586}:
\begin{quotation}
  Von Stund' an sollen Raum für sich und Zeit für sich völlig zu
  Schatten herabsinken und nur noch eine Art Union der beiden soll
  Selbständigkeit bewahren.\footnote{Henceforth space by itself, and
    time by itself, are doomed to fade away into mere shadows, and
    only a kind of union of the two will preserve an independent
    reality.}
\end{quotation}
In particular, there is no longer an invariant notion of simultaneity
between events, so instead of affine subspaces of simultaneity, we
have lightcones at every spacetime event $a$: the \textbf{lightcone}
$\LL_a$ of $a$ being defined as those events which are a zero proper distance
away from $a$:
\begin{equation}
  \LL_a = \left\{ b \in \Af^4 \middle | \Delta(b-a) = 0\right\}.
\end{equation}

The kinematical group of Minkowski spacetime is the \textbf{Poincaré
  group} and consists of those affine transformations which preserve
the proper distance between events.  It embeds in $\GL(5,\RR)$ as
those matrices
\begin{equation}
  \label{eq:poin-in-gl5}
  \begin{pmatrix}
    L & v \\
    0 & 1
  \end{pmatrix}
\end{equation}
where $L^T \eta L = \eta$ and $v \in \RR^4$.  Matrix multiplication
shows that the Poincaré group is isomorphic to the semidirect product
$\Ort(3,1) \ltimes \RR^4$, where $\Ort(3,1)$ is the \textbf{Lorentz
  group}.  Acting on an event $(x,1)$ with the matrix in
equation~\eqref{eq:poin-in-gl5}, we obtain the event $(Lx + v, 1)$,
which is the effect of a Lorentz transformation $(x \mapsto L x)$ and a
(spatiotemporal) translation $x \mapsto x + v$; that is,
\begin{equation}
  \begin{pmatrix}
    L & v \\
    0 & 1
  \end{pmatrix}=
    \begin{pmatrix}
    I & v \\
    0 & 1
  \end{pmatrix}
    \begin{pmatrix}
    L & 0 \\
    0 & 1
  \end{pmatrix}.
\end{equation}
The Lie algebra of the Poincaré group embeds in $\gl(5,\RR)$ as those
matrices of the form
\begin{equation}
  \begin{pmatrix}
    X & v \\
    0 & 0
  \end{pmatrix}
\end{equation}
where $X^T \eta + \eta X = 0$ and $v \in \RR^4$.  Introducing a basis
$L_{AB} = - L_{BA}, P_A$, where now $A,B = 0,1,2,3$, by
\begin{equation}
  \begin{pmatrix}
    X & v \\
    0 & 0
  \end{pmatrix} = \tfrac12 X^{AB} L_{AB} + v^A P_A,
\end{equation}
it is easy to calculate the nonzero Lie brackets:
\begin{equation}\label{eq:poincare-brackets}
  \begin{split}
    [L_{AB},L_{CD}] &= \eta_{BC} L_{AD} - \eta_{AC} L_{BD} - \eta_{BD} L_{AC} + \eta_{AD} L_{BC}\\
    [L_{AB},P_C] &= \eta_{BC} P_A - \eta_{AC} P_B.
  \end{split}
\end{equation}
To ease comparison with the Galilei algebra
\eqref{eq:gal-algebra-brackets}, we will let $P_A = \{H = P_0, P_a\}$ and
$L_{AB} = \{B_a = L_{0a}, L_{ab}\}$, relative to which the brackets become
\begin{equation}
  \label{eq:poincare-kla-brackets}
  \begin{split}
    [L_{ab},L_{cd}] &= \delta_{bc} L_{ad} - \delta_{ac} L_{bd} -  \delta_{bd} L_{ac} + \delta_{bd} L_{ac} \\
    [L_{ab}, B_b] &= \delta_{bc} B_a - \delta_{ac} B_b\\
    [L_{ab}, P_b] &= \delta_{bc} P_a - \delta_{ac} P_b\\
    [B_a, B_b] &= L_{ab}\\
    [B_a, P_b] &= \delta_{ab} H\\
    [B_a, H] &= P_a.
  \end{split}
\end{equation}
We see that again $L_{ab}$ span an $\so(3)$ subalgebra relative to
which $B_a,P_a$ transform according to the three-dimensional vector
representation and $H$ transforms according to the one-dimensional
scalar representation.  What sets the Poincaré and Galilei algebras
apart are the Lie brackets which do not involve the $L_{ab}$: the last
bracket in equation~\eqref{eq:gal-algebra-brackets} and the last three
brackets in equation~\eqref{eq:poincare-kla-brackets}.

\subsection{Lie algebra of symmetries}
\label{sec:lie-algebra-symm}

Minkowski spacetime is a lorentzian manifold, diffeomorphic to $\RR^4$
with lorentzian metric
\begin{equation}
  g = - dt^2 + dx^2 + dy^2 + dz^2 = \eta_{\mu\nu} dx^\mu dx^\nu
\end{equation}
relative to cartesian coordinates $x^\mu = (t,x,y,z)$.  The Poincaré
Lie algebra is isomorphic to the Lie algebra of Killing vector fields
of the metric $g$.  Let $\xi = \xi^\mu \d_\mu$ denote a vector field
of Minkowski spacetime.  It is a Killing vector field if $\eL_\xi g =
0$, which translates into
\begin{equation}
  \eta_{\rho\nu} \d_\mu\xi^\rho + \eta_{\mu\rho} \d_\nu\xi^\rho= 0,
\end{equation}
or, defining $\xi_\mu = \eta_{\mu\rho}\xi^\rho$, into Killing's
equation\footnote{Solutions of equation~\eqref{eq:mink-killing-eqn}
  are the Noether charges for point symmetries of the geodesic
  equation.  Indeed, if we consider the variational problem with
  lagrangian $\eL = \tfrac12 \eta_{\mu\nu} \dot x^\mu \dot x^\nu$ and
  ask which point transformations $\delta x^\mu = \xi^\mu(x)$ leave
  $\eL$ invariant, we find that $\xi^\mu$ must satisfy
  equation~(\ref{eq:mink-killing-eqn}).}:
\begin{equation}\label{eq:mink-killing-eqn}
  \d_\mu \xi_\nu + \d_\nu \xi_\mu = 0.
\end{equation}
Notice that $\d_\mu\d_\nu\xi_\rho$ is clearly
symmetric in $\mu\leftrightarrow\nu$ and, from Killing's equation, also
skewsymmetric in $\nu \leftrightarrow\rho$.  Therefore
$\d_\mu\d_\nu\xi_\rho = 0$ and hence $\xi_\mu = \Lambda_{\mu\nu} x^\nu +
a_\mu$.  Re-inserting this into Killing's equation, we find that
$\Lambda_{\mu\nu} = - \Lambda_{\nu\mu}$ and we may write the general
solution of Killing's equation as
\begin{equation}
  \xi = \tfrac12 \Lambda^{\mu\nu} \xi_{L_{\mu\nu}} + a^\mu \xi_{P_\mu},
\end{equation}
where
\begin{equation}
  \xi_{L_{\mu\nu}}= x_\nu \d_\mu - x_\mu \d_\nu \qquad\text{and}\qquad
  \xi_{P_\mu} = \d_\mu.
\end{equation}
One can check that these vector fields obey the opposite (i.e.,
negative) brackets of those of the Poincaré Lie algebra:

\begin{equation}
  \label{eq:opposite-poincare}
  \begin{split}
    [\xi_{L_{\mu\nu}}, \xi_{L_{\rho\sigma}}] &= - \eta_{\nu\rho}
    \xi_{L_{\mu\sigma}}+ \eta_{\mu\rho} \xi_{L_{\nu\sigma}}+ \eta_{\nu\sigma}
    \xi_{L_{\mu\rho}} - \eta_{\mu\sigma} \xi_{L_{\nu\rho}}\\
    [\xi_{L_{\mu\nu}}, \xi_{P_\rho}] &= - \eta_{\nu\rho} \xi_{P_\mu} + \eta_{\mu\rho} \xi_{P_\nu}.
  \end{split}
\end{equation}
The fact that we have an antihomomorphism of Lie algebras might seem
counter-intuitive, but we will see that it is natural in the
context of homogeneous spaces, where the group action is induced from
left multiplication in the group.  The infinitesimal generators of
left multiplication are the right-invariant vector fields whose Lie
brackets are opposite to those of the left-invariant vector fields
defining the Lie algebra.

In contrast, Galilei spacetime is a non-lorentzian geometry:
there is no invariant metric, but rather an invariant Newton--Cartan
structure.\footnote{Some authors (e.g., \cite{Duval:2014uoa}) refer to
  this structure as a ``weak'' Newton--Cartan structure, reserving the
  unqualified name for the structure which results by an additional
  choice of an adapted connection; that is a connection relative to
  which the clock one-form and the spatial cometric are parallel.}
Relative to cartesian coordinates $(x,y,z,t)$, the clock defines a
one-form $\tau = dt$.  Indeed, as shown in
Figure~\ref{fig:clock-fibration}, the clock is the linear projection
$\RR^4 \to \RR$ taking $b-a$ to $\tau(b-a)$.  This is nothing but the
derivative of the projection $\pi : \Af^4 \to \Af^1$, which in this
model of the affine space is given by $\pi(x,y,z,t) = t$; in other
words, $dt$.  We will see later that in a general Newton--Cartan
manifold, the clock one-form need not be exact or even closed.  The
ruler defines an invariant symmetric $(2,0)$-tensor field
$\lambda = \d_x \otimes \d_x + \d_y \otimes \d_y + \d_y \otimes \d_y$.
Interpreting $\lambda$ as a symmetric bilinear form on one-forms, we
notice that $\lambda$ is degenerate along $dt$.  It is often called
the ``spatial cometric''.  In analogy with a lorentzian spacetime, let
us say that a vector field $\xi$ is ``Killing'', if it preserves the
clock one-form $\tau$ and the spatial cometric $\lambda$; that is,
\begin{equation}\label{eq:gal-killing}
  \eL_\xi \tau = 0 \qquad\text{and}\qquad \eL_\xi \lambda = 0.
\end{equation}
For Galilei spacetime, and introducing coordinates
$x^a = (x,y,z)$, the general solution of
equations~\eqref{eq:gal-killing} is given by
\begin{equation}
  \xi = \alpha \d_t + v^a(t) \d_a + \tfrac12 T^{ab}(t) (x_b \d_a - x_a\d_b),
\end{equation}
where $\alpha \in \RR$ and where $v^a$ and $T^{ab} = - T^{ba}$ are
smooth functions of $t$.  In contrast to the Lie algebra
of isometries of a lorentzian manifold, the Lie algebra of symmetries
of the (weak) Newton--Cartan structure of Galilei spacetime is
infinite-dimensional, and is known as the \textbf{Coriolis algebra}
\cite{Duval:1993pe}.  It contains (the opposite of) the Galilei
algebra as a subalgebra, spanned by
\begin{equation}
  \xi_H = \d_t, \qquad \xi_{P_a} = \d_a, \qquad \xi_{B_a}= t \d_a
  \qquad\text{and}\qquad \xi_{L_{ab}} = - x_a \d_b + x_b \d_a.
\end{equation}
Had we considered (strict) Newton--Cartan structures, including the
adapted connection as part of the data, then the Lie algebra of
symmetries would be finite-dimensional.\footnote{The reason is that
Newton--Cartan structures are Cartan geometries and the infinitesimal
automorphisms of a Cartan geometry form a finite-dimensional Lie
algebra.}

\section{Symmetry}
\label{sec:symmetries}

In Section~\ref{sec:motivation} we discussed two models of the
universe: the Galilei and Minkowski spacetimes.  Both are
four-dimensional affine space homogeneous under the action of a
kinematical Lie group: the Galilei and Poincaré groups, respectively.
In this section we will discuss the notion of a kinematical Lie group
more formally and will discuss the classification of kinematical Lie
algebras.

\subsection{Kinematical Lie algebras}
\label{sec:kinem-lie-algebr}

In a landmark paper \cite{Bacry:1968zf} written more than half a
century ago, Bacry and Lévy-Leblond asked themselves the question of
which were the possible kinematics, rephrasing the question
mathematically as the classification of kinematical Lie algebras.
A careful comparison of the Poincaré and Galilei algebras we
met in Section~\ref{sec:motivation} suggests the following
definition for four-dimensional spacetimes.\footnote{Strictly
  speaking, the definition is for spatially isotropic spacetimes.
  There are generalisations where the rotational subalgebra $\r$ in
the definition is replaced by a Lorentz subalgebra.  Such homogeneous
spaces do occur in nature.  Indeed, as shown in
\cite{Gibbons:2019zfs}, the blow-up of spatial infinity of Minkowski
spacetime is a homogeneous space of the Poincaré group with lorentzian
isotropy.  There are other homogeneous spaces of the Poincaré group
occurring at the asymptotic infinities of Minkowski spacetime, as
discussed in \cite{Figueroa-OFarrill:2021sxz}.}

\begin{definition}\label{def:KLA}
  A \textbf{kinematical Lie algebra} is a ten-dimensional (real) Lie
  algebra $\k$ with generators $L_{ab}=-L_{ba}, B_a, P_a, H$ with $a,b=1,2,3$
  satisfying the following conditions:
  \begin{itemize}
  \item the generators $L_{ab}$ span an $\so(3)$-subalgebra $\r$ of $\g$:
    \begin{equation}\label{eq:gen-kla-1}
      [L_{ab},L_{cd}] = \delta_{bc} L_{ad} - \delta_{ac} L_{bd} -  \delta_{bd} L_{ac} + \delta_{bd} L_{ac},
    \end{equation}
  \item the generators $B_a,P_a$ transform as vectors under $\r$:
    \begin{equation}\label{eq:gen-kla-2}
      \begin{split}
        [L_{ab}, B_b] &= \delta_{bc} B_a - \delta_{ac} B_b\\
        [L_{ab}, P_b] &= \delta_{bc} P_a - \delta_{ac} P_b\\
      \end{split}
    \end{equation}
  \item and the generator $H$ transforms as a scalar:
    \begin{equation}\label{eq:gen-kla-3}
      [L_{ab},H] = 0.
    \end{equation}
  \end{itemize}
\end{definition}
In addition, Bacry and Lévy-Leblond initially also imposed that the
Lie brackets should be invariant under parity $P_a \mapsto -P_a$ and
time-reversal $H \mapsto -H$; although they did point out that those
restrictions were ``by no means compelling'' and indeed twenty years
later, Bacry and Nuyts \cite{MR857383} lifted those conditions
arriving at a classification of four-dimensional kinematical Lie
algebras.  This classification was recovered using deformation theory
in \cite{Figueroa-OFarrill:2017ycu} and extended to arbitrary
dimension in \cite{Figueroa-OFarrill:2017tcy,Andrzejewski:2018gmz}.
The definition of kinematical algebra in $d+1$ dimensions is formally
as the one above, except that $a,b=1,\dots, d$ and the subalgebra $\r$
spanned by $L_{ab}$ is now isomorphic to $\so(d)$.  The case of $d=1$
corresponds to the Bianchi classification of three-dimensional real
Lie algebras \cite{Bianchi,MR1900159}, here re-interpreted as
kinematical Lie algebras for two-dimensional spacetimes.  The cases of
$d=2$ and $d=3$ are the most complicated due to the existence of
$\epsilon_{ab}$ and $\epsilon_{abc}$ which are  $\r$-invariant and can
thus appear in the Lie brackets and, indeed, there are kinematical Lie
algebras in dimension $2+1$ and $3+1$ which have no higher-dimensional
analogues.  We will refer the interested reader to the papers cited
above and will concentrate here on those kinematical Lie algebras
which exist in generic dimensions.

Before we state the classification, let us make an important remark.
Although the notation for the generators of a kinematical Lie algebra
suggests a physical interpretation: namely, $L_{ab}$ generate
rotations, $B_a$ boots, $P_a$ spatial translations and $H$ temporal
translations, it would be imprudent to take this very seriously.  The
physical interpretation of the generators can only be determined once
we realise them geometrically as vector fields in a spacetime.  In the
two examples we have seen in Section~\ref{sec:motivation}, it is
indeed the case that the generators can be interpreted as above, but
this is certainly not true in most cases.

One way to approach the classification is to write down the most
general $\r$-invariant Lie brackets for the generators $B_a,P_a,H$ and
impose the Jacobi identity. The Jacobi identity cuts out an algebraic
variety $\eJ$ in the vector space of possible brackets: i.e., the
vector space of linear maps $\wedge^2 W \to \k$, where $W\subset \k$
is the vector subspace spanned by $B_a, P_a, H$. Two points in $\eJ$
define isomorphic kinematical Lie algebras if and only if they are
related by a change of basis in $W$. We take care of this ambiguity by
quotienting $\eJ$ by the action of the subgroup of $\GL(W)$ which
commutes with the action of $\r$. In practice, one selects a unique
representative for each isomorphism class of kinematical Lie algebras.

Table~\ref{tab:KLAs} lists the kinematical Lie algebras in generic
dimension $d+1$.  For $d \leq 2$, there are some degeneracies (e.g., if
$d=2$, the Galilei algebra $\g$ is isomorphic to the Carroll algebra
$\car$), but for general $d$ the table below lists non-isomorphic
kinematical Lie algebras and for $d>3$ the table is complete.  The
table lists the nonzero Lie brackets except for the common ones in
every kinematical Lie algebra.  It also uses a shorthand notation
omitting indices.  The only $\r$-invariant tensor which can appear is
$\delta_{ab}$ and hence there is an unambiguous way to add indices.
For example, $[H,\B] = \B + \P$ unpacks as $[H,B_a] = B_a + P_a$,
whereas $[\B,\P] = H + \L$ stands for $[B_a, P_b]= \delta_{ab} H +
L_{ab}$, et cetera.  There is no standard notation for all the kinematical
Lie algebras, so we have made some choices.

\begin{table}[h!]
  \centering
  \caption{Kinematical Lie algebras in generic dimension}
  \label{tab:KLAs}
  \rowcolors{2}{red!10}{yellow!10}
  \begin{tabular}{>{$}l<{$}|*{5}{>{$}l<{$}}|l} \toprule
    \multicolumn{1}{c|}{Name} &  \multicolumn{5}{c|}{Nonzero Lie brackets in addition to \eqref{eq:gen-kla-1}--\eqref{eq:gen-kla-3}} & \multicolumn{1}{c}{Comments} \\\midrule
       \s & & & & & & \\
       \g & [H,\B] = - \P & & & & & \\
    \n^0    & [H,\B] = \B + \P & [H, \P] = \P & & & & \\
    \n^+_\gamma  & [H,\B] = \gamma \B & [H,\P] = \P & & & & $\gamma \in [-1,1]$ \\
    \n^-_\chi & [H,\B] = \chi \B + \P & [H,\P] = \chi \P - \B &  & & & $\chi \geq 0$ \\
    \car & & & & [\B,\P] = H & & \\
    \choice{\iso(d,1)}{\iso(d+1)} & [H,\B] = -\varepsilon \P & &  [\B,\B]= \varepsilon \L & [\B,\P] = H & & $\varepsilon = \pm 1$ \\
    \so(d+1,1) & [H,\B] = \B & [H,\P] = -\P & &  [\B,\P] = H + \L & & \\
    \choice{\so(d,2)}{\so(d+2)} & [H,\B] = -\varepsilon \P & [H,\P] = \varepsilon \B &  [\B,\B]= \varepsilon \L & [\B,\P] = H &  [\P,\P] = \varepsilon \L & $\varepsilon = \pm 1$ \\ \bottomrule
  \end{tabular}
\end{table}

We now describe each of the algebras in turn:
\begin{itemize}
\item The Lie algebra $\s$ is the \textbf{static} kinematical Lie
  algebra: all additional brackets are zero.  Therefore every
  kinematical Lie algebras is a deformation of $\s$.

\item The Galilei algebra is denoted $\g$ and we have denoted by
  $\n^0$ a closely related algebra.  In $\g$ and $\n^0$, the adjoint
  action of $H$ is not diagonalisable over the complex numbers, but
  has a nontrivial Jordan block:
  \begin{equation}
    \ad_H^{\g}
    \begin{pmatrix} \B \\ \P \end{pmatrix} = \begin{pmatrix}
      0 & -1 \\ 0 & 0
    \end{pmatrix} \begin{pmatrix} \B \\ \P
    \end{pmatrix}\qquad\text{and}\qquad
    \ad_H^{\n^0}
    \begin{pmatrix} \B \\ \P \end{pmatrix} = \begin{pmatrix}
      1 & 1 \\ 0 & 1
    \end{pmatrix} \begin{pmatrix} \B \\ \P
    \end{pmatrix}.
  \end{equation}

\item There are two one-parameter families of algebras: $\n^+_\gamma$,
  with $\gamma \in [-1,1]$, which for $\gamma = -1$ is one of the two
  \textbf{Newton--Hooke} algebras; and $\n^-_\chi$, with
  $\chi \geq 0$, which for $\chi = 0$ is the other Newton--Hooke
  algebra. These two families correspond to the cases where the
  adjoint action of $H$ is diagonalisable over the complex numbers: in
  $\n^+_\gamma$, the eigenvalues are real, whereas in $\n^-_\chi$ they
  are complex: \begin{equation}
  \ad_H^{\n^+}
  \begin{pmatrix} \B \\ \P \end{pmatrix} = \begin{pmatrix}
    \gamma & 0 \\ 0 & 1
  \end{pmatrix} \begin{pmatrix} \B \\ \P
  \end{pmatrix}\qquad\text{and}\qquad
  \ad_H^{\n^-}
  \begin{pmatrix} \B \\ \P \end{pmatrix} = \begin{pmatrix}
    \chi & 1 \\ -1 & \chi
  \end{pmatrix} \begin{pmatrix} \B \\ \P
  \end{pmatrix}.
\end{equation}

\item The Carroll algebra is denoted $\car$.

\item The Poincaré algebra is $\iso(d,1)$ and the euclidean algebra is
  $\iso(d+1)$.

\item The remaining algebras are semisimple (for $d\geq2$) and consist
  of $\so(d+2)$, $\so(d+1,1)$ and $\so(d,2)$.  Finite-dimensional
  semisimple Lie algebras are rigid, so they cannot be deformed
  further.  However they can be contracted.  Not all the kinematical
  Lie algebras in the table can be obtained as contractions of the
  simple ones: those which can are the Poincaré, euclidean, (both)
  Newton--Hooke, Galilei, Carroll and static algebras.  These are
  precisely the algebras which admit parity and time-reversal
  automorphisms; that is, the ones originally classified in
  \cite{Bacry:1968zf}.
\end{itemize}

\subsection{Aristotelian Lie algebras}
\label{sec:arist-lie-algebr}

A closely related family of Lie algebras are the \textbf{aristotelian
  algebras}, defined just like in Definition~\ref{def:KLA}, but
dropping the boosts.

\begin{definition}\label{def:ALA}
  An \textbf{aristotelian Lie algebra} is a real Lie
  algebra $\a$ with generators $L_{ab}=-L_{ba}, P_a, H$, with
  $a,b=1,\dots, d$, satisfying the following conditions:
  \begin{itemize}
  \item the generators $L_{ab}$ span an $\so(d)$-subalgebra $\r$ of $\g$:
    \begin{equation}\label{eq:aristo-gen-1}
      [L_{ab},L_{cd}] = \delta_{bc} L_{ad} - \delta_{ac} L_{bd} -  \delta_{bd} L_{ac} + \delta_{bd} L_{ac},
    \end{equation}
  \item the generators $P_a$ transform as vectors under $\r$:
    \begin{equation}\label{eq:aristo-gen-2}
        [L_{ab}, P_b] = \delta_{bc} P_a - \delta_{ac} P_b
    \end{equation}
  \item and the generator $H$ transforms as a scalar:
    \begin{equation}\label{eq:aristo-gen-3}
      [L_{ab},H] = 0.
    \end{equation}
  \end{itemize}
\end{definition}

Aristotelian Lie algebras are easy to classify in any dimension and
the result is contained in
\cite[Appendix~B]{Figueroa-OFarrill:2018ilb} and summarised in the
Table~\ref{tab:ALAs}, which lists the nonzero Lie brackets in addition
to those fixed by the definition.  We omit aristotelian Lie algebras
which do not exist in general dimension.

\begin{table}[h!]
  \centering
  \caption{Aristotelian Lie algebras}
  \label{tab:ALAs}
  \rowcolors{2}{red!10}{yellow!10}
  \begin{tabular}{>{$}l<{$}|*{2}{>{$}l<{$}}|l}\toprule
    \multicolumn{1}{l|}{Name~~~~~~~} & \multicolumn{2}{c|}{Nonzero Lie brackets}& \multicolumn{1}{c}{Comments}\\\midrule
    \iso(d) \oplus \RR & & & \\
    \fsim(d) & [H,P_a] = P_a & &\\
    \choice{\so(d,1)\oplus \RR}{\so(d+1)\oplus \RR} & & [P_a,P_b] = \varepsilon L_{ab} & $\varepsilon = \pm 1$\\
    \bottomrule
  \end{tabular}
\end{table}

Let us describe each of the aristotelian Lie algebras in turn:
\begin{itemize}
\item The aristotelian Lie algebra with no additional nonzero Lie
  brackets, which we could term the ``static'' aristotelian Lie
  algebra, is isomorphic to $\iso(d) \oplus \RR$, with $\iso(d)$
  spanned by $L_{ab}, P_a$ and the one-dimensional Lie subalgebra
  spanned by $H$, which is central.

\item If instead of being central, we think of $H$ as dilatations, we
  obtain a Lie algebra isomorphic to the similitude algebra of
  $d$-dimensional euclidean space: $\fsim(d)$.  This is also denoted
  $\co(d) \ltimes \RR^d$, where $\co(d) = \so(d) \oplus \RR$ is the
  extension of the rotation algebra by dilatations and $\RR^d$
  transforms as a vector under rotations but with nonzero conformal
  weight.

\item If $H$ remains central, but now the translations do not commute,
  we obtain trivial central extensions of $\so(d,1)$ or $\so(d+1)$.
\end{itemize}

\subsection{Central  extensions}
\label{sec:central-extensions}

Central extensions of Lie algebras arise naturally in Physics.  In
quantum Physics they arise due to the fact that the state space of a
quantum system is a projective space (the space of rays of a Hilbert
space) so that the action of a group $\Ggr$ on the projective space
may only lift to a projective representation on the Hilbert space, and
hence an honest representation of a one-dimensional central extension
of $\Ggr$.  In classical Physics they arise due to the fact that
homogeneous symplectic manifolds of a Lie group $\Ggr$ are (up to
covering) coadjoint orbits of $\Ggr$ or perhaps a one-dimensional
central extension of $\Ggr$, as we will discuss in
Section~\ref{sec:coadjoint-orbits}.

Mathematically, a central extension of a Lie algebra $\k$ is a special
case of a Lie algebra extension.  A Lie algebra $\widetilde\k$ is said
to be a an \textbf{extension} of a Lie algebra $\k$ by a Lie algebra
$\a$ if they fit in an exact sequence of Lie algebras
\begin{equation}
  \label{eq:extension}
  \begin{tikzcd}
    0 \arrow[r] & \a \arrow[r] & \widetilde\k \arrow[r] & \k \arrow[r]
    & 0,
  \end{tikzcd}
\end{equation}
This is equivalent to the following conditions: $\widetilde\k = \k
\oplus \a$ as a vector space, $\a$ is an ideal of
$\widetilde\k$ (i.e., $[\widetilde\k,\a] \subset \a$) and the quotient
Lie algebra $\widetilde\k/\a$ is isomorphic to $\k$.  If $\a$ is
central, so that $[\a,\widetilde\k]=0$, then we have a \textbf{central
extension}.  Notice that $\k$ is not necessarily a Lie subalgebra of
$\widetilde\k$.  If that is the case, the sequence is said to be split
and we have that $\widetilde\k$ is the semidirect product of $\k$ with
$\a$.  A special case of semidirect products are the trivial
extensions, when $\widetilde\k = \k \oplus \a$ as a Lie algebra; that
is, $\k$ and $\a$ are subalgebras (actually ideals) and $[\k,\a] =
0$.

Whereas every Lie algebra admits trivial extensions, the only
kinematical Lie algebras in Table~\ref{tab:KLAs} admitting nontrivial
central extensions (in dimension $d>2$) are the static ($\s$),
Newton--Hooke ($\n^\pm$) and Galilei ($\g$) algebras.  To describe
them, we introduce a new generator $Z$ with $[Z,-] = 0$ and modify the
Lie brackets of the kinematical Lie algebra by
$[B_a,P_b] = \delta_{ab} Z$.  The central extension of the Galilei
algebra is called the \textbf{Bargmann algebra}.

A one-dimensional extension (not necessarily central) of a
kinematical Lie algebra is called a \textbf{generalised Bargmann
  algebra}.  Apart from the central extensions listed above and the
trivial extensions, there is a small list (some with parameters).
Those for which $[B_a,P_b] = \delta_{ab} Z$ are deformations of the
central extension of the static kinematical Lie algebra and have been
classified in \cite{Figueroa-OFarrill:2017ycu} (for $d=3$) and in
\cite{Figueroa-OFarrill:2017tcy} (for $d>3$).  Those for which
$[B_a,P_b]=0$ are listed here for the first time.

Table~\ref{tab:gen-bargmann} lists the (nontrivial) generalised
Bargmann algebras in dimension $d>2$.  In the table $\k$ stands for
the kinematical Lie algebra being extended and the brackets listed are
the ones which involve the additional generator $Z$, so they are
either new or modifications of the brackets in $\k$.  The (nonzero)
parameter $\alpha$ in the last three rows is effective: different
values of $\alpha$ give non-isomorphic Lie algebras.

\begin{table}[h!]
  \setlength{\tabcolsep}{3pt}
  \centering
  \caption{Generalised Bargmann algebras in $d>2$}
  \label{tab:gen-bargmann}
  \setlength{\extrarowheight}{2pt}
  \rowcolors{2}{red!10}{yellow!10}
  \begin{tabular}{>{$}l<{$}|*{2}{>{$}l<{$}}|l}
    \k  & \multicolumn{2}{c|}{Brackets involving $Z$} &  \multicolumn{1}{c}{Comments} \\
    \toprule
    \s  & [\B,\P] = Z & & \\
    \n^+ & [\B,\P] = Z & & \\
    \n^- & [\B,\P] = Z & & \\
    \g & [\B,\P] = Z & & \\
    \midrule
    \n^+_\gamma & [\B,\P] = Z & [H,Z] = (\gamma+1) Z & $\gamma\in(-1,1]$\\
    \n^0 & [\B,\P] = Z & [H,Z] = 2 Z & \\
    \n^-_\chi & [\B,\P] = Z & [H,Z] = 2 \chi Z & $\chi > 0$\\
    \midrule
    \s & & [H,Z] = Z & \\
    \g & & [H,Z] = Z & \\
    \n^+_\gamma & & [H,Z] = \alpha Z & $\gamma \in [-1,1]$ and $\alpha \neq 0$\\
    \n^0 & & [H,Z] = \alpha Z & $\alpha \neq 0$\\
    \n^-_\chi & & [H,Z] = \alpha Z & $\chi \geq 0$ and $\alpha \neq 0$\\
    \bottomrule
  \end{tabular}
\end{table}

Aristotelian Lie algebras (if $d>2$) admit no nontrivial central
extensions: there are no $\r$-invariant cochains, let alone cocycles.
They do, however, admit nontrivial non-central extensions, which are
listed in Table~\ref{tab:ext-aristo}, which lists the aristotelian Lie
algebra being extended and the brackets involving the additional
generator $Z$.  Again the (nonzero) parameter $\alpha$ is effective.

\begin{table}[h!]
  \setlength{\tabcolsep}{3pt}
  \centering
  \caption{One-dimensional extensions of aristotelian Lie algebras in $d>2$}
  \label{tab:ext-aristo}
  \setlength{\extrarowheight}{2pt}
  \rowcolors{2}{red!10}{yellow!10}
  \begin{tabular}{>{$}l<{$}|>{$}l<{$}|l}
    \multicolumn{1}{c|}{$\a$}  & \multicolumn{1}{c|}{Brackets involving $Z$} &  \multicolumn{1}{c}{Comments} \\
    \toprule
    \iso(d) \oplus \RR & [H,Z] = Z & \\
    \fsim(d) & [H,Z] = \alpha Z & $\alpha \neq 0$ \\
    \choice{\so(d,1)\oplus \RR}{\so(d+1)\oplus \RR} & [H,Z] = Z & \\
    \bottomrule
  \end{tabular}
\end{table}

Changing notation: $(H,Z) \mapsto (D,H)$, the Lie algebras in
Table~\ref{tab:ext-aristo} are examples of Lifshitz Lie algebras (see,
e.g., \cite{Figueroa-OFarrill:2022kcd}).  The extension of $\fsim(d)$
is the original Lifshitz algebra, where the parameter $\alpha$ is
typically denoted $z$:
\begin{equation}
  [D,\P] = \P  \qquad\text{and} \qquad[D,H] = z H,
\end{equation}
in addition to the brackets
\eqref{eq:aristo-gen-1}--\eqref{eq:aristo-gen-3} common to all
aristotelian Lie algebras.

\section{Geometry}
\label{sec:geometries}

In this section we discuss non-lorentzian geometries.  In the spirit
of Klein's Erlangen Programme, we start by discussing the homogeneous
spacetimes associated to the kinematical Lie algebras discussed in
Section~\ref{sec:symmetries}.  We will see that these spaces fall
into different families depending on the structure which the
kinematical group preserves: metric (riemannian or lorentzian),
Newton--Cartan, carrollian or aristotelian.  Each of these geometries
is a Cartan geometry modelled on a homogeneous spacetime and we will
discuss them in turn.  In a sense, all we are doing is extending to
the non-lorentzian context the standard sequence of ideas:
\begin{equation}
  \text{Poincaré symmetry} \longrightarrow \text{Minkowski spacetime}
  \longrightarrow \text{lorentzian geometry}.
\end{equation}
Of course, Minkowski spacetime is not the only homogeneous space of
the Poincaré group, so the passage from the Poincaré group to
Minkowski spacetime requires a choice, whereas the passage from
Minkowski spacetime to lorentzian geometry is more or less forced.

\subsection{Homogeneous spaces}
\label{sec:homogeneous-spaces}

In this section we review the basic notions of homogeneous geometry.

\subsubsection{Group actions on manifolds}
\label{sec:group-acti-manif}

Let $\Ggr$ be a Lie group.  A (linear) representation of $\Ggr$ on a
vector space $V$ is a Lie group homomorphism $\rho: \Ggr \to \GL(V)$;
that is, $\rho$ is a smooth map and a group homomorphism
$\rho(ab) = \rho(a)\rho(b)$ for all $a,b \in \Ggr$.  We are also
interested in nonlinear realisations\footnote{In Physics it is
  customary to reserve the name ``nonlinear realisation'' only to
  transitive actions (see later), when $M$ is diffeomorphic to a coset
  space $\Ggr/\Hgr$.  A manifold admitting a transitive action of a
  Lie group is the nonlinear analogue of an \emph{irreducible}
  representation.  In the same way that it is useful to consider
  representations which are not necessarily irreducible, we shall
  consider nonlinear realisations where the action is not necessarily
  transitive.} of $\Ggr$ on a manifold $M$.  It would be tempting by
analogy with the case of a linear representation to define a nonlinear
realisation as a Lie group homomorphism $\rho : \Ggr \to \Diff(M)$,
except for the fact that the diffeomorphism group $\Diff(M)$ of a
manifold is not typically a Lie group.  Instead we define nonlinear
realisations as \emph{actions}. One has to distinguish between left
and right actions; although it is easy to go between them.  By a
\textbf{(left) action} of $\Ggr$ on a manifold $M$ we mean a smooth
map $\alpha : \Ggr \times M \to M$, written simply as $\alpha(g,p) = g
\cdot p$, satisfying two properties:
\begin{itemize}
\item for all $g_1,g_2 \in \Ggr$ and $p \in M$, $(g_1 g_2)\cdot p = g_1
  \cdot (g_2 \cdot p)$; and

\item for all $p \in M$, $e \cdot p = p$ where $e \in \Ggr$ is the
  identity element.
\end{itemize}
If we fix $g \in \Ggr$, $\alpha(g,-): M \to M$ is a diffeomorphism which
we typically denote $\alpha_g$.  On the other hand, if we fix $p \in
M$, we get a map $\alpha(-,p): \Ggr \to M$ known as the \textbf{orbit
  map}, as its image is the orbit of $p$ under $\Ggr$.

Let $\g$ denote the Lie algebra of $\Ggr$.  An action of $\Ggr$ on $M$ gives
rise to a Lie algebra antihomomorphism $\xi: \g \to \eX(M)$, assigning to
every $X \in \g$ a vector field $\xi_X$ and such that for all $X,Y \in
\g$, $[\xi_X,\xi_Y] = - \xi_{[X,Y]}$, where the bracket on the LHS is
the Lie bracket of vector fields and that on the RHS is the bracket on
$\g$.  The vector fields in the image of $\xi$ are called the
\textbf{fundamental vector fields} of the group action.
Although one can redefine the fundamental vector fields in such a way
that the new map $\g \to \eX(M)$ is a Lie algebra homomorphism, it
turns out not to be natural, as we will see shortly.

A group action $\alpha : \Ggr \times M \to M$ is said to be
\textbf{effective} if the only element $g \in \Ggr$ which acts trivially
(i.e., which obeys $g \cdot p = p$ for all $p \in M$) is the identity
element.  A weaker condition is for the action to be \textbf{locally
  effective}, which says that the elements of $\Ggr$ which act trivially
form a discrete subgroup of $\Ggr$.  This is equivalent to the map $\xi:
\g \to \eX(M)$ being injective, so that no nonzero element in $\g$ is
sent to the zero vector field.

A group action $\alpha : \Ggr \times M \to M$ is said to be
\textbf{transitive} if given any two points $p,q \in M$, there is some
$g \in \Ggr$ with $q = g \cdot p$.  Equivalently, if the $\Ggr$-orbit of any
point is the whole manifold.  This is the analogue for nonlinear
realisations of irreducibility for linear representations.  A linear
representation of $\Ggr$ on $V$ is irreducible if there are no proper
subspaces of $V$ which are stable under $\Ggr$.  Similarly, an action of
$\Ggr$ on $M$ is transitive if there are no proper submanifolds of $M$
stable under the action of $\Ggr$.

A manifold $M$ is said to be a \textbf{homogeneous space} of a Lie
group $\Ggr$ if $\Ggr$ acts transitively on $M$.  The
\textbf{stabiliser subgroup} of a point $p \in M$ is the subgroup
$\Hgr \subset \Ggr$ which fixes $p$:
$\Hgr = \left\{g \in \Ggr \middle | g \cdot p = p \right\}$.  It is a
closed subgroup of $\Ggr$.  Its Lie algebra $\h$ consists of those
fundamental vector fields which vanish at $p$.  If $M$ is a
homogeneous space of $\Ggr$, then the stabiliser subgroups of all of
its points are conjugate in $\Ggr$.  Indeed, let $\Hgr_p$ denote the
stabiliser subgroup of $p \in M$ and $\Hgr_q$ that of $q \in M$.
Since $\Ggr$ acts transitively, there is some $g \in \Ggr$ such that
$q = g \cdot p$ and hence $h \in \Hgr_q$ if and only if
$h = g h' g^{-1}$ for some $h' \in \Hgr_p$. Often one picks an
``origin'' $o \in M$ and lets $\Hgr$ denote the stabiliser subgroup of
$o$.  Then $M$ is diffeomorphic to the space of left cosets
$\Ggr/\Hgr$.  This is why homogeneous spaces are often referred to as
\emph{coset spaces} or \emph{coset manifolds}.  Of course the choice
of origin is immaterial, since from the point of $\Ggr$ all points in
a homogeneous space ``look the same''.

It is not just that $M$ and $\Ggr/\Hgr$ are diffeomorphic, but that they are
$\Ggr$-equivariantly so: the diffeomorphism $M \to \Ggr/\Hgr$ intertwines
between the left action of $\Ggr$ on $M$ and the left action of $\Ggr$ on
$\Ggr/\Hgr$ which is induced from left multiplication in $\Ggr$:
if $g'\Hgr \in \Ggr/\Hgr$ and $g \in \Ggr$, we have that $g \cdot g'\Hgr =
(gg')\Hgr$.
Now recall that the vector fields which generate left multiplication
on $\Ggr$ are the right-invariant vector fields and they satisfy the
opposite Lie algebra.  This explains why it is natural for the map
$\xi : \g \to \eX(M)$ to be an antihomomorphism.

\subsubsection{Linear isotropy representation and invariant tensors}
\label{sec:line-isotr-repr}

Let $M$ be a homogeneous space of $\Ggr$ and $\Hgr \subset \Ggr$ the stabiliser
of the origin $o \in M$.  Since every $h \in \Hgr$ preserves $o$, the
derivative at $o$ of the diffeomorphism $\alpha_h : M \to M$ defines a
linear transformation $\lambda(h)$ of the tangent space $T_oM$.  Since
$\alpha$ is an action, in particular, $\alpha_{h_1} \circ \alpha_{h_2}
= \alpha_{h_1h_2}$ for all $h_1,h_2 \in \Hgr$ and, by the chain rule,
$\lambda : \Hgr \to \GL(T_oM)$ is a representation, known as the
\textbf{linear isotropy representation}.

The linear isotropy representation plays a very important rôle in
determining the $\Ggr$-invariant tensor fields on a homogeneous space
$M$.  An important result, which is a special case of the
\emph{fundamental principle of holonomy} (see, e.g.,
\cite[Para.~10.19]{Besse} in the riemannian case, but holds more
generally for any connection), states that there is a one-to-one
correspondence between $\Hgr$-invariant tensors on $T_oM$ and
$\Ggr$-invariant tensor fields on $M$.  Briefly, it goes as follows.
If $\Phi$ is a $\Ggr$-invariant tensor field on $M$, its value at the
origin is a tensor $\Phi_o$ on $T_oM$ which is invariant under the
linear isotropy representation of $\Hgr$.  Conversely, given an
$\Hgr$-invariant tensor $\Phi_o$ on $T_oM$ we may extend it to a
tensor field on $M$ via the $\Ggr$ action.  Its value $\Phi_p$ at
$p \in M$ is defined by picking $g \in \Ggr$ with $g \cdot o = p$ and
acting on $\Phi_o$ with $g$: $\Phi_p = g \cdot \Phi_o$.  The problem
is that there is typically not a unique $g \in \Ggr$ connecting $o$ to
$p$, so which one do we choose?  It turns out that the choice is
immaterial: if $g' \in \Ggr$ is any other such element, then
$g' = g h$ for some $h \in \Hgr$ and precisely because $\Phi_o$ is
$\Hgr$-invariant, $g'\cdot \Phi_o = g \cdot \Phi_o$ and it does not
matter whether we use $g$ or $g'$ to calculate $\Phi_p$.

If in addition, $\Hgr$ is a connected subgroup with Lie algebra $\h
\subset \g$, then $\Ggr$-invariant tensor fields on $M$ are in one-to-one
correspondence with $\h$-invariant tensors on $T_oM$.  Determining the
$\h$-invariant tensors is a reasonably simple linear algebra problem
in most cases.

\subsubsection{Klein pairs}
\label{sec:klein-pairs}

Let $M$ be a homogeneous space of $\Ggr$ with typical stabiliser $\Hgr$.
Let $\g$ and $\h$ denote the Lie algebras of $\Ggr$ and $\Hgr$,
respectively.  Then we may associate to $M$ the \textbf{Klein pair}
$(\g,\h)$.  Not every pair $(\g,\h)$ consisting of a Lie algebra $\g$
and a Lie subalgebra $\h$ is a Klein pair.  It has to be
\emph{geometrically realisable}, which says that there exists some Lie
group $\Ggr$ with Lie algebra $\g$ such that the connected subgroup $\Hgr$
generated by $\h$ is closed.   As explained, for example, in
\cite[Appendix~B]{Figueroa-OFarrill:2018ilb}, there is a one-to-one
correspondence between (effective, geometrically realisable) Klein
pairs and simply-connected homogeneous spaces of $\Ggr$.  (See
\cite[Appendix~B.3]{Figueroa-OFarrill:2018ilb} for a simple example of
a Klein pair which is not geometrically realisable.)  Paraphrasing
slightly, (effective, geometrically realisable) Klein pairs classify
homogeneous spaces up to covering, in the same way that Lie algebras
classify Lie groups up to covering.

There is a notion of isomorphism between Klein pairs which is crucial
in classifications.  We say that two Klein pairs $(\g_1,\h_1)$ and
$(\g_2, \h_2)$ are \emph{isomorphic}, if there is a Lie algebra
isomorphism $\varphi: \g_1 \to \g_2$ with $\varphi(\h_1)= \h_2$.
Isomorphic Klein pairs, if geometrically realisable, give rise to
locally isomorphic homogeneous spaces.

Let $M$ be a homogeneous $\Ggr$-space with Klein pair $(\g,\h)$.  We say
that the Klein pair is \textbf{reductive} if there exists a
complementary subspace $\m$ with $\g = \h \oplus \m$ which is stable
under the restriction to $\Hgr$ of the adjoint action of $\Ggr$ on $\g$.  If
$\Hgr$ is connected, reductivity says that $[\h,\m] \subset \m$.  In
the reductive case, the vector space isomorphism $T_oM \cong \m$
intertwines between the linear isotropy representation on $T_oM$ and
the restriction to $\Hgr$ of the $\Ggr$-adjoint representation on
$\m$.  In the non-reductive case, there is a vector space isomorphism
$T_oM \cong \g/\h$, where the quotient vector space $\g/\h$ is
naturally a representation of $\Hgr$.  In practice we work with
$\g/\h$ by working with $\g$ and just dropping any terms belonging to
$\h$ at the end.

A reductive Klein pair $(\g = \h \oplus \m,\h)$ is said to be
\textbf{symmetric} if $[\m,\m] \subset \h$.  Symmetric Klein pairs are
the infinitesimal description (up to coverings) of symmetric spaces.

It may be convenient to write the reductive and symmetry conditions in
a basis.  Let $X_i$ denote a basis for $\h$.  The Klein pair $(\g,\h)$
is reductive if we can complete to a basis $X_i, Y_I$ for $\g$ such
that $[X_i, Y_I] = c_{iI}{}^J Y_J$; that is, no $X_i$ appear in the
RHS.  For a reductive Klein pair, $[Y_I,Y_J] = c_{IJ}{}^i X_i+
c_{IJ}{}^K Y_K$ in general, but if it is symmetric then $c_{IJ}{}^K =
0$ and hence $[Y_I,Y_J] = c_{IJ}{}^i X_i$.

\subsubsection{Exponential coordinates}
\label{sec:expon-coord}

Let us now discuss coordinates on homogeneous spaces, but first we
review exponential coordinates on a Lie group.

In the neighbourhood of any point $g$ in a Lie group $\Ggr$, we have
exponential coordinates associated to every choice of basis for the
Lie algebra $\g$.  Recall that the exponential map $\exp: \g \to \Ggr$,
which for a matrix Lie group is just the matrix exponential, is a
diffeomorphism between a neighbourhood of $0 \in \g$ and a
neighbourhood of the identity $e \in \Ggr$.  Let $X_1,\dots,X_n$ be a
basis for $\g$ and consider $\exp(x^1 X_1 + \cdots x^n X_n) \in \Ggr$.
The $(x^1,\dots,x^n)$ are local coordinates for $\Ggr$ centred at the
identity, which has coordinates $(0,\dots,0)$.  We may now use left
(or right) multiplication to give coordinates in a neighbourhood of
any other $g \in \Ggr$; for example, $g \exp(y^1 X_1 + \cdots y^n X_n)$
give local coordinates for $\Ggr$ near $g$.  On overlaps, the change of
coordinates between these exponential coordinates is real analytic,
which shows that Lie groups are not just smooth but actually real
analytic manifolds.

Now let us consider a homogeneous space $M \cong \Ggr/\Hgr$ with Klein pair
$(\g,\h)$.  Recall that the identification of $M$ with $\Ggr/\Hgr$ implies a
choice of origin $o \in M$ (corresponding to the identity coset) with
stabiliser $\Hgr$.  Let us choose a vector space complement to $\h$ in
$\g$ and write $\g = \h \oplus \m$.  In the reductive case, we can
(and will) choose $\m$ so that $[\h,\m] \subset \m$, but in general
this may not be possible.  Choosing a basis $Y_1,\dots,Y_m$ for $\m$
we obtain local coordinates near the origin on $M$ by
$\exp(x^1 Y_1 + \cdots + x^m Y_m) \cdot o$ or, having identified $M$
with $\Ggr/\Hgr$, by $\exp(x^1 Y_1 + \cdots + x^m Y_m)\Hgr$.  In general we can
only hope for local coordinates. Indeed, a coset representative is a
choice of section of the principal $\Hgr$-bundle $\Ggr \to \Ggr/\Hgr$ and
principal bundles admit sections if and only if they are trivial.

In effect, what we are doing is choosing a \textbf{coset representative}
(here $\exp(x^1 Y_1 + \cdots + x^m Y_m) \in \Ggr$) for each coset in
$\Ggr/\Hgr$ near the identity coset.  This is a locally defined smooth map
$M \to \Ggr$, which is only defined in a neighbourhood of the origin, and
we may use it to pull back differential forms on $\Ggr$ to $M$.  Every
Lie group $\Ggr$ has a distinguished $\g$-valued one-form $\vartheta \in
\Omega^1(\Ggr,\g)$: the left-invariant Maurer--Cartan one-form.  If we
identify $\g = T_eG$ with the tangent space at the identity, then
$\vartheta_g : T_g \Ggr \to T_e \Ggr$ is simply the differential of left
multiplication by $g^{-1}$.  We may use a (local) coset representative
$L : M \to \Ggr$ to pull back $\vartheta$ to a local one-form
$L^*\vartheta$ on $M$ which, for $\Ggr$ a matrix group, has the simpler
expression
\begin{equation}
  L^*\vartheta = L^{-1}dL.
\end{equation}
Although this is strictly speaking only valid for $\Ggr$ a matrix group,
one does not go wrong by assuming we are in a matrix group for
calculations provided that in the end we express the final result in a
way that makes sense for a general Lie group.  For example, it follows
from the above expression that $L^*\vartheta$ is left-invariant, since
if we multiply $L(x)$ by a constant group element $g$ on the left, it
remains invariant:
\begin{equation}
  (gL)^{-1} d(gL) = L^{-1} g^{-1} g dL = L^{-1} dL.
\end{equation}
Similarly, differentiating again, we find that
\begin{equation}
  d(L^{-1}dL) = dL^{-1} \wedge dL = -L^{-1}dL \wedge L^{-1} dL =
  -\tfrac12 [L^{-1}dL, L^{-1}dL],
\end{equation}
which is the Maurer--Cartan structure equation.  Notice that in the
equation above we wrote the term $L^{-1}dL \wedge L^{-1}dL$ which
involves matrix multiplication as a commutator $\tfrac12 [L^{-1}dL,
L^{-1}dL]$, which makes sense (as the Lie bracket in the Lie algebra
$\g$) for $\Ggr$ any group, not necessarily a matrix
group.\footnote{Whereas Ado's theorem says that any finite-dimensional
real Lie algebra is a matrix Lie algebra, the similar result for Lie
groups is false.  The simplest counterexample to the putative Lie
group version of Ado's theorem is the universal cover of
$\SL(2,\RR)$. See, for example, Graeme Segal's lectures in
\cite{MR1356712}.}

Suppose that $\Ggr/\Hgr$ is reductive, so that $\g = \h \oplus \m$
with $[\h,\m]\subset \m$.  Then we can split the pull-back of the
Maurer--Cartan form into its components along $\h$ and $\m$:
\begin{equation}
  L^{-1} dL = (L^{-1}dL)_\h + (L^{-1}dL)_\m = \omega + \theta,
\end{equation}
where $\omega$, the component along $\h$ is a connection one-form and
$\theta$, the component along $\m$, is a soldering form.  Indeed,
under right multiplication by a local $\Hgr$ transformation $L \mapsto
L h^{-1}$,
\begin{equation}
  L^{-1}dL \mapsto (Lh^{-1})^{-1} d(Lh^{-1}) = h (L^{-1}dL) h^{-1} + h dh^{-1} =
  h\omega h^{-1} - dh h^{-1} + h \theta h^{-1},
\end{equation}
so that comparing the $\h$ and $\m$ components, we arrive at
\begin{equation}
  \omega \mapsto  h\omega h^{-1} - dh h^{-1} \qquad\text{and}\qquad
  \theta \mapsto h \theta h^{-1}.
\end{equation}
If $\Ggr/\Hgr$ is not reductive, there is no natural split of the
Maurer--Cartan one-form.  We can still project to $\g/\h$ to obtain a
soldering form, but there is no uniquely defined component along $\h$
and, moreover, no such component can be chosen in such a way that
results in a connection.

Let us consider in this light the examples in
Section~\ref{sec:motivation}: the Galilei and Minkowski spacetimes.
Both spacetimes are homogeneous spaces of the translation subgroup: in
fact, they are principally homogeneous spaces since every point has
trivial stabiliser.  However we wish to view them as homogeneous
spaces of their kinematical Lie groups: the Galilei and Poincaré
groups, respectively.  Let us start with Minkowski spacetime, which
should be more familiar.  We work now in general dimension $d+1$.

\subsubsection{Minkowski spacetime as a homogeneous space}
\label{sec:mink-spac}

The Poincaré algebra is given in
equation~\eqref{eq:poincare-brackets}.  Let $\h$ be the span of
$L_{AB}$ and $\m$ the span of $P_A$, where $A,B =0,\dots,d$.  Then it
follows that the Klein pair $(\g,\h)$ is reductive.  Here we do have a
global coset representative $L(x) = \exp(x^A P_A)$, which gives global coordinates
to Minkowski spacetime.  We can pull-back the Maurer--Cartan form and
we get
\begin{equation}
  L^{-1} dL = dx^A P_A.
\end{equation}
We see that here $L^{-1}dL$ takes values in $\m$.  This is very
special.  In general for a reductive Klein pair $(\g,\h)$, the
pull-back of the Maurer--Cartan one-form takes values in $\g$: so it
has an $\h$-component and an $\m$-component.  The $\h$-component is a
connection one-form whereas the $\m$-component is a soldering form
(i.e., a coframe or an inverse vielbein).  Here we see that the
connection one-form is absent.  We can explain this as follows.  The
linear isotropy representation of $\h$ on $\m$ admits an invariant
symmetric inner product $\eta \in \odot^2\m^*$, where $\odot$
denotes the symmetric tensor product, with entries $\eta(P_A,P_B) =
\eta_{AB}$ of lorentzian signature.  We can apply this to $L^{-1}dL$
to obtain a Poincaré-invariant metric on Minkowski spacetime:
\begin{equation}
  \eta(L^{-1}dL, L^{-1}dL) = \eta_{AB}dx^A dx^B,
\end{equation}
which is nothing else but the standard Minkowski metric in flat
coordinates.  Of course, relative to flat coordinates, the connection
one-form (relative to the coordinate frame) vanishes, which explains
why there is no $\h$-component in $L^{-1}dL$.

The action of the Poincaré group on Minkowski spacetime relative to
these coordinates is easy to work out, since it is induced by left
multiplication in the Poincaré group.  Translations just shift the
coordinates:\footnote{This is not usually so simple, particularly in
  exponential coordinates the way we have defined them.  In some
  examples, calculations are simpler in modified exponential
  coordinates where we take product of exponentials instead of a
  single exponential.}
\begin{equation}
  \exp(a^A P_A) \exp(x^A P_A) = \exp((x^A+a^A) P_A),
\end{equation}
so that
$\exp(a^A P_A) \cdot (x^0,\dots,x^d) = (x^0 + a^0, \dots, x^d + a^d)$.
Lorentz transformations act linearly on the coordinates.  If
$h \in \Hgr$, then
\begin{equation}
  h \exp(x^A P_A) = h \exp(x^A P_A) h^{-1} h = \exp(x^A \Ad_h P_A ) h,
\end{equation}
where we have introduced the notation $\Ad$ for the restriction to
$\Hgr$ of the adjoint representation of $\Ggr$.  Since $\m$ is stable
under the action of $\Hgr$, $\Ad_h P_A \in \m$, so we can write it
as $\Ad_h P_A = P_B h^B{}_A$ and hence
\begin{equation}
  h \exp(x^A P_A) = \exp(h^B{}_A x^A h P_B) h.
\end{equation}
Acting on the ``origin'' of Minkowski spacetime or on the identity
coset $e\Hgr = \Hgr$, we have that
\begin{equation}
  h \exp(x^A P_A) \Hgr = \exp(h^B{}_A x^A h P_B) \Hgr,
\end{equation}
using that $h\Hgr = \Hgr$, since $\Hgr$ is a subgroup.  Therefore
Lorentz transformations in Minkowski spacetime are linear relative to
the exponential coordinates.  This is a general fact about reductive
homogeneous spaces $\Ggr/\Hgr$: in exponential coordinates, $\Hgr$
acts linearly.  It is important to realise that this is a
coordinate-dependent statement and, moreover, only applies to the
reductive situation.  It is the linear isotropy representation (on the
tangent space at the origin) which, as the name belies, is always
linear, regardless of reductivity.

\subsubsection{Galilei spacetime as a homogeneous space}
\label{sec:galil-spac-as}

Let us now consider Galilei spacetime, which is described by a
Klein pair $(\g,\h)$ where $\g$ is the Lie algebra spanned by
$L_{ab}, B_a, P_a, H$, for $a,b = 1,\dots,d$, and whose brackets are
given by equation~\eqref{eq:gal-algebra-brackets} and $\h$ is the
subalgebra spanned by $L_{ab},B_a$.  We choose the reductive
complement $\m$ to be the span of $P_a, H$.  We choose exponential
coordinates $(t, x^a)$ via the coset representative
\begin{equation}
  L(t,x) = \exp(t H + x^a P_a).
\end{equation}
Here again the pull-back of the Maurer--Cartan one-form has no
$\h$-component:
\begin{equation}
  L^{-1} dL = dt H + dx^a P_a.
\end{equation}
The action of the Galilei group is again easy to work out using left
multiplication: translations again shift the exponential coordinates
$\exp(s H + v^a P_a) \cdot (t,x^a) = (t + s, x^a + v^a)$, whereas
rotations and boosts act as follows.  Let $R \in \Ggr$ be a rotation;
that is, an element of the $\SO(d)$ subgroup generated by the
$L_{ab}$.  Then
\begin{equation}
  R \exp( t H + x^a P_a) =   R \exp( t H + x^a P_a) R^{-1} R = \exp (t
  H + x^a \Ad_R P_a) R,
\end{equation}
where we have used that $H$ is a scalar and hence commutes with the
rotations.  Again $\Ad_R P_a = P_b R^b{}_a$ and hence acting on the
identity coset we read off the action of rotations on the exponential
coordinates: $R \cdot (t, x^a) = (t, R^b{}_a x^a)$.  Now let us
consider the boosts.  Let $h := \exp(v^a B_a)$.  Then, as before,
\begin{equation}
  h \exp(t H + x^a P_a) = \exp( \Ad_h (t H + x^a P_a)) h.
\end{equation}
We work out the term inside the exponential:
\begin{equation}
  \Ad_h (t H + x^a P_a) = \Ad_{\exp(v^b B_b)} (t H + x^a P_a) = \exp(
  v^b \ad_{B_b})(t H + x^a P_a),
\end{equation}
where $\ad_{B_b} H := [B_b, H] = P_b$ and $\ad_{B_b} P_a = [B_b, P_a]
= 0$.  Substituting, we find
\begin{equation}
  h (t H + x^a P_a) h^{-1} = (t H + (x^a + t v^a) P_a).
\end{equation}
Acting on the identity coset again we see that boosts act on the
exponential coordinates by $h \cdot (t, x^a) = (t, x^a + t v^a)$,
which are precisely the Galilei boosts we saw in
Section~\ref{sec:galilei-spacetime}.

To determine the Galilei-invariant tensors in Galilei spacetime $\Ggr/\Hgr$,
we need to determine the $\Hgr$-invariant tensors of the linear isotropy
representation.  Canonically dual to the basis $H,P_a$ for $\m$ we
have a basis $\eta,\pi^a$ for $\m^*$.  We need to work out the linear
isotropy representation of $\Hgr$ on both $\m$ and $\m^*$ and hence on
tensors.  The linear isotropy representation is such that the $L_{ab}$
generate rotations and $B_a$ generate boosts.  It is easy to
determine the tensors invariant under rotations.   First of all $H \in
\m$ is invariant, but also its dual $\eta$.  A classical theorem of
Weyl's \cite[Theorem~2.11.A]{MR1488158} says that every $\SO(d)$
invariant tensor of the $d$-dimensional vector representation can be
constructed out of $\delta_{ab}$, its inverse and $\epsilon_{a_1\dots
  a_d}$.  Concerning the boosts, we have that
\begin{equation}
  \begin{aligned}\relax
    B_a \cdot H &= P_a\\
    B_a \cdot P_b & =0
  \end{aligned}
  \qquad\text{and}\qquad
  \begin{aligned}\relax
    B_a \cdot \eta &= 0\\
    B_a \cdot \pi^b & = -\delta^b_a \eta,
  \end{aligned}
\end{equation}
where we have used that the action of $B_a$ on $\m$ is induced by the
restriction of adjoint representation $B_a \cdot X = \ad_{B_a} X =
[B_a, X]$ for $X \in \m$, whereas that on $\m^*$ is induced by the
restriction of the coadjoint representation $B_a \cdot \alpha =
\ad^*_{B_a} \alpha = - \alpha \circ \ad_{B_a}$ for $\alpha \in \m^*$.
We see that $\eta \in \m^*$ is $\Hgr$-invariant and so is $\delta^{ab}
P_a \otimes P_b$.  Applying $\eta$ to the pull-back of the
Maurer--Cartan one-form we obtain a Galilei-invariant one-form on
Galilei spacetime, namely, the clock one-form
\begin{equation}
  \tau := \eta(L^{-1}dL) = \eta (dt H + dx^a P_a) = dt.
\end{equation}
The vielbein dual to the soldering form $dt H + dx^a P_a$ is given by
$\frac{\d}{\d t}$ and $\frac{\d}{\d x^a}$.  The Galilei-invariant
tensor field corresponding to $\delta^{ab}
P_a \otimes P_b$ is then $\delta^{ab}\frac{\d}{\d x^a} \otimes
\frac{\d}{\d x^b}$, which is the spatial cometric on Galilei
spacetime we saw in Section~\ref{sec:galilei-spacetime}.

\subsubsection{Summary}
\label{sec:summary}

We may summarise the above discussion as follows:
\begin{itemize}
\item Homogeneous spaces of a group $\Ggr$ are described infinitesimally
  by a Klein pair $(\g,\h)$ where $\g$ is the Lie algebra of $\Ggr$ and
  $\h$ a Lie subalgebra generating a closed subgroup $\Hgr$ of $\Ggr$.  The
  homogeneous space can be identified with the coset space $\Ggr/\Hgr$
  consisting of left $\Hgr$-cosets $g\Hgr$ in $\Ggr$.  The action of $\Ggr$ on
  $\Ggr/\Hgr$ is induced from left multiplication on $\Ggr$.

\item As a vector space, $\g = \h \oplus \m$, where if possible we
  choose $\m$ in such a way that $[\h,\m] \subset \m$.  If this is
  possible, we say that $(\g,\h)$ is reductive.

\item Every choice of basis $X_1,\dots,X_m$ for $\m$ gives rise to
  exponential coordinates near the identity coset of $\Ggr/\Hgr$ corresponding to a
  (locally defined) coset representative $L : \Ggr/\Hgr \to \Ggr$, where $L(x)
  = \exp(x^1 X_1 + \cdots x^m X_m)$.   The action of $\Ggr$ on
  the exponential coordinates can be calculated in principle simply by
  left multiplying in $\Ggr$: $g L(x) = L(g\cdot x) h(g,x)$ for some
  $h(g,x) \in \Hgr$.  In the reductive case, the action of $h \in
  \Hgr$ on the exponential coordinates is linear.

\item We can use the coset representative to pull-back the
  Maurer--Cartan one-form to $\Ggr/\Hgr$.  This results in a locally defined
  one-form with values in $\g$.  In the reductive case, it decomposes
  into an $\h$-connection and a soldering form.  In the non-reductive
  case, the $\h$-component is \emph{not} a connection, but the
  projection to $\g/\h$ is still a soldering form.

\item In the reductive case, the representation of $\Hgr$ on $\m$ is
  called the the linear isotropy representation.  In the
  non-reductive case, the linear isotropy representation is carried
  by the quotient vector space $\g/\h$.  In practice we work with
  $\g/\h$ by calculating brackets in $\g$ and then dropping from the
  RHS anything belonging to $\h$.

\item $\Ggr$-invariant tensor fields on $\Ggr/\Hgr$ are in one-to-one
  correspondence with $\Hgr$-invariant tensors on $\g/\h$ (or on $\m$ in
  the reductive situation).  If $\Hgr$ is connected, this is the same as
  $\h$-invariant tensors, which are typically simple to determine, at
  least if of small rank.

\end{itemize}

\subsection{Homogeneous kinematical spacetimes}
\label{sec:homog-kinem-spac}

A homogeneous kinematical spacetime is a homogeneous space of a
kinematical group of the right dimension.  Recall from
Definition~\ref{def:KLA} (but now for arbitrary dimension) that a
kinematical Lie algebra for $(d+1)$-dimensional spacetimes consists of
a subalgebra $\r \cong \so(d)$ with generators $L_{ab}$, two copies of
the vector representation with generators $B_a, P_a$ and an additional
scalar generator $H$.  Suppose that $\g$ is such a kinematical Lie
algebra.  A kinematical Klein pair for a $(d+1)$-dimensional spacetime
takes the form $(\g,\h)$, where $\h \subset \g$ is a Lie subalgebra
spanned by $L_{ab}$ and $V_a = \alpha B_a + \beta P_a$ for some
$\alpha,\beta \in \RR$.

The determination of such Klein pairs was done in
\cite{Figueroa-OFarrill:2018ilb}, whose results we summarise in
Table~\ref{tab:spacetimes}, where we have excluded some spacetimes
which only exist for $d=1,2$.  We have chosen a basis for $\g$ in such
a way that $\h$ is always spanned by $L_{ab}$ and (the new) $B_a$.
This facilitates comparison of the different homogeneous spacetimes,
but also obscures the isomorphisms between some of the Lie algebras.  For
example, the kinematical Lie algebras for Minkowski ($\mathsf{M}$) and
anti~de~Sitter--Carroll ($\mathsf{AdSC}$) spacetimes are isomorphic to the Poincaré
algebra.  The explicit isomorphism is the identity on the rotation and
time-translation generators, but exchanges boosts and spatial momenta:
\begin{equation}\label{eq:Mink-AdSC-iso}
 \L^{\mathsf{M}} = \L^{\mathsf{AdSC}}, \quad H^{\mathsf{M}} =
 H^{\mathsf{AdSC}}, \quad \B^{\mathsf{M}} = \P^{\mathsf{AdSC}}
 \quad\text{and}\quad \P^{\mathsf{M}} = - \B^{\mathsf{AdSC}}.
\end{equation}
This illustrates the need to specify a geometric realisation before
assigning a physical/geometric meaning to the generators of a
kinematical Lie algebra, since what is a translation in Minkowski
spacetime is a carrollian boost in anti~de~Sitter--Carroll.

Similarly, the kinematical Lie algebras for hyperbolic space ($\mathsf{H}$), de
Sitter spacetime $(\mathsf{dS})$ and the lightcone ($\mathsf{LC}$) are
isomorphic (to the Lorentz algebra in one dimension higher).  The
explicit isomorphisms are again the identity on the rotation and
time-translation generators:
\begin{equation}
  \L^{\mathsf{dS}} =   \L^{\mathsf{H}} =   \L^{\mathsf{LC}}
  \quad\text{and}\quad   H^{\mathsf{dS}} =   H^{\mathsf{H}} =
  H^{\mathsf{LC}},
\end{equation}
but now
\begin{equation}
  \P^{\mathsf{dS}} = \B^{\mathsf{H}} \quad\text{and}\quad \B^{\mathsf{dS}} = - \P^{\mathsf{H}},
\end{equation}
and
\begin{equation}
  \B^{\mathsf{LC}} = \tfrac1{\sqrt2} (\B^{\mathsf{dS}} -
  \P^{\mathsf{dS}}) = -\tfrac1{\sqrt2} (\B^{\mathsf{H}} +
  \P^{\mathsf{H}}) \quad\text{and}\quad \P^{\mathsf{LC}} =
  \tfrac1{\sqrt2} (\B^{\mathsf{dS}} + \P^{\mathsf{dS}}) = \tfrac1{\sqrt2} (\B^{\mathsf{H}} -
  \P^{\mathsf{H}}).
\end{equation}

\begin{table}[h!]
  \centering
  \caption{Homogeneous ($d+1$)-dimensional  (spatially isotropic) kinematical spacetimes}
  \label{tab:spacetimes}
  \rowcolors{2}{red!10}{yellow!10}
  \resizebox{\textwidth}{!}{
    \begin{tabular}{l|>{$}l<{$}|*{5}{>{$}l<{$}}}\toprule
      \multicolumn{1}{c|}{Name} & \multicolumn{1}{c|}{Klein pair} & \multicolumn{5}{c}{Nonzero Lie brackets in addition to $[\L,\L] = \L$, $[\L, \B] = \B$, $[\L,\P] = \P$} \\\midrule
      Minkowski & (\iso(d,1),\so(d,1)) & [H,\B] = -\P & & [\B,\B] = \L & [\B,\P] = H &\\
      de~Sitter & (\so(d+1,1),\so(d,1)) & [H,\B] = -\P & [H,\P] = -\B & [\B,\B]= \L & [\B,\P] = H & [\P,\P]= - \L \\
      anti~de~Sitter & (\so(d,2),\so(d,1))  & [H,\B] = -\P & [H,\P] = \B & [\B,\B]= \L & [\B,\P] = H & [\P,\P] = \L \\\midrule
      euclidean & (\iso(d+1),\so(d+1)) &[H,\B] = \P & & [\B,\B] = -\L & [\B,\P] = H &  \\
      sphere &  (\so(d+2),\so(d+1)) & [H,\B] = \P & [H,\P] = -\B & [\B,\B]= -\L & [\B,\P] = H & [\P,\P]= - \L  \\
      hyperbolic &  (\so(d+1,1),\so(d+1)) & [H,\B] = \P & [H,\P] = \B & [\B,\B]= -\L & [\B,\P] = H & [\P,\P] = \L \\\midrule
      Galilei & (\g,\iso(d)) & [H,\B] = -\P & & & & \\
      de~Sitter--Galilei & (\n^+_{\gamma =-1},\iso(d))  & [H,\B] = -\P & [H,\P] = -\B & & & \\
      torsional de~Sitter--Galilei & (\n^+_{\gamma\in(-1,1)},\iso(d)) & [H,\B] = -\P & [H,\P] = \gamma\B + (1+\gamma)\P & & & \\
      torsional de~Sitter--Galilei & (\n^0,\iso(d)) & [H,\B] = -\P & [H,\P] = \B + 2\P & & & \\
      anti~de~Sitter--Galilei & (\n^-_{\chi=0},\iso(d)) & [H,\B] =  -\P & [H,\P] = \B & & & \\
      torsional anti~de~Sitter--Galilei & (\n^-_{\chi>0},\iso(d)) & [H,\B] = -\P & [H,\P] = (1+\chi^2) \B + 2\chi \P & & & \\\midrule
      Carroll  & (\car,\iso(d)) & & & & [\B,\P] = H & \\
      de~Sitter--Carroll & (\iso(d+1), \iso(d)) & & [H,\P] = -\B & & [\B,\P] = H & [\P,\P] = -\L \\
      anti~de~Sitter--Carroll & (\iso(d,1), \iso(d)) & & [H,\P] = \B & & [\B,\P] = H & [\P,\P] = \L  \\
      lightcone & (\so(d+1,1), \iso(d)) & [H,\B] = \B & [H,\P] = -\P & & [\B,\P] = H + \L & \\\bottomrule
    \end{tabular}
  }
\end{table}

Table~\ref{tab:spacetimes} is divided into sections corresponding to
the class of geometry the spacetime describes: lorentzian, riemannian,
galilean and carrollian.  They can be distinguished by the type of
invariant tensor fields or, as explained in
Section~\ref{sec:line-isotr-repr}, by the $\h$-invariant tensors of
the linear isotropy representation on $\g/\h$.  We shall now go
through the table in some detail.  Further details can be found in
\cite{Figueroa-OFarrill:2018ilb,Figueroa-OFarrill:2019sex}.

\subsubsection{Homogeneous lorentzian spacetimes}
\label{sec:homog-lor-spac}

The first section of the table consists of those homogeneous
kinematical spacetimes admitting an invariant lorentzian metric.  They
admit an $\h$-invariant lorentzian inner product on $\g/\h$.  The
dimension of the kinematical Lie group for a ($d+1$)-dimensional
homogeneous spacetime is $\tfrac12 (d+1)(d+2)$, hence if the
kinematical Lie group is acting via isometries, the geometry is
maximally symmetric.  In lorentzian signature they are Minkowski,
de~Sitter and anti~de~Sitter spacetimes.  Geometrically, there is a
one-parameter (the scalar curvature) family of both de~Sitter and
anti~de~Sitter spacetimes, but as homogeneous spacetimes they are
isomorphic.  The parameter is simply the scale of the invariant
metric.  These spacetimes are symmetric spaces and the stabiliser
subalgebra $\h \cong \so(d,1)$ in all cases.

\subsubsection{Homogeneous riemannian "spacetimes"}
\label{sec:homog-riem-spac}

The second section of the table consists of homogeneous spaces of
kinematical Lie groups which admit an invariant riemannian metric.
They can hardly be considered as spacetimes, so we will not mention
them again.  The same dimension arguments as for the lorentzian
spacetimes imply that these riemannian homogeneous spaces are
maximally symmetric, so they are the euclidean and hyperbolic spaces
and the round sphere.  Again the curvature of the sphere and
hyperbolic space is a choice of additional structure on the
homogeneous spaces.  All round spheres are described by the same Klein
pair $(\so(d+2),\so(d+1))$, for example.  These riemannian spaces are
symmetric spaces and the stabiliser subalgebra $\h \cong \so(d+1)$ in
all cases.

\subsubsection{Homogeneous galilean spacetimes}
\label{sec:homog-galil-spac}

The third section of the table consists of homogeneous spaces of
kinematical Lie groups admitting an invariant galilean structure: a
clock one-form and a spatial cometric.  The clock one-form comes from
an $\h$-invariant covector in $(\g/\h)^*$, the dual of the linear
isotropy representation.  The spatial cometric comes from an
$\h$-invariant symmetric bivector in $\odot^2(\g/\h)$, the symmetric
square of the linear isotropy representation.  Apart from Galilei
spacetime, discussed in \ref{sec:galil-spac-as}, which is the
non-relativistic limit of Minkowski spacetime, there are two
one-parameter families of spacetimes.  One family is the
de~Sitter--Galilei family with parameter $\gamma \in [-1,1]$.  For
$\gamma = -1$, it is the non-relativistic limit of de~Sitter
spacetimes and hence a symmetric space associated to one of the
Newton--Hooke algebras.  For any $\gamma \in (-1,1)$, the spacetime is
reductive but not symmetric and associated to the kinematical Lie
algebra $\n^+_\gamma$.  The notation notwithstanding, the spacetime
with $\gamma = 1$ is not associated to $\n^+_{\gamma=1}$ but instead
to the Lie algebra $\n^0$, which is obtained as a (singular) limit
$\lim_{\gamma\to 1}\n^+_\gamma$.  This limit is analogous to a
contraction, but it is not a contraction in that the Lie algebras
$\n^+_\gamma$ are not isomorphic for different values of
$\gamma \in [-1,1]$.  The canonical invariant connection (see, e.g.,
\cite{MR0059050}) has torsion proportional to $1+\gamma$ and hence the
spacetimes for $\gamma \neq -1$ may be thought of as torsional
de~Sitter--Galilei spacetimes.  The other family is the
anti~de~Sitter--Galilei family with parameter $\chi \geq 0$.  For
$\chi = 0$, it is the non-relativistic limit of anti~de~Sitter
spacetime and hence a symmetric space, associated to the other
Newton--Hooke algebra.  For any $\chi > 0$, it is a reductive,
non-symmetric homogeneous spacetime.  Again the canonical invariant
connection has torsion (proportional to $\chi$) and these spacetimes
are therefore called torsional anti~de~Sitter--Galilei spacetimes.
The limit $\chi \to \infty$ of the torsional anti~de~Sitter--Galilei
spacetimes coincides with the limit $\gamma \to 1$ of the torsional
de~Sitter--Galilei spacetime.  In all cases, the stabiliser subalgebra
$\h \cong \iso(d)$.


\subsubsection{Homogeneous carrollian spacetimes}
\label{sec:homog-carr-spac}

The fourth section of the table consists of four homogeneous
kinematical spacetimes admitting an invariant carrollian structure: a
nowhere-vanishing vector field and a ``spatial metric''.  The vector
field comes from an invariant vector in the linear isotropy
representation $\g/\h$ and the spatial metric comes from an invariant
symmetric bilinear form in $\odot^2(\g/\h)^*$.  Three of these
carrollian spacetimes are limits of the lorentzian spacetimes in the
Table: Carroll spacetime (of Minkowski) and de~Sitter--Carroll and
anti~de~Sitter--Carroll spacetimes (of de~Sitter and anti~de~Sitter,
respectively).  They are symmetric spaces.  The fourth carrollian
spacetime is the lightcone in Minkowski spacetime one dimension
higher.  It is the only non-reductive homogeneous spacetime in the
table.  In all cases, the stabiliser subalgebra $\h \cong \iso(d)$,
but despite being abstractly isomorphic to the stabiliser subalgebra
of the galilean spacetimes, their images under the linear isotropy
representation are not conjugate subalgebras of $\gl(\g/\h)$, which
explains why they have different invariants and hence why the
geometries are different: carrollian instead of galilean.

\subsubsection{Homogeneous aristotelian spacetimes}
\label{sec:homog-arist-spac}

Although not in the Table, there are also aristotelian spacetimes
which are homogeneous spaces of Lie groups of the aristotelian Lie
algebras in Table~\ref{tab:ALAs}.  The stabiliser subalgebra is always
the rotational subalgebra $\r \cong \so(d)$ and hence there is a
unique Klein pair for each aristotelian Lie algebra.  In the order
given in Table~\ref{tab:ALAs}, they are the static aristotelian
spacetime, the torsional static aristotelian spacetime and the product
of the round $d$-dimensional sphere or $d$-dimensional hyperbolic
space with the real line.  All are reductive and all but the torsional
static spacetime, whose canonical connection has torsion, are
symmetric.

\subsection{Non-lorentzian geometries}
\label{sec:non-lorentz-geom}

As we saw in Section~\ref{sec:homog-kinem-spac}, the homogeneous
spatially isotropic kinematical spacetimes come in several families
depending on their invariant tensors.  We shall ignore the riemannian
case in what follows, since they do not admit an interpretation as
spacetimes (e.g., the boosts are actually rotations).  We shall
now describe the (Cartan) geometries modelled on the homogeneous
spacetimes.  It turns out that all of the geometries we consider:
lorentzian, galilean, carrollian and aristotelian are examples of
$G$-structures; that is, they are defined by distinguished vielbeins
transforming under a subgroup of the general linear group.  The prime
example is lorentzian geometry, where the distinguished vielbeins
transform under local Lorentz transformations on overlaps and can
subsequently be interpreted as the (pseudo) orthonormal frames
relative to a lorentzian metric.

\subsubsection{Basic notions about $G$-structures}
\label{sec:basic-notions-about}

Now consider an $n$-dimensional manifold $M$.  Let $p \in M$.
A \textbf{frame at $p$} is an isomorphism $u: \RR^n \to T_p M$ of
vector spaces.  The images under $u$ of the standard basis
$(\e_1,\dots,\e_n)$ of $\RR^n$ give a basis $(u(\e_1),\dots,u(\e_n))$
for the tangent space at $p$.  If $u,u'$ are two frames at $p$, then
$h:= u^{-1}\circ u' \in \GL(n,\RR)$, which we may rewrite as
$u' = u \circ h$.  This defines a right action of $\GL(n,\RR)$ on the
set $F_p$ of frames at $p$, which is free (if $u \circ h = u$, then
$h$ is the identity) and transitive (any two frames are
related by some $h \in \GL(n,\RR)$).  The disjoint union
$F(M) = \bigsqcup_{p\in M} F_p$ is the total space of the
\textbf{frame bundle} of $M$: a smooth right principal
$\GL(n,\RR)$-bundle, whose (local) sections are called \textbf{moving
  frames} or \textbf{vielbeins}.

Let $\Ggr \subset \GL(n,\RR)$ be a Lie subgroup.  A \textbf{$\Ggr$-structure} on
$M$ is a principal $\Ggr$-subbundle $P \subset F(M)$ of the frame bundle:
this amounts to restricting to a collection of frames such for any two
frames $u,u'$ at $p$ in this collection, $u^{-1} \circ u' \in \Ggr$.  The
existence of a $\Ggr$-structure is not guaranteed: there are topological
obstructions.  For example, if $(M,g)$ is a Lorentzian manifold, we can
always pick pseudo-orthonormal frames and it follows that if $u,u'$
are pseudo-orthonormal frames at $p$, $u^{-1}\circ u' \in \Ort(d,1)
\subset \GL(d+1,\RR)$.  The Lorentz group $\Ort(d,1)$ is not connected: it
has four connected components: depending on whether or not the
temporal or spatial orientations are preserved.  This then leads to
topological obstructions (temporal and/or spatial orientability) to
further reduce the structure group from $\Ort(d,1)$ to its connected
component $\SO(d,1)_0$ (i.e., the proper orthochronous Lorentz group)
or some group in between.

Associated to every $\Ggr$-structure $\pi: P \to M$ there is a
\textbf{soldering form} $\theta \in \Omega^1(P;\RR^n)$, which is a
$\RR^n$-valued one-form on the total space $P$.  Let $u$ be a frame at $p$
and suppose that $X_u \in T_u P$  is a tangent vector to $P$ at $u$.
Then $\theta_u(X_u) = u^{-1}(\pi_* X_u)$.  In other words,
$\theta_u(X_u)$ is the coordinate vector of $\pi_* X_u \in T_p M$
relative to the frame $u$.  Given a vielbein
(i.e., a local section $s : U \to P$ on some open subset $U \subset
M$) we can use it to pull back $\theta$ to $U$.  Since $\theta$ is
$\RR^n$-valued, so is its pull-back and we may write it as a linear
combination of the standard basis of $\RR^n$: $s^*\theta = \vartheta^i
\e_i$ for some one-forms $\vartheta^i$ defined only on $U$.  Then
$(\vartheta^1,\dots,\vartheta^n)$ is often called the \emph{inverse
  vielbein}, but of course it is simply the canonically dual coframe
to the vielbein.

The soldering form is the fundamental object which allows to relate
the representation theory of $\Ggr$ to the geometry of any manifold with
a $\Ggr$-structure.  For more details about this in the context of
non-lorentzian geometry, please see
\cite[Section~2]{Figueroa-OFarrill:2020gpr}.

\subsubsection{Lorentzian geometry}
\label{sec:lorentzian-geometry}

Let us see how these ideas play out in the familiar case of lorentzian
geometry.

Let $(\g= \h \oplus \m,\h)$ be a reductive Klein pair for any one of
the homogeneous lorentzian manifolds in Table~\ref{tab:spacetimes}.
In all cases, $\h \cong \so(d,1)$ and the infinitesimal linear
isotropy representation $\lambda : \h \to \gl(\m)$ preserves a
lorentzian inner product $\eta$, say, on $\m$; that is, for all
$X \in \h$ and $Y_1,Y_2\in \m$, we have that
\begin{equation}
  \eta(\lambda_X Y_1, Y_2) + \eta(Y_1, \lambda_X Y_2) = 0.
\end{equation}
Every choice of basis for $\m$ defines an isomorphism $\m \to
\RR^{d+1}$ which may be used to transport the lorentzian inner product
to $\RR^{d+1}$.  Choosing a pseudo-orthonormal basis for $\m$ brings
the inner product on $\RR^{d+1}$ to be the standard one with diagonal
matrix with entries $(-1,1,\dots,1)$ and embeds $\h \subset
\gl(d+1,\RR)$ as the standard Lorentz algebra.

Let $M$ be a ($d+1$)-dimensional manifold with a
$G=\Ort(d,1)$-structure. Then $M$ is covered by open subsets
$\{U_\alpha\}$ and  each $U_\alpha$ we have an inverse vielbein
$\vartheta_\alpha$ taking values in $\RR^{d+1}$ and such that on
nonempty overlaps $U_{\alpha\beta}:= U_\alpha \cap U_\beta$, the
inverse vielbeins are related by local $\Ort(d,1)$-transformations
$h_{\alpha\beta}: U_{\alpha\beta} \to \Ort(d,1)$.  Using the
$\Ort(d,1)$-invariant lorentzian inner product $\eta$ on $\RR^{d+1}$,
we can define on each $U_\alpha$ a local lorentzian metric
\begin{equation}
  g_\alpha := \eta(\vartheta_\alpha,\vartheta_\alpha).
\end{equation}
But because $\eta$ is $\Ort(d,1)$ invariant, these local metrics agree on
overlaps and hence they glue to a lorentzian metric $g$ on $M$.  This
shows that a lorentzian metric on an ($d+1$)-dimensional manifold $M$
is equivalent to a $\Ggr$-structure on $M$ with $G=\Ort(d,1)$.

This generalises in the sense that if an $n$-dimensional manifold $M$
admits a $\Ggr$-structure with $\Ggr \subset \GL(n,\RR)$, then every (nonzero)
$\Ggr$-invariant tensor of $\RR^n$ gives rise to a (nowhere-vanishing)
global tensor field on $M$: it is defined locally using the (inverse)
vielbeins, but the $\Ggr$-invariance guarantees that these local tensor
fields glue on overlaps.

\subsubsection{Newton--Cartan geometry}
\label{sec:newt-cart-geom}

Let $(\g= \h \oplus \m,\h)$ be a reductive Klein pair for any one of
the homogeneous galilean manifolds in Table~\ref{tab:spacetimes}.  In
all cases, $\h \cong \iso(d)$ and the infinitesimal linear isotropy representation
$\lambda : \h \to \gl(\m)$ preserves a covector in $\m^*$ and a
symmetric bivector in $\odot^2\m$.  Choose basis $(P_0,P_1,\dots,P_d)$
for $\m$ and canonical dual basis $\pi^0,\pi^1,\dots,\pi^d$ for
$\m^*$, relative to which the invariant covector is $\pi^0$ and the
invariant symmetric bivector is $P_1^2 + \cdots + P_d^2 = \delta^{ab}
P_a P_b$.  The subgroup $\Ggr \subset \GL(d+1,\RR)$ which preserves these
tensors consists of matrices of the form
\begin{equation}\label{eq:gal-structure-group}
  \begin{pmatrix}
    1 & \bzero^T \\ \bv & A
  \end{pmatrix} \qquad\text{with}\qquad \bv \in
  \RR^d, A \in \Ort(d).
\end{equation}

A \textbf{(weak) Newton--Cartan structure} on a ($d+1$)-manifold $M$
is a $\Ggr$-structure with $\Ggr \subset \GL(d+1,\RR)$ the subgroup given by the
matrices in equation~\eqref{eq:gal-structure-group}.  The
$\Ggr$-invariant tensors give rise to global (nowhere-vanishing) tensor
fields on $M$: the clock one-form $\tau \in \Omega^1(M)$ defined
locally on $U_\alpha$ by $\tau_\alpha := \pi^0(\vartheta_\alpha)$
relative to the inverse vielbein $\vartheta_\alpha$.  Similarly the
spatial cometric is given locally on $U_\alpha$ by $\lambda_\alpha :=
\delta^{ab} (E_\alpha)_a (E_\alpha)_b$, where $E_\alpha$ is the
vielbein dual to $\vartheta_\alpha$.  These symmetric bivectors glue
to give a symmetric $(2,0)$-tensor field $\lambda$ on $M$.
Equivalently, one could define a (weak) Newton--Cartan structure on
$M$ by specifying a nowhere vanishing one-form $\tau \in \Omega^1(M)$
and a corank-1 positive-semidefinite symmetric $(2,0)$-tensor field
$\lambda \in \Gamma(\odot^2TM)$ with the property that $\lambda(\tau,
-) = 0$.

A \textbf{Newton--Cartan structure} is obtained by enhancing a weak
Newton--Cartan structure with an \textbf{adapted connection}: an
affine connection relative to which $\tau$ and $\lambda$ are
parallel.  Such connections were studied initially in
\cite{MR175340,MR334831}.  As explained, e.g., in
\cite[Section~2]{Figueroa-OFarrill:2020gpr}, every $\Ggr$-structure has
an \emph{intrinsic torsion} which is the part of the torsion tensor of
an adapted connection which is independent of the connection.

This is not something one is familiar with from lorentzian geometry,
since the Fundamental Theorem of lorentzian (or, more generally,
pseudo-riemannian) geometry states that there exists a unique
torsion-free adapted (here, metric) connection.  So that intrinsic
torsion of a lorentzian geometry is always zero.

However for a Newton--Cartan structure this is not the case.  As first
shown in \cite{MR334831}, the intrinsic torsion of a Newton--Cartan
connection can be identified with $d\tau \in \Omega^2(M)$, for $\tau$
the clock-one form.  Hence the intrinsic torsion need not \emph{a
  priori} be zero.  Also shown in \cite{MR175340,MR334831} is that
specifying the torsion does not uniquely determine the adapted
connection: there is contorsion, which is measured by an arbitrary
two-form.

A study of how the bundle of two-forms decomposes under the action of
the structure group reveals that there are three\footnote{This is in
  generic dimension $d+1$: if $d=1$ then there are only two classes and
  if $d=4$ and assuming that $M$ is orientable, there are five
  classes.  See \cite[Appendix~B]{Figueroa-OFarrill:2020gpr}.}
classes of Newton--Cartan structures
\cite[Theorem~6]{Figueroa-OFarrill:2020gpr}:
\begin{itemize}
\item \textbf{torsionless} (NC): $d\tau = 0$;
\item \textbf{twistless torsional} (TTNC): $d\tau \wedge \tau = 0$;
  and
\item \textbf{torsional} (TNC): $d\tau \wedge \tau \neq 0$.
\end{itemize}
These classes first appeared in \cite{Christensen:2013lma} (see
Table~I in that paper) in the context of Lifshitz holography.

The homogeneous examples in Table~\ref{tab:spacetimes} are all
such that $d\tau = 0$, but there are homogeneous examples of all three
kinds \cite{Grosvenor:2017dfs}.

A rich source of (weak) Newton--Cartan structures arise as null
reductions of lorentzian manifolds
\cite{PhysRevD.31.1841,Julia:1994bs}.  Let $(N,g)$ be a lorentzian
manifold with a null nowhere-vanishing Killing vector $\xi$ and
suppose that $\xi$ is complete so that it integrates to a
one-parameter $\Gamma$ subgroup of isometries of $N$.  Let us assume
that the action of $\Gamma$ on $N$ is such that the quotient
$M := N/\Gamma$ is smooth making the projection $\pi : N \to M$ into
a smooth submersion.  Then $M$ inherits from $N$ a (weak)
Newton--Cartan structure as follows.    The Killing one-form
$\xi^\flat$ dual to $\xi$ is the pull-back via $\pi$ of a clock
one-form $\tau \in \Omega^1(M)$
\begin{equation}
  \xi^\flat = \pi^*\tau,
\end{equation}
which is nowhere vanishing since $\xi$ is.  We define the ruler
$\lambda$ as follows.  Clearly it is enough to know what
$\lambda(\alpha,\beta)$ is for any two one-forms
$\alpha,\beta \in \Omega^1(M)$.  Give two such one-forms $\alpha,\beta$,
let $X_\alpha,X_\beta \in \eX(N)$ be vector fields on $N$ which are
metrically dual to the pull-backs
$\pi^*\alpha,\pi^*\beta \in \Omega^1(N)$.  Then
$\lambda(\alpha,\beta)$ is the function on $M$ whose pull-back to $N$
agrees with the inner product $g(X_\alpha,X_\beta)$.  It follows that
$\lambda(\tau,-) = 0$ and hence that $(M,\tau,\lambda)$ is a (weak)
Newton--Cartan structure.

\subsubsection{Carrollian geometry}
\label{sec:carrollian-geometry}

Not all carrollian spacetimes in Table~\ref{tab:spacetimes} are
reductive: the lightcone is not.  So in order to treat all cases
together we will be working with a Klein pair $(\g,\h)$ and simply
define the infinitesimal linear isotropy representation $\lambda :\h
\to \gl(\g/\h)$.  In all cases, $\h \cong \iso(d)$, but this is a
different (i.e., non-conjugate) Lie subalgebra of $\gl(\g/\h)$ than
the one in the galilean examples. This means that $\g/\h$ has
different $\h$-invariant tensors in this case.  The $\h$-invariant
tensors are now a vector in $\g/\h$ and a symmetric bilinear form in
$\odot^2(\g/\h)^*$.  We can choose basis $(\overline P_0,\overline
P_1,\dots,\overline P_d)$ for $\g/\h$, where $\overline P_A = P_A \mod
\h$, and canonical dual basis $(\pi^0,\pi^1,\dots,\pi^d)$ for
$(\g/\h)^*$, relative to which the invariant tensors are $P_0$ and
$\delta_{ab}\pi^a\pi^b = (\pi^1)^2 + \dots + (\pi^d)^2$.  The subgroup
$\Ggr \subset \GL(d+1,\RR)$ which preserves these two tensors consists of
matrices of the form
\begin{equation}
  \label{eq:car-structure-group}
  \begin{pmatrix}
    1 & \bv^T\\ \bzero & A
  \end{pmatrix} \qquad\text{with}\qquad \bv \in
  \RR^d, A \in \Ort(d).
\end{equation}
This group is abstractly isomorphic to the one with matrices
\eqref{eq:gal-structure-group}, but of course they are not conjugate
in $\GL(d+1,\RR)$ since they have different invariants.
The connected component of the group $\Ggr$ (where $A \in \SO(d)$) also
leaves invariant $\pi^0 \wedge \pi^1 \wedge \cdots \wedge \pi^d \in
\wedge^{d+1}(\g/\h)^*$.

A \textbf{(weak) carrollian structure} on a ($d+1$)-dimensional
manifold $M$ is a $\Ggr$-structure with $\Ggr \subset \GL(d+1,\RR)$ the
subgroup consisting of the matrices in
equation~\eqref{eq:car-structure-group}.  The $\Ggr$-invariant tensors
give rise to a (nowhere-vanishing) vector field $\xi \in \eX(M)$ and a
positive-semidefinite corank-$1$ symmetric $(0,2)$-tensor field $h \in
\Gamma(\odot^2T^*M)$ with the property that $h(\xi,-)=0$.  If $M$ is
simply connected, then the structure group can be further reduced to
the connected component $G_0$ and hence there is also a ``volume''
form $\mu \in \Omega^{d+1}(M)$.  Even if the structure group does not
reduce, we still have a locally defined volume form $\mu_\alpha$ on
each $U_\alpha$ and they can be chosen so that they may change by a
sign on overlaps.

A \textbf{carrollian structure} is a weak carrollian structure
enhanced by an adapted connection.  As in the case of a Newton--Cartan
structure, the torsion of the adapted connection does not characterise
the connection uniquely: the contorsion here is measured by a section
of the subbundle $\odot^2 \Ann \xi \subset \odot^2 T^*M$, where
$\Ann\xi \subset T^*M$ is the bundle of one-forms which annihilate the
vector field $\xi$.  The intrinsic torsion is now given by $\eL_\xi
h$, the Lie derivative of $h$ along $\xi$ (see
\cite[Proposition~8]{Figueroa-OFarrill:2020gpr}) and studying the
decomposition of the bundle $\odot^2\Ann \xi$ under the action of the
structure group results in four\footnote{This is for $d>1$: if $d=1$
  there are only two classes, as for the $d=1$ Newton--Cartan
  geometries, consistent with the fact that in $1+1$ dimensions,
  there is no real distinction between carrollian and Newton--Cartan
  structures.} classes of carrollian structures
\cite[Theorem~10]{Figueroa-OFarrill:2020gpr}:
\begin{itemize}
\item \textbf{totally geodesic}: $\eL_\xi h = 0$;
\item \textbf{minimal}: $\eL_\xi \mu = 0$, where $\mu$ is the (possibly only locally defined) volume form;
\item \textbf{totally umbilical}: $\eL_\xi h = f h$ for some $f \in  C^\infty(M)$; and
\item \textbf{generic}, if none of the above are satisfied.
\end{itemize}

The names have been chosen in analogy with the theory of hypersurfaces
in riemannian geometry.  This is more than an analogy in that, as
shown in \cite{Duval:2014uoa,Hartong:2015xda} a natural source of
carrollian manifolds are null hypersurfaces in lorentzian manifolds.
Indeed if $N \subset M$ is a null hypersurface in a lorentzian
manifold $(M,g)$, then $h$ is the pull-back of $g$ to $N$ and $\xi$ is
the null vector field (tangent to $N$) whose integral curves are the
null geodesic generators of $N$.  Then $\eL_\xi h$ is the null second
fundamental form of the hypersurface and the names above coincide with
the classification of hypersurfaces based on their second fundamental
form.  The minimality condition is equivalently but more commonly
rephrased as the vanishing of the trace of the Weingarten map.  There
is also a null Weingarten map for null hypersurfaces and it is
traceless if and only if the carrollian structure is minimal.  Classic
references on null hypersurfaces are \cite{MR886772,MR1777311} and in
the present context \cite{Hartong:2015xda,Figueroa-OFarrill:2020gpr}.

The homogeneous carrollian spacetimes in Table~\ref{tab:spacetimes}
can be realised as null hypersurfaces in the maximally symmetric
lorentzian manifolds in Table~\ref{tab:spacetimes}, but in one
dimension higher: Carroll spacetime and the lightcone are null
hypersurfaces in Minkowski spacetime, whereas de~Sitter--Carroll and
anti~de~Sitter--Carroll spacetimes are null hypersurfaces in de~Sitter
and anti~de~Sitter spacetimes, respectively.

The symmetric carrollian spacetimes in Table~\ref{tab:spacetimes}
(i.e., all but the lightcone) are totally geodesic, whereas the
lightcone is totally umbilical.  In fact, being homogeneous, the
function $f \in C^\infty(M)$ in the definition of totally umbilical is
a constant.  We are not aware of homogeneous examples of minimal
and/or generic carrollian structures, but they should exist.

\subsubsection{Aristotelian geometry}
\label{sec:arist-geom}

Aristotelian geometries are also describable in terms of
$\Ggr$-structures, where $\Ggr \subset \GL(d+1,\RR)$ is the
intersection of any two of the groups defining a galilean, carrollian
or lorentzian structures.  Comparing the matrices in
equations~\eqref{eq:gal-structure-group} and
\eqref{eq:car-structure-group}, we see that $\Ggr \cong \Ort(d)$
consists of matrices of the form
\begin{equation}\label{eq:ari-structure-group}
  \begin{pmatrix}
    1 & \bzero^T \\ \bzero & A
  \end{pmatrix} \qquad\text{with}\qquad A \in \Ort(d).
\end{equation}

Choosing basis $P_0,P_1,\dots,P_d$ for $\RR^{d+1}$ and canonical dual
basis $\pi^0,\pi^1,\dots,\pi^d$, we see that $P_0$ and $\pi^0$ are
invariant and so are $\delta^{ab} P_a P_b = P_1^2 + \cdots + P_d^2$
and $\delta_{ab} \pi^a \pi^b = (\pi^1)^2 + \cdots + (\pi^d)^2$.

A \textbf{(weak) aristotelian structure} on a ($d+1$)-dimensional
manifold $M$ is a $\Ggr$-structure with $\Ggr \subset \GL(d+1,\RR)$ the
subgroup of matrices of the form given in
equation~\eqref{eq:ari-structure-group}.  The $\Ggr$-invariant tensors
described above give rise to the following: a vector field $\xi$, a
one-form $\tau$, a symmetric $(0,2)$-tensor field $h$ and a symmetric
$(2,0)$-tensor field $\lambda$ in such a way that $(\tau,\lambda)$ and
$(\xi,h)$ are simultaneously a (weak) Newton--Cartan and (weak)
carrollian structure.  The details about the classification of
aristotelian $\Ggr$-structures via their intrinsic torsion can be found
in \cite[Section~5]{Figueroa-OFarrill:2020gpr} and a recent discussion
of aristotelian geometry in the context of fractons can be found in
\cite[Section~5]{Bidussi:2021nmp}.

\subsection{Coadjoint orbits}
\label{sec:coadjoint-orbits}

In this section we describe the method of coadjoint orbits in order to
write down particle actions.

\subsubsection{Adjoint and coadjoint actions}
\label{sec:adjo-coadj-acti}

Let $\Ggr$ be a Lie group and let $\g$ be its Lie algebra, whose dual
vector space is denoted $\g^*$.  We identify $\g$ with the tangent
space $T_e\Ggr$ to $\Ggr$ at the identity.  The identity is fixed
under conjugation by any $g \in \Ggr$ and therefore the differential
of conjugation by $g$ defines a group homomorphism
$\Ad : \Ggr \to \GL(\g)$ known as the \textbf{adjoint representation}.
For a matrix group $\Ggr$, if $g \in \Ggr$ and $X \in \g$, the adjoint
action is simply matrix conjugation $\Ad_g X = g X g^{-1}$.  The
adjoint representation on $\g$ induces the \textbf{coadjoint
  representation} $\Ad^*: \Ggr \to \GL(\g^*)$: if $g \in \Ggr$ and
$\alpha \in \g^*$, we have that
$\Ad^*_g \alpha = \alpha \circ \Ad_{g^{-1}}$.  In other words, for all
$X \in \g$, and using dual pairing notation:
\begin{equation}
  \left<\Ad_g^* \alpha, X\right> = \left<\alpha, \Ad_{g^{-1}} X\right>.
\end{equation}
Infinitesimally, we have the adjoint $\ad : \g \to \gl(\g)$ and
coadjoint $\ad^* : \g \to \gl(\g^*)$ representations of the Lie
algebra, defined by $\ad_X Y = [X,Y]$ and $\ad^*_X \alpha = -\alpha
\circ \ad_X$ for all $X,Y \in \g$ and $\alpha \in \g^*$.  This last
condition can be expressed in dual pairing notation as
\begin{equation}\label{eq:inf-coadjoint}
  \left<\ad^*_X \alpha, Y\right> = - \left<\alpha, [X,Y]\right>.
\end{equation}

\subsubsection{The symplectic structure on a coadjoint orbit}
\label{sec:sympl-struct-coadj}

Let $\eO_\alpha$ be the \textbf{coadjoint orbit} of $\alpha \in \g^*$;
that is,
\begin{equation}
  \eO_\alpha =  \left\{ \Ad^*_g \alpha ~ \middle | ~ g \in \Ggr\right\}.
\end{equation}
A fundamental property of coadjoint orbits is that they admit a
$\Ggr$-invariant symplectic structure, given by the
Kirillov--Kostant--Souriau symplectic form $\omega_{\text{KKS}}$.
There are several ways to describe this symplectic form.  Perhaps the
simplest description is in terms of the corresponding Poisson
brackets.  Every $X \in \g$ defines a linear function $\ell_X$ on
$\g^*$ by $\ell_X(\alpha) = \left<\alpha,X\right>$ for all $\alpha \in
\g^*$.  We may restrict the $\ell_X$ to smooth functions on the
coadjoint orbit $\eO_\alpha$.  Their differentials $d\ell_X$ span the
cotangent space to the orbit at any point in the orbit.  Therefore the
Poisson bivector $\Pi_{\text{KKS}}$ dual to the symplectic form (and
hence the symplectic form itself) is uniquely determined by its value
on the $d\ell_X$.  These are given by the Lie algebra itself:
\begin{equation}
  \Pi_{\text{KKS}}(d\ell_X, d\ell_Y) = \left\{\ell_X,\ell_Y \right\}_{\text{KKS}} = \ell_{[X,Y]},
\end{equation}
and hence the Jacobi identity follows from that of the Lie algebra.  The
Jacobi identity for the Poisson brackets is equivalent to the closure
of the $2$-form inverse of the Poisson bivector.  The functions
$\ell_X$ are hamiltonians for the $\Ggr$-action on $\eO_\alpha$: the
hamiltonian vector fields $\left\{ \ell_X,- \right\}$ generate the
infinitesimal action of $\Ggr$ on $\eO_\alpha$.  To show this, let
$\zeta_X \in \eX(\eO_\alpha)$ be the vector fields which generate the
$\Ggr$ action; that is, the value at $\alpha$ of the vector field
$\zeta_X$ is given by $(\zeta_X)_\alpha = \ad^*_X \alpha$.  It follows that
the derivative of the function $\ell_Y$ along the vector field
$\zeta_X$ evaluated at $\alpha$ is given by
\begin{equation}
  (\zeta_X)_\alpha (\ell_Y) = \left<\alpha, [X,Y] \right> = \ell_{[X,Y]}(\alpha).
\end{equation}

A different description, in the spirit of
Section~\ref{sec:line-isotr-repr}, is via the holonomy principle for
homogeneous spaces.  A $\Ggr$-invariant $2$-form
$\omega \in \Omega^2(\eO_\alpha)$ determines and is determined by a
$\Ggr_\alpha$-invariant
$\omega_\alpha \in \wedge^2 T^*_\alpha \eO_\alpha$, where
$\Ggr_\alpha \subset \Ggr$ is the stabiliser of $\alpha$.  Every $X
\in \g$ defines a vector field on $\eO_\alpha$ and, evaluating it at
$\alpha$, gives a tangent vector there.  This defines a linear map $\g
\to T_\alpha\eO_\alpha$, sending $X \in \g$ to $\ad^*_X\alpha$ which, since
$\eO_\alpha$ is an orbit, is surjective.  The kernel of this map is
the Lie algebra $\g_\alpha$ of the stabiliser group $\Ggr_\alpha$ of
$\alpha$.  This shows that $T_\alpha \eO_\alpha$ is isomorphic to
$\g/\g_\alpha$.  Denoting the quotient map $\g \to \g/\g_\alpha$ by
$X \mapsto \overline X$, we have that
\begin{equation}
  \omega_\alpha(\overline X, \overline Y) := \left<\alpha, [X,Y]\right>.
\end{equation}
It is not hard to check that the RHS only depends on $X,Y$ modulo
$\g_\alpha$ and that $\omega_\alpha$ is non-degenerate.  The resulting
$\Ggr$-invariant $2$-form, denoted $\omega_{\text{KKS}}$ is closed:
indeed, its pull-back $\pi^*\omega_{\text{KKS}} \in \Omega^2(\Ggr)$
to $\Ggr$ under the orbit map $\pi: \Ggr \to \eO_\alpha$ sending $g
\mapsto \Ad^*_g \alpha$ is not just closed but in fact exact:
\begin{equation}\label{eq:KKS-exact-pullback}
  \pi^*\omega_{\text{KKS}} = -d \left<\alpha, \vartheta\right>,
\end{equation}
where $\vartheta$ is the left-invariant Maurer--Cartan one-form on
$\Ggr$.  This shows that $\omega_{\text{KKS}}$ is a $\Ggr$-invariant
symplectic form on $\eO_\alpha$.

Finally, and perhaps the most conceptual reason why coadjoint orbits
are symplectic manifolds is that they arise by symplectic reduction
from the canonical symplectic structure on the cotangent bundle
$T^*\Ggr$, which is the phase space of the Lie group $\Ggr$ thought of
as a configuration space. Any diffeomorphism of $\Ggr$ induces a
diffeomorphism of $T^*\Ggr$ which preserves the symplectic form. In
particular, the symplectic form on $T^*\Ggr$ is invariant under the
diffeomorphisms induced from left- and right-multiplications in
$\Ggr$. The existence of left- and right-invariant vector fields on
Lie groups says that $\Ggr$ is parallelisable and hence that $T^*\Ggr$
is trivial: that is, $T^*\Ggr \cong \Ggr \times \g^*$. There are two
natural trivialisations: one using left-multiplication and the other
using right-multiplication.  Let us use left-multiplication to
trivialise $T^*\Ggr$ and hence identify it with $\Ggr \times \g^*$.
The cartesian projection $\Ggr \times \g^* \to \g^*$ defines a
function $\mu : T^*\Ggr \to \g^*$ which is $\Ggr$-equivariant: it
intertwines between the action of $\Ggr$ on $T^*\Ggr$ induced by
left-multiplication and the coadjoint action of $\Ggr$ on $\g^*$.
Pick $\alpha \in \g^*$ and consider $\mu^{-1}(\alpha)$.  These are all
the points in $\Ggr \times \g^*$ of the form $(g,\alpha)$ for any $g \in
\Ggr$ and hence it is a copy of $\Ggr$.  This copy of $\Ggr$ in
$T^*\Ggr$ is preserved by the stabiliser $\Ggr_\alpha$ of $\alpha$.
Quotienting gives the symplectic quotient
$\mu^{-1}(\alpha)/\Ggr_\alpha$ which is a symplectic manifold
diffeomorphic to $\Ggr/\Ggr_\alpha$ or, equivalently, to the coadjoint
orbit $\eO_\alpha$.  The resulting symplectic form on $\eO_\alpha$ is
uniquely characterised by the fact that its pull-back to
$\mu^{-1}(\alpha)$ agrees with the restriction to $\mu^{-1}(\alpha)$
of the canonical symplectic form on $T^*\Ggr$ and a calculation shows
that this is again $\omega_{\text{KKS}}$.

In summary, coadjoint orbits of a group $\Ggr$ are homogeneous
symplectic manifolds of $\Ggr$.  There is a partial converse to this
result, which roughly speaking says that all homogeneous symplectic
manifolds are coadjoint orbits.  More precisely, one has the following
``folkloric'' theorem, proved recently in \cite{Beckett:2022wvo}.

\begin{mainthm}
  Let $\Ggr$ be a connected Lie group and $(M,\omega)$ a
  simply-connected homogeneous symplectic manifold of $\Ggr$.  Then
  there exists a covering $\pi: (M,\omega) \to (\eO,\omega_{\text{KKS}})$,
  with $\eO$ a coadjoint orbit of a one-dimensional central extension
  of $\Ggr$, such that $\pi^*\omega_{\text{KKS}} = \omega$.
\end{mainthm}

\subsubsection{Elementary classical systems}
\label{sec:elem-class-syst}

Homogeneous symplectic manifolds of a Lie group $\Ggr$ are the
\textbf{elementary classical systems} with symmetry $\Ggr$, or perhaps
more colloquially, the \textbf{elementary particles} with symmetry
$\Ggr$ in the nomenclature of Souriau \cite{Souriau}. The above
theorem implies that they are locally symplectomorphic to coadjoint
orbits of $\Ggr$ or possibly a one-dimensional central\footnote{The
  extension has to be central since the coadjoint orbit of a
  non-central extension of a Lie group $\Ggr$ does not admit a natural
  action of $\Ggr$.}  extension of $\Ggr$.  As shown by
Souriau\cite{Souriau}, whether or not we need to centrally extend the
group comes down to the symplectic cohomology of $\Ggr$, which we now
describe briefly.

A smooth function $\theta: \Ggr \to \g^*$ allows us to define an
affinisation of the coadjoint representation
\begin{equation}\label{eq:affine-action}
  g \cdot \alpha := \Ad_g^*\alpha + \theta(g).
\end{equation}
This defines an affine action $g_1 \cdot (g_2 \cdot \alpha) = (g_1g_2)
\cdot \alpha$ precisely when $\theta$ obeys the cocycle condition
\begin{equation}
  \theta(g_1g_2) = \Ad_{g_1}^* \theta(g_2) + \theta(g_1).
\end{equation}
Differentiating $\theta$ at the identity gives a linear map $d_e\theta
: \g \to \g^*$ and hence a bilinear form $c$ on the Lie algebra
defined by
\begin{equation}
  c(X,Y) = \left<(d_e\theta)(X),Y\right>.
\end{equation}
We say that $\theta$ is a \emph{symplectic cocycle} if it satisfies
the cocycle condition and $c(X,Y) = - c(Y,X)$.  In that case,
$c \in \wedge^2\g^*$ is a Chevalley--Eilenberg cocycle and hence
defines a central extension of $\g$, which is trivial if and only if
there exists $\beta \in \g^*$ such that
$c(X,Y) = - \left<\beta, [X,Y]\right>$.  A symplectic cocycle defines
a class in the \emph{symplectic cohomology}
$H_{\text{symp}}^1(\Ggr,\g^*)$, which is trivial if and only if there
exists $\beta \in \g^*$ such that $\theta(g) = \Ad_g^* \beta - \beta$.
In that case, the affine coadjoint action~\eqref{eq:affine-action}
becomes equivalent to the linear coadjoint representation, being
simply the conjugation of the linear coadjoint representation by the
constant translation $\alpha \mapsto \alpha - \beta$.

As shown by Souriau, every time a Lie group $\Ggr$ acts symplectically
on a symplectic manifold $(M,\omega)$ and assuming that the
fundamental vector fields are hamiltonian, we get a class in the
symplectic cohomology of $\Ggr$.  Indeed, consider the linear map
$\varphi: \g \to C^\infty(M)$, sending $X \to \varphi_X$, where the
hamiltonian vector field $\left\{-, \varphi_X \right\}$ generates the
infinitesimal action of $\Ggr$.  Dual to $\varphi$ we have
the\footnote{\emph{moment} in the original French and also in part of
  the symplectic geometry literature} \emph{momentum map}
$\mu : M \to \g^*$, defined by $\left<\mu,X\right> = \varphi_X$,
originally introduced by Souriau.  The symplectic cocycle
$\theta: G \to \g^*$ measures the failure of the moment map to be
equivariant relative to the coadjoint representation:
\begin{equation}\label{eq:symplectic-cocycle}
  \theta(g) := \Ad_g^* \mu(p) - \mu(g\cdot p)
\end{equation}
for any $p \in M$.  Surprisingly, perhaps, assuming that $M$ is
connected, this does not depend on $p$.  If the symplectic cohomology
class of $\theta$ is trivial, so that $\theta(g) = \Ad_g^*\beta -
\beta$ for some fixed $\beta \in \g^*$, we can translate $\mu \mapsto
\mu - \beta$ in such a way that the translated $\mu$ is equivariant:
\begin{equation}
  \mu(g\cdot p) - \beta = \Ad_g^* (\mu(p) - \beta).
\end{equation}
This modifies the functions $\varphi_X$ by constants $\varphi_X
\mapsto \varphi_X - \left<\beta,X\right>$, which do not change the
hamiltonian vector fields.

Suppose that $\Ggr$ acts transitively on $M$, so that $M$ is one orbit
of $\Ggr$.  If the class of $\theta$ (given in
\eqref{eq:symplectic-cocycle}) in symplectic cohomology vanishes, then
the image of the moment map $\mu : M \to \g^*$ is a coadjoint orbit of
$\Ggr$.  One can show that $\mu$ is a covering map and hence $M$
covers a coadjoint orbit of $\Ggr$.  If $\Ggr$ has vanishing
symplectic cohomology, all elementary systems with symmetry $\Ggr$ are
(up to covering) coadjoint orbits of $\Ggr$.  In contrast, if the
symplectic cohomology of $\Ggr$ is not zero, then some elementary
systems with symmetry $\Ggr$ do not cover coadjoint orbits of $\Ggr$
but of a one-dimensional central extension of $\Ggr$.

For example, the symplectic cohomology of the Poincaré group vanishes,
so that Poincaré-invariant classical elementary systems (i.e.,
particles) are classified by the coadjoint orbits of the Poincaré
group. In contrast, the Galilei group does have nontrivial symplectic
cohomology and hence Galilei-invariant particles are classified by
coadjoint orbits of the Bargmann group, the one-dimensional central
extension of the Galilei group already discussed (at the Lie algebraic
level) in Section~\ref{sec:central-extensions}.

\subsubsection{Coadjoint orbits from geodesic motion}
\label{sec:coadj-orbits-from}

In the context of lorentzian geometry we can understand the emergence
of the coadjoint orbits as follows.  Suppose that $(M,g)$ is a
lorentzian manifold and let $\gamma$ be an affinely parametrised
geodesic for the Levi-Civita connection; i.e., a solution of
$\frac{D\dot\gamma}{dt} = 0$.  If $\xi \in \eX(M)$ is a Killing
vector field, then the inner product $g(\xi,\dot\gamma)$ is constant
along the geodesic.  Let $\Ggr$ be the isometry group and let $\g$ be
its Lie algebra.  Then to every $X \in \g$ we associate a Killing
vector field $\xi_X$ and therefore every geodesic defines a
momentum $\mu$ in $\g^*$; namely,
the linear map $\mu: \g \to \RR$ defined by $\left<\mu,X\right> =
g(\xi_X,\dot\gamma)$.  For every $a \in \Ggr$, let $\phi_a : M \to M$ be
the corresponding isometry and suppose that $\gamma$ is a geodesic.
Then $\phi_a \circ \gamma$ is also a geodesic and its momentum is
given by $\Ad^*_a \mu$, where $\mu$ is the momentum of $\gamma$.  The
collection of momenta corresponding to all geodesics which are related
to $\gamma$ by an isometry define the coadjoint orbit of the momentum
of $\gamma$.

As an example, consider affinely parametrised geodesics in Minkowski
spacetime $(\mathsf{M},\eta)$.  In flat coordinates $x^\mu$, they are given by
straight lines $x^\mu(\lambda) = a^\mu + \lambda k^\mu$.  Therefore
$\dot\gamma = \dot x^\mu \d_\mu =  k^\mu\d_\mu$ and the momentum $\mu$
is given by the linear function sending
\begin{equation}
  P_\mu \mapsto \eta(\d_\mu, \dot\gamma) = \eta(\d_\mu, k^\nu \d_\nu) = k^\mu
  \eta_{\mu\nu} = k_\mu
\end{equation}
and
\begin{equation}
  L_{\mu\nu} \mapsto \eta(x_\mu\d_\nu - x_\nu\d_\mu, \dot\gamma) =
  \eta(x_\mu\d_\nu - x_\nu\d_\mu, k^\rho\d_\rho) = a_\mu k_\nu - a_\nu k_\mu,
\end{equation}
which are the linear and relativistic angular momenta, respectively,
of the particle. The quadratic function
$P^2 = \eta^{\mu\nu}P_\mu P_\nu$ on $\g^*$ which takes the value $k^2$
on the momentum $\mu$ is constant on the coadjoint orbit and
corresponds to $-m^2$, where $m$ is the particle mass. Acting with the
translations in the Poincaré group on the geodesic, we can set
$a^\mu = 0$ and acting with the Lorentz transformations which preserve
the origin, we can bring $k^\mu$ to any desired point on the
mass-shell $k^2 = -m^2$. For $m\neq 0$, we can take
$k^\mu = (m,0,0,0)$, and for $k^2 = 0$ we can take
$k^\mu = (1,0,0,1)$, for instance.

As an example, consider the geodesic traced by a massive particle with
mass $m$ in the rest frame, whose momentum is $\mu = m \pi^0$ relative
to the basis $\pi^\mu,\lambda^{\mu\nu}$ for $\g^*$ canonically dual to
the basis $P_\mu, L_{\mu\nu}$ for $\g$.  Relative to this basis, the
infinitesimal coadjoint action is given by
\begin{equation}
  \begin{split}
    \ad^*_{L_{\mu\nu}} \pi^\rho &= \delta^\rho_\nu \pi_\mu - \delta^\rho_\mu \pi_\nu\\
    \ad^*_{P_\mu} \pi^\rho &= \lambda^\rho{}_\mu\\
    \ad^*_{L_{\mu\nu}} \lambda^{\alpha\beta} &= \delta^\alpha_\nu  \lambda_\mu{}^\beta - \delta^\alpha_\mu \lambda_\nu{}^\beta - \delta^\beta_\nu \lambda_\mu{}^\alpha + \delta^\beta_\mu \lambda_\nu{}^\alpha\\ 
    \ad^*_{P_\mu} \lambda^{\alpha\beta} &= 0,
  \end{split}
\end{equation}
where we have raised and lowered indices with $\eta_{\mu\nu}$. This
allows us to determine the stabiliser subalgebra of $\mu = m \pi^0$,
which is seen to be spanned by $L_{12},L_{13}, L_{23},P_0$: that is,
by the infinitesimal generators of rotations and time translations.
These generate a subgroup of the Poincaré group isomorphic to
$\SO(3) \times \RR$. The coadjoint orbit $\eO$ is a homogeneous space
of the Poincaré group with Klein pair $(\iso(3,1),\so(3)\oplus \RR)$
and can be identified with the cotangent bundle of (one sheet of) the
mass-shell hyperboloid. The homogeneous space with Klein pair
$(\iso(3,1), \so(3))$ is, in the language of Souriau\cite{Souriau},
the \emph{evolution space} $\eE$ of the free massive particle. It is a
principal bundle over the coadjoint orbit $\eO$ with structure group
the one-dimensional group generated by time translations (in Minkowski
spacetime). If we let $\varpi : \eE \to \eO$ denote the bundle
projection, the pullback $\sigma := \varpi^*\omega_{\text{KKS}}$ of
the symplectic structure defines a pre-symplectic structure on
$\eE$. It is a closed degenerate two-form on $\eE$ whose kernel $\ker
\sigma$ defines a rank-one integrable distribution whose leaves are
the trajectories of the massive particle. In Souriau's language, but
going back to Lagrange, the coadjoint orbit is the \emph{space of
  motions} of the massive particle.

\subsubsection{Particle actions from coadjoint orbits}
\label{sec:part-acti-from}

The trajectory of a free particle defines a point in the space of
motions but a curve in the evolution space.  Therefore if we wish to
define a variational problem whose extrema are the trajectories of a
free particle, the lagrangian should be defined on the evolution
space.  We shall assume that just like the space of motions, the
evolution space is also a homogeneous space of the symmetry group
$\Ggr$ under discussion.  In the example above, $\Ggr$ is the Poincaré
group and for the massive spinless particle, it is indeed the case
that both the space of motions and the evolution space are homogeneous
spaces of $\Ggr$.

In this more general discussion, we shall assume that the space of
motions is a coadjoint orbit $\eO_\alpha$ of $\Ggr$ and we shall let
$\Ggr_\alpha \subset \Ggr$ denote the stabiliser subgroup of $\alpha$.
We shall let $\varpi : \eE \to \eO_\alpha$ be the projection sending a
point $p \in \eE$ to the unique trajectory passing through $p$, which
is a point in the space of motions.  This projection is
$\Ggr$-equivariant, so that $\varpi(g \cdot p) = \Ad_g^* \varpi(p)$.
Choosing a point $o \in \eE$ with $\varpi(o) =\alpha$, we have a a
commuting triangle
\begin{equation}
  \begin{tikzcd}
    \Ggr \arrow[r,"\widehat\pi"] \arrow[dr,"\pi"] & \eE
    \arrow[d,"\varpi"]\\
    & \eO_\alpha,
  \end{tikzcd}
\end{equation}
where $\widehat\pi :\Ggr \to \eE$ and $\pi : \Ggr \to \eO_\alpha$ are
the orbit maps: $\widehat\pi(g) = g \cdot o$ and $\pi(g) = \Ad^*_g
\alpha$, respectively.   Commutativity of the triangle says that $\pi
= \varpi \circ \widehat\pi$.  We will let $\Ggr_o \subset \Ggr$ be the
stabiliser subgroup of $o \in \eE$ and we observe that $\Ggr_o \subset
\Ggr_\alpha$. Indeed, if $g \in \Ggr_o$, then
\begin{equation}
  \alpha = \varpi(o) = \varpi(g \cdot o) = \Ad_g^* \varpi(o) = \Ad_g^* \alpha,
\end{equation}
where we have used the equivariance of $\varpi$.  Let
$\sigma = \varpi^*\omega_{\text{KKS}} \in \Omega^2(\eE)$ and $\omega =
\pi^*\omega_{\text{KKS}} \in \Omega^2(G)$.  Then the commutativity of
the above triangle implies that
\begin{equation}
  \widehat\pi^*\sigma = \widehat\pi^* \varpi^* \omega_{\text{KKS}} =
  (\varpi \circ \widehat\pi)^*\omega_{\text{KKS}} =
  \pi^*\omega_{\text{KKS}} = \omega.
\end{equation}

Now let $I \subset \RR$ be an interval with parameter $\lambda$ and let
$\gamma : I \to \eE$ be a curve in the evolution space passing through
$o$. It is a physical trajectory if and only if
$\dot\gamma \in \ker \sigma$, so that $\varpi(\gamma(\lambda))$ is
constant and equal to $\varpi(o) = \alpha$. We now set up a
variational problem whose extremals are precisely such curves.

Any curve $\gamma : I \to \eE$ may be lifted  to a curve
$\widehat\gamma :I \to \Ggr$ in the group so that
$\widehat\gamma(\lambda) \cdot o = \gamma(\lambda)$.  This lift is not
unique, since we may multiply on the right with any $h : I \to
\Ggr_o$.  Indeed,
\begin{equation}
  (\widehat\gamma h)(\lambda) \cdot o = \left(\widehat\gamma(\lambda)
  h(\lambda)\right)\cdot o = \widehat\gamma(\lambda) \cdot h(\lambda)
\cdot o = \widehat\gamma(\lambda) \cdot o = \gamma(\lambda).
\end{equation}
Recall that $\widehat\pi^*\sigma = \omega = \pi^*\omega_{\text{KKS}}$
and hence by equation~\eqref{eq:KKS-exact-pullback}, it is exact:
\begin{equation}
  \widehat\pi^*\sigma = -d \left<\alpha, \vartheta\right>.
\end{equation}
We may define an action functional for $\gamma: I
\to \eE$ by lifting the curve to the group $\widehat\gamma : I \to
\Ggr$ and defining
\begin{equation}\label{eq:particle-functional}
  S[\widehat\gamma] := \int_I \left<\alpha, \widehat\gamma^*\vartheta\right>.
\end{equation}
At first sight it seems that this depends on the lift
$\widehat\gamma$, but notice that under a gauge transformation
$\widehat\gamma \mapsto \widehat\gamma h$, the above action transforms
as
\begin{equation}
  S[\widehat\gamma h] = S[\widehat\gamma] + S[h],
\end{equation}
where $S[h]$ is a constant.  We thus conclude that the variational
problem for the action functional~\eqref{eq:particle-functional} is
independent on the lift and, therefore, it defines an action
functional for curves $\gamma : I \to \eE$.

Varying the action functional we find, using the Maurer--Cartan
structure equation $d\vartheta = - \tfrac12 [\vartheta,\vartheta]$, that
\begin{equation}
  \delta S[\widehat\gamma] = - \int_I \left<\alpha,
    \left[\widehat\gamma^{-1}\dot{\widehat\gamma},
    \widehat\gamma^{-1}\delta\widehat\gamma\right]\right> d\lambda = \int_I
  \left<\ad^*_{\widehat\gamma^{-1}\dot{\widehat\gamma}} \alpha,
    \widehat\gamma^{-1}\delta\widehat\gamma\right> d\lambda,
\end{equation}
where we have used equation~\eqref{eq:inf-coadjoint}.  This vanishes
for all variations if and only if
$\ad^*_{\widehat\gamma^{-1}\dot{\widehat\gamma}} \alpha = 0$, so that
$\widehat\gamma^{-1}\dot{\widehat\gamma} =
\vartheta(\dot{\widehat\gamma}) \in \g_\alpha$.  We claim that this is
equivalent to $\dot{\widehat\gamma} \in \ker \omega$.  Indeed,
\begin{equation}
  \begin{split}
    \imath_{\dot{\widehat\gamma}}\omega &= -
    \imath_{\dot{\widehat\gamma}} d \left<\alpha, \vartheta\right>\\
    &= \tfrac12 \imath_{\dot{\widehat\gamma}} \left<\alpha,
      [\vartheta,\vartheta]\right>\\
    &= \left<\alpha, [\vartheta(\dot{\widehat\gamma}),
      \vartheta]\right>\\
    &= - \left<\ad^*_{\vartheta(\dot{\widehat\gamma})} \alpha, \vartheta\right>,
  \end{split}
\end{equation}
which vanishes if and only if $\vartheta(\dot{\widehat\gamma}) \in
\g_\alpha$.  Finally, we observe that $\dot{\widehat\gamma} \in \ker
\omega$ if and only if $\dot\gamma \in \ker \sigma$.

In summary, free particle motion with momentum in the coadjoint orbit
$\eO_\alpha$ defines a curve in the evolution space which is an
extremal of the action functional \eqref{eq:particle-functional}.  In
the following section we will see several examples of this
construction, but before doing that let us make an important remark.

As observed in Section~\ref{sec:homog-kinem-spac}, the same
kinematical Lie group might have inequivalent homogeneous spacetimes.
For example, the Poincaré group has both Minkowski and
anti~de~Sitter--Carroll as homogeneous spaces.  Since coadjoint orbits
are a property of the group, their interpretation as the space of
motions of a particle in a homogeneous space requires additional
information.  In this example, the Poincaré coadjoint orbit $\eO$ of
$\alpha = m \pi^0$ can be interpreted as the space of motions of a
spinless particle of mass $m$ in Minkowski spacetime.  What is its
interpretation in terms of anti~de~Sitter--Carroll spacetime?  The
evolution space $\varpi: \eE \to \eO$ is also common to both Minkowski
and anti~de~Sitter--Carroll spacetimes, but it admits projections to
both spacetimes.  In terms of their Klein pairs, with $\g$ standing
for the Poincaré algebra and $\h_{\eO} = \left<L_{ab},H\right>$,
$\h_{\eE} = \left<L_{ab}\right>$, $\h_{\mathsf{M}} = \left<L_{ab},
  B_a\right>$ and $\h_{\mathsf{AdSC}} = \left<L_{ab}, P_a\right>$, we
have the following maps:
\begin{equation}
  \begin{tikzcd}
    & \eE \arrow[ld] \arrow[rd] \arrow[d] & \\
    \mathsf{M} & \eO & \mathsf{AdSC}
  \end{tikzcd}
  \qquad\qquad
  \begin{tikzcd}
    & (\g,\h_{\eE}) \arrow[ld] \arrow[rd] \arrow[d] & \\
    (\g,\h_{\mathsf{M}}) & (\g,\h_{\eO}) & (\g, \h_{\mathsf{AdSC}}).
  \end{tikzcd}
\end{equation}
A point in the coadjoint orbit $\eO$ lifts to a curve in the evolution
space $\eE$ and this projects to a curve in Minkowski spacetime
$\mathsf{M}$ or to a curve in anti~de~Sitter--Carroll spacetime
$\mathsf{AdSC}$.  The curve in Minkowski spacetime corresponds to the
trajectory of a massive spinless particle, since that is how we
arrived at this space of motions.  We may similarly interpret the
curve in anti~de~Sitter--Carroll spacetime as the trajectory of a
carrollian particle.  We will see this example in detail in
Section~\ref{sec:comp-mink-mathsf}.

\section{Dynamics}
\label{sec:dynamics}

In this section we will construct dynamical systems living in
some of the homogeneous kinematical spacetimes introduced in
Section~\ref{sec:homog-kinem-spac}.  We will focus on the construction
of particle actions in various dimensions for some of these
spacetimes.  We will use the techniques of nonlinear
realisations \cite{Coleman:1969sm,Callan:1969sn} and the coadjoint
orbit method \cite{MR1066693,Souriau} described in
Section~\ref{sec:coadjoint-orbits}.  Although we are in the context of
particle dynamics, the language of nonlinear realisations borrows from
its original use in quantum field theory.

In the current context and at its most basic, a nonlinear realisation
of a Lie group $\Ggr$ is a smooth transitive action of $\Ggr$ on a
manifold $M$.  As we discussed in
Section~\ref{sec:homogeneous-spaces}, once we choose an origin $o \in
M$, we get a diffeomorphism $M \cong \Ggr/\Hgr$, with $\Hgr$ the
stabiliser of the origin.  This diffeomorphism is $\Ggr$-equivariant,
intertwining between the $\Ggr$ action on $M$ and the $\Ggr$ action on
$\Ggr/\Hgr$ given simply by left multiplication.  Another choice of
origin would select a different subgroup $\Hgr$, but any two such
subgroups are conjugate in $\Ggr$ and hence the choice is immaterial.
Let us choose an origin once and for all and hence a description of
$M$ as the coset space $\Ggr/\Hgr$.  As discussed in
Section~\ref{sec:klein-pairs}, such a nonlinear realisation of $\Ggr$
is described (up to coverings) infinitesimally by a Klein pair
$(\g,\h)$, where $\g$ is the Lie algebra of $\Ggr$ and $\h \subset \g$
the Lie subalgebra corresponding to the subgroup $\Hgr$.  In the
language of nonlinear realisations, the subalgebra $\h$ generates some
of the \emph{unbroken} symmetries.

One of the fundamental assumptions in the pioneering papers
\cite{Coleman:1969sm,Callan:1969sn} on nonlinear realisations is that
the group $\Ggr$ is compact, connected and semisimple.  Since $\Hgr$
is a closed subgroup, it is also compact and hence any
finite-dimensional representation of $\Hgr$ is completely reducible
into simple (i.e., irreducible) representations.  In particular, we
can consider the restriction to $\Hgr$ of the adjoint representation
of $\Ggr$ on $\g$.  The subalgebra $\h$ is a subrepresentation and
since $\Hgr$ is compact, it has a complementary subrepresentation $\g
= \h \oplus \m$, where $\m$ is isomorphic to the tangent space of the
homogeneous space $M$ at the origin, not just as a vector space but as
a representation space of $\Hgr$.  (Recall that $\Hgr$ acts on $T_oM$
via the linear isotropy representation.)  In other words, the Klein pair
$(\g,\h)$ is reductive.  The subspace $\m$ is said to generate the
\emph{broken} symmetries and its elements are often referred as
\emph{Goldstone bosons}.

Of course, as we have already seen, kinematical groups are certainly
not compact and seldom semisimple, so its Klein pairs $(\g,\h)$ need
not be reductive. However, as we saw in
Section~\ref{sec:homog-kinem-spac}, with one notable exception (the
lightcone), the Klein pairs of the kinematical spacetimes are
reductive and hence we can talk unambiguously about broken and
unbroken generators.  In general, the broken generators are
equivalence classes in $\g/\h$.

Spacetimes are not the only nonlinear realisations of kinematical
groups that we will be interested in.  In a sense, these describe the
vacua.  When discussing particle dynamics, the mere existence of the
particle in the spacetime breaks the symmetry further.  The resulting
nonlinear realisation can often be interpreted as the evolution space
(in the sense of Souriau \cite{Souriau}) of the particle dynamical
system.  In the case of elementary particles (again in the sense of
Souriau), the evolution space fibres over a coadjoint orbit of the
kinematical group.  And indeed, one of the approaches to elementary
particles in a given homogeneous spacetime $\Ggr/\Hgr$ is to classify
coadjoint orbits of $\Ggr$, pass to their evolution spaces and project
onto the spacetime.  This method was illustrated in
Section~\ref{sec:part-acti-from}, resulting in an explicit expression
for the particle action functional \eqref{eq:particle-functional}.

In practical terms, the method we will follow in this section is the
following.
\begin{enumerate}
\item We consider nonlinear realisations of a kinematical (or
  closely related) group $\Ggr$ on the evolution space of the
  dynamical system corresponding to a particle propagating on a
  homogeneous spacetime.  Let the evolution space be the coset
  manifold $\Ggr/\Hgr$ where $\Hgr$ a subgroup of unbroken
  symmetries.
\item We then choose a basis of the Lie algebra $\h$ and extend it to
  a basis for $\g$.  If the Klein pair $(\g,\h)$ is reductive, we will
  choose the basis in such a way that the split $\g = \h \oplus \m$ is
  preserved by the adjoint action of $\Hgr$.  Even if $(\g,\h)$ is not
  reductive, we will choose a vector space complement $\m$.
\item This choice of basis for $\m$ gives local coordinates for
  $\Ggr/\Hgr$ near the origin via (modified) exponential coordinates,
  as discussed in Section~\ref{sec:expon-coord}.  In effect what these
  local coordinates do is define a local coset representative $g:
  \Ggr/\Hgr \to \Ggr$, which in principle is only defined in a
  neighbourhood of the origin.
\item  We pull back the left-invariant Maurer--Cartan one form on
  $\Ggr$ via the local coset representative and obtain a $\g$-valued
  one-form $\Omega = g^{-1} dg$.  Being $\g$-valued it may be
  decomposed into a component along $\h$ and a component along the
  chosen complement $\m$: $\Omega = \Omega_{\Hgr} +
  \Omega_{\Ggr/\Hgr}$.
\item In order to obtain a particle lagrangian with lowest order in
  derivatives, we take a linear combination of the $\Hgr$-invariant
  components of $\Omega$ and pull them back to the interval
  parametrising the worldline of the particle.  In some cases, for
  example that of the one-dimensional Schwarzian particle discussed in
  Section~\ref{sec:schwarzian}, we may consider instead
  $\Hgr$-invariant quadratic expressions in the components of the
  Maurer--Cartan form.
\end{enumerate}

We shall have ample opportunity to see how this process works in
practice in a number of examples, to which we now turn.

\subsection{Relativistic Particle lagrangians}
\label{sec:rel-particles}

Here, we give a short exposition of the lagrangians of free spinless
particles built on the Poincaré group for spacetime dimensions $>3$
using nonlinear realisations, see e.g.~\cite{Gomis:2006xw}.  Free
particles can be timelike, lightlike and tachyonic due to the causal
structure of Minkowski spacetime.  The Poincaré algebra in $d+1$
spacetime dimensions, denoted $\mathfrak{iso}(d,1)$, is given in
\eqref{eq:poincare-brackets}, where now $A,B = 0,\dots, d$ and $\eta_{AB}$
is the mostly-plus Minkowski metric.  We separate the time and space
indices according to $A = (0,a)$, with $a =1,\dots, d$.  We may also
consider lightcone coordinates, where $A = (+,-,i)$ with
$i =1,\dots,d-1$.

\subsubsection{Massive particle}
\label{sec:massive-rel}

We begin with the construction from nonlinear realisations. For a
massive particle in the rest frame, the momentum eigenvalues take the
form $p_A=(m,0,0,\dots,0)$ with $SO(d)$ stabiliser in the Lorentz
group generated by rotations\footnote{The stabiliser in the Poincaré
  group contains also the time translations $P_0$.}. The Klein pair is
$(\mathfrak{iso}(d,1),\mathfrak{so}(d))$, which describes the
evolution space of a spinless massive particle in Minkowski spacetime,
as described in Section~\ref{sec:coadj-orbits-from} in the special
case of $d=3$.

The local subgroup $H$ of the nonlinear realisation is thus $SO(d)$
and we write the coset representative for the evolution space as
\begin{align}\label{eq:gP-massive}
  g = \underbrace{e^{x^A P_A}}_{g_0} b,
\end{align}
where $g_0$ is a coset representative of Minkowski spacetime thought
of as the coset space $\text{Poincaré}/\text{Lorentz}$ and $b=e^{v^a
  B_a}$ ($a=1,\dots, d$) is a general boost generated by those
boost generators $B_a := L_{0a}$ of the Lorentz group which are broken
due to the presence of a massive particle in Minkowski spacetime.

The pull-back of the Maurer--Cartan form of the Poincaré group is
\begin{equation}\label{eq:MC-one-form}
  \Omega = g^{-1} dg = b^{-1} \Omega_0 b + b^{-1} db = \Omega_{(P)}^A
  P_A + \tfrac12 \Omega^{AB}_{(M)} L_{AB},
\end{equation}
where $\Omega_0 = dx^A P_A = g_0^{-1}dg_0$ is the Maurer--Cartan form
of Minkowski spacetime.

For a relativistic massive particle in another background, for example
$\mathsf{AdS}$, the form of $\Omega_0$ will differ: it would be the pull-back
of the Maurer--Cartan form via a local coset representative for
$\mathsf{AdS}$.

The explicit form of the Maurer--Cartan forms are obtained by
computing the adjoint representation of the boost $b$ on the
generators of space-time translation $P_A$:
\begin{equation}\label{eq:adjrep}
 b^{-1}  P_A b = \Phi_A{}^B(v^{a})P_B,
 \end{equation}
and
\begin{equation}\label{eq:adjrep1}
b^{-1} db= \tfrac12 \eta^{AB}\Phi_A{}^C(v^a)  \,d \Phi_B{}^D(v^a) L_{CD},
 \end{equation}
where $\Phi_A{}^B(v^a)$ is the fundamental representation of the
Lorentz group, which here depends on the $d$ boost parameters $v^a$.
Explicitly,
\begin{equation}
  \begin{split}
    b^{-1} P_0 b &= cosh\|v\| P_0 - \frac{\sinh\|v\|}{\|v\|} v^a P_a\\
    b^{-1} P_a b &= P_a - \frac{1 - \cosh\|v\|}{\|v\|^2} v_a v^b P_b -
    \frac{\sinh\|v\|}{\|v\|} v_a P_0,
  \end{split}
\end{equation}
where $\|v\|^2 = \delta_{ab}v^a v^b$.

We want to construct a lagrangian in terms of the pull-back of the
Maurer--Cartan forms subject to two conditions: it should have the
lowest possible number of derivatives and it should be invariant under
the unbroken $SO(d)$ subgroup of the Lorentz group.  One can therefore
choose any component of the Maurer--Cartan form which is invariant
under rotations.  In this example, we can take the component
$\Omega_{(P)}^0$ along $P_0$.  Therefore we take as lagrangian\footnote{The cases of
  two- and three-dimensional spacetimes are special.  In three
  dimensions we could also add a Wess--Zumino term associated to the
  $\so(2)$ rotational component of the Maurer--Cartan form
  \cite{Schonfeld:1980kb, Mezincescu:2010gb, Gomis:2012ki}; whereas in
  two dimensions, the absence of spatial rotations implies that we can
  use any component of the Maurer--Cartan form in the lagrangian.} the
pull-back of $\Omega_{(P)}^0$ to the world-line of the particle: a
curve $\gamma(\tau)$ parametrised by $\tau$.

The action of a free massive particle is given by
\begin{equation}\label{eq:lag-massive-spinless-mink}
  I[t,x^a,v^a]=-m\int \gamma^* \Omega_{(P)}^0 = -m \int d\tau \,\dot
  x^A \Phi_A{}^0(v^a) = - m \int \left(\cosh\|v\| \dot t -
    \frac{\sinh\|v\|}{\|v\|} v_a \dot x^a \right) d\tau.
\end{equation}
The vector $\Phi_A{}^0(v^a)$ is a timelike unit Lorentz vector and
therefore $\eta^{AB}\Phi_A{}^0 \Phi_B{}^0=-1$.  The lagrangian depends
on the $d+1$ spacetime coordinates $t, x^a$ and the $d$ boost
parameters $v^a$.  The action constructed by nonlinear realisations
could be interpreted as a canonical action \cite{Gomis:2012ki}. In
fact the momentum is
\begin{equation}
  p_A =\frac{\partial L}{\partial\dot x^A}= -m \Phi_A{}^0(v^a),
\end{equation}
so that
\begin{equation}
  p_0 = -m \cosh\|v\| \qquad\text{and}\qquad p_a = m  \frac{\sinh\|v\|}{\|v\|}v_a.
\end{equation}
and the action becomes
\begin{equation}\label{eq:nlrcoadj}
  I[x^A,v^a]= \int d\tau \,p_A(v^a)\dot x^A.
\end{equation}
This action is invariant under reparametrisations and the reduced
physical space (e.g., by choosing $x^0 = \tau$) has a symplectic
structure.  For other orbits the structure of the action is the same,
the only difference will be the form of the constraint of $\Phi_A{}^0(v^a)$.
This form of the action is recovered in the coadjoint orbit approach,
see for example \cite{Gomis:2021irw} and below.

Now if we regard $p_A$ as $d+1$ independent degrees of freedom,
we can rewrite the action as
\begin{equation}\label{eq:particlepaction}
  I[x^A,p_A]= \int d\tau\left(p_A \dot x^A- \tfrac{\gamma}{2}\left(\eta^{AB}{p_A}{p_B}+m^2)\right)\right),
\end{equation}
which is the canonical action of a massive spinless relativistic free
particle.  Using the equations of motion of ${\Phi_A}^0$ and $\gamma$, we obtain
\begin{equation}\label{eq:inversehiggs}
  \Phi_A{}^0=-\frac{\dot x_A}{\sqrt{-\dot x^2}}.
\end{equation}
Note that the previous relation can also be obtained from the
the vanishing the Maurer--Cartan form $\Omega_{(P)}^a$ associated to
the broken translations
\begin{equation}\label{eq:IH00}
  \Omega_{(P)}^a = dx^A \Phi_A{}^a=0,
\end{equation}
which is known as the inverse Higgs mechanism \cite{Ivanov:1975zq}.

Substituting back \eqref{eq:inversehiggs} in the canonical action and
using the equations of motion of $p_A$ and $\gamma$, we obtain the
geometrical action
\begin{equation}
  I[x^A]=-m \int d\tau{\sqrt{-\dot x^2}}.
\end{equation}
The quantisation of the mass-shell constraint for a massive
particle gives the wave equation
\begin{equation}
  \left(\Box - m^2\right)\Phi(t,\vec x)=0.
\end{equation}

\subsubsection{Relativistic Massless particle}
\label{sec:massless-rel}

Let us now consider a spinless massless particle.  In the standard
frame the momentum takes the form $p_A=(1,0,0,\dots, 1)$, whose
stabiliser in the Lorentz group is isomorphic to the euclidean group
$ISO(d-1)$, with null rotations playing the rôle of euclidean
translations.  The Klein pair for the evolution space is therefore
$(\mathfrak{iso}(d,1),\mathfrak{iso}(d-1))$.  It is useful to work
in a lightcone frame, associated to the lightcone coordinates:
$x^+,x^-,x^i$ with $x^\pm=\tfrac1{\sqrt2}(x^d \pm x^0)$ and transverse
coordinates $x^i$ with $i=1,\dots,d-1$.  The Poincaré algebra in
a lightcone frame has the following nonzero brackets:
\begin{equation}\label{eq:PoincareLC}
  \begin{aligned}
    \left[L_{ij}, L_{kl}\right] &= \delta_{jk} L_{il} - \delta_{ik} L_{jl} - \delta_{jl} L_{ik} + \delta_{il}L_{jk}\\
    \left[L_{ij}, L_{\pm k}\right] &= \delta_{jk} L_{\pm i} - \delta_{ik} L_{\pm j}\\
    \left[L_{+-}, L_{\pm i}\right] &= \pm L_{\pm i}\\
    \left[ L_{+i}, L_{-j} \right] &= -\delta_{ij}L_{+-} - L_{ij}
  \end{aligned}
  \qquad\qquad
  \begin{aligned}
    \left[L_{+-}, P_\pm\right] &= \pm P_\pm\\
    \left[L_{ij}, P_k\right] &= \delta_{jk} P_i - \delta_{ik} P_j\\
    \left[ L_{\pm i}, P_\mp \right] &= - P_i\\
    \left[ L_{\pm i}, P_j \right] &= \delta_{ij} P_\pm,
  \end{aligned}
\end{equation}
where we have used that $\eta_{ij}=\delta_{ij}$ and $\eta_{+-} = 1$.  In
this case, the Klein pair is $(\iso(d,1),\iso(d-1))$ with
$\iso(d-1)$ spanned by $L_{ij}, L_{+i}$ and hence it is not reductive:
indeed, $[L_{+i}, L_{-j}]$ always has a rotational component.

The coset space describing the evolution space is now
$ISO(d,1)/ISO(d-1)$ and we can choose a local coset representative
\begin{equation}
\label{eq:gP-massless}
g = e^{x^A P_A} b \,=g_0 b,
\end{equation}
where now $b=e^{v^i L_{-i}} e^{u L_{+-}}$ with $L_{-i},L_{+-}$ are the
broken boost generators.

In order to compute the Maurer--Cartan form we need the analogue of \eqref{eq:adjrep} for this new
form of $b$.  Computing the adjoint representation in this case, the
result is a Lorentz matrix $\Phi_A{}^B(v^i,u)$ where now $A,B$ are
lightcone indices. The translational component of the
Maurer--Cartan form invariant under $ISO(d-1)$ is $\Omega^+_{(P)}$
and, therefore, the invariant action is
\begin{equation}
  I[x^A,v^i,u]= \int d\tau \,\dot x^A \Phi_A{}^+(v^i,u),
\end{equation}
where now $\Phi_A{}^+(v^i,u)$ is a Lorentz vector with vanishing norm.
The momenta are
\begin{equation}
  p_A=\frac{\partial L}{\partial\dot x^A}= \Phi_A{}^+(v^i,u)
\end{equation}
and the action becomes
\begin{equation}\label{eq:nlrcoadj1}
  I[x^A,v^i,u]= \int d\tau \,p_A(v^i,u)\dot x^A.
\end{equation}
In this case, $p_A$ are the components of a null vector, but if we
consider $p_A$ as $d+1$ independent variables, we need to introduce a
Lagrange multiplier $\gamma$ to implement the constraint $p^2 = 0$ so
that the action becomes
\begin{equation}\label{particlepaction}
  I[x^A,p_A]= \int d\tau\left(p_A\dot x^A - \tfrac{1}{2}\gamma \eta^{AB}{p_A}{p_B}\right),
\end{equation}
which is the canonical action of a massless spinless relativistic particle.

\subsubsection{The coadjoint orbit method for relativistic particles}
\label{sec:coadjoint-rel}

We note that the component $\Omega_{(P)}^0$ of the Maurer--Cartan form
that we used to construct the action for the massive spinless particle
is nothing but the pairing of the Maurer--Cartan form with the
momentum vector.  Indeed, as was done inn
Section~\ref{sec:coadj-orbits-from} in four dimensions, canonically
dual to the basis $L_{AB}, P_A$ of the Poincaré algebra $\g$, we have
the basis $\lambda^{AB},\pi^A$ for the dual $\g^*$.  Then the momentum
for a spinless massive particle in the restframe is $p_A \pi^A = m
\pi^0$, where $p_A = (m,0,\dots,0)$, and hence
\begin{equation}
  \Omega_{(P)}^0 = \left< \pi^0, \Omega_{(P)}\right>,
\end{equation}
which, using equation~\eqref{eq:MC-one-form}, can be rewritten as
\begin{equation}
\Omega_{(P)}^0 = \left< \pi^0, b^{-1}\left( dx^A P_A\right) b \right>
= \left< \pi^0, \Ad_{b^{-1}} \left( dx^A P_A \right)\right> =
\left<\Ad^*_b \pi^0 , dx^A P_A\right>,
\end{equation}
where we have used that the action of the Lorentz group element
$b^{-1}$ on the vector representation is the adjoint action of
$b^{-1}$ as an element of the Poincaré group, which a semi-direct
product of the Lorentz group and the vector representation. Similarly,
$\Ad^*_b$ is the coadjoint action on the dual space.

Writing $x^A P_A =\X$ we therefore have the equivalent form of the
lagrangian~\eqref{eq:inversehiggs}, see for example
\cite{Gomis:2021irw},
\begin{equation}
\label{eq:Lpair}
L = \langle \pi , \dot{\X} \rangle\,,
\end{equation}
where $\pi$ is an arbitrary element of the orbit of $m \pi^ 0$.  This
orbit can be parametrised with the boost parameters $v^a$ that then
appear algebraically in the lagrangian.  The momenta can be thought as
elements of the dual of the Lie algebra.  The derivative in $\dot{\X}$
denotes the derivative with respect to the parameter of the world-line
and so we have explicitly carried out the pull-back.

For other orbits like the massless case, we take as element of  the dual Lie algebra
$\pi^0 + \pi^d$ or for the tachyonic case, the element $\pi^d$.  The
action of the Lorentz group on the space of momenta take the
form~\eqref{eq:Lpair} is universal in all cases.

\subsubsection{Comparing Minkowski and $\mathsf{AdSC}$ particles}
\label{sec:comp-mink-mathsf}

As discussed at the end of Section~\ref{sec:part-acti-from}, coadjoint
orbits are intrinsic to the group $\Ggr$ and the same kinematical
group might give rise to different kinematical spacetimes: e.g.,
Minkowski and anti~de~Sitter--Carroll ($\mathsf{AdSC}$) are both
homogeneous spacetimes of the Poincaré group.

In Section~\ref{sec:massive-rel} we derived the lagrangian for a
massive spinless particle in Minkowski spacetime.  Consider a curve
\begin{equation}
  \gamma(\tau) = e^{tH} e^{x^aP_a} e^{v^a B_a}
\end{equation}
in the evolution space, where $t,x^a, v^a$ are functions of $\tau$.
Then the lagrangian corresponding to the coadjoint orbit of $m \pi^0
\in \g^*$ is given in equation~\eqref{eq:lag-massive-spinless-mink} by
\begin{equation}
  L = -m \left( \cosh\|v\| \dot t - \frac{\sinh\|v\|}{\|v\|} v_a \dot
    x^a  \right).
\end{equation}
We would like to interpret this as a particle action in
$\mathsf{AdSC}$.  As shown in equation~\eqref{eq:Mink-AdSC-iso}, what
is a translation in Minkowski is a carrollian boost in
anti~de~Sitter--Carroll.  This suggests considering the same curve but
written in different coordinates adapted to $\mathsf{AdSC}$:
\begin{equation}
  \gamma(\tau) = e^{u H} e^{y^a B_a} e^{w^a P_a}
\end{equation}
for some functions $u,y^a,w^a$ of $\tau$.  We now interpret $u,y^a$ as
local coordinates on $\mathsf{AdSC}$ and $w^a$ as parametrising the
carrollian boosts which are broken due to the presence of a particle
in $\mathsf{AdSC}$.  The explicit change of coordinates from
$(t,x^a,v^a)$ to $(u,y^a,w^a)$ is given by
\begin{equation}
  \begin{aligned}
    u &= t - \frac{\tanh\|v\|}{\|v\|} x^a v_a\\
    y^a &= v^a\\
    w^a &= x^a + \left( \frac{1-\cosh\|v\|}{\cosh\|v\|} \right)
    \frac{x_b v^b}{\|v\|^2} v^a
  \end{aligned}
  \qquad\text{with inverse}\qquad
  \begin{aligned}
    t &= u + \frac{\sinh\|y\|}{\|y\|} w_a y^a\\
    x^a &= w^a - \frac{1 -\cosh\|y\|}{\|y\|^2} w_b y^b y^a\\
    v^a &= y^a.
  \end{aligned}
\end{equation}
We can then perform the change of variables in the lagrangian to
arrive at the following lagrangian for a particle in $\mathsf{AdSC}$:
\begin{equation}
  L = -m \left( \cosh\|y\| \dot u  + \frac{\sinh\|y\|}{\|y\|} w_a \dot
  y^a + \left( 1 - \frac{\sinh\|y\|}{\|y\|} \right) \frac{w_b y^b y_a \dot y^a}{\|y\|^2}\right).
\end{equation}
The canonical momenta are
\begin{equation}
  \begin{split}
      p_0 &= \frac{\partial L}{\partial \dot u} = - m \cosh\|y\|\\
      p_a &= \underbrace{\frac{w_b y^b}{\|y\|^2} y_a}_{p_a^\parallel} + \underbrace{\frac{\sinh\|y\|}{\|y\|}
      \left( \delta_{ab} - \frac{y_ay_b}{\|y\|^2} \right) w^b}_{p_a^\perp}.
  \end{split}
\end{equation}
We see that the ``spatial'' momentum $p_a$ breaks up into a
longitudinal component $p_a^\parallel$ along $y^a$ and a transverse
component $p_a^\perp$.  The Euler--Lagrange equation for $u$ says that
$\|y\|$ is constant, whereas the Euler--Lagrange equation for $w^a$
says that if $\|y\| \neq 0$, then $y^a$ is constant.  Hence a massive
particle in $\mathsf{AdSC}$ does not move.

\subsection{Non-relativistic particle lagrangians}
\label{sec:nonrel-particles}

In this section we will consider non-relativistic particle
lagrangians.  We will start by considering the centrally extended
Newton--Hooke algebra $\widetilde\n^-$.  The Newton--Hooke algebra
$\n^-$ was defined in Section~\ref{sec:kinem-lie-algebr} and
corresponds to $\chi = 0$ in the family of kinematical Lie algebras
$\n^-_\chi$ in Table~\ref{tab:KLAs}.  Its central extension is listed
in Table~\ref{tab:gen-bargmann}.  We will introduce an additional
parameter in the Lie brackets to allow us to take a limit to the
Bargmann Lie algebra which is the universal central extension of the
Galilei algebra $\g$.

The centrally extended Newton--Hooke Lie algebra $\widetilde\n^-$ is
spanned by $L_{ab}, B_a, P_a, H, Z$ where $L_{ab}$ span the $\so(d)$
rotational subalgebra.  The Lie brackets are the generic kinematical
Lie brackets of equations~\eqref{eq:gen-kla-1}, \eqref{eq:gen-kla-2}
and \eqref{eq:gen-kla-3} together with the following nonzero brackets:
\begin{equation}
  \label{eq:nh-bargmann}
  [H,B_a] = P_a, \qquad [H,P_a] = \tfrac1{R^2} B_a
  \qquad\text{and}\qquad [B_a, P_b] = \delta_{ab} Z,
\end{equation}
with $Z$ central.  Notice that for any nonzero real number $R$, these
Lie algebras are all isomorphic, but if we take the limit $R \to
\infty$ we obtain the Bargmann algebra in
Section~\ref{sec:kinem-lie-algebr} after $H \mapsto -H$.

\subsubsection{Massive particle}

In this section we will see that a massive spinless particle with
symmetry algebra $\widetilde\n^-$ is equivalent to a $d$-dimensional
harmonic oscillator.  In order to see that, we will consider the
coset space with Klein pair $(\widetilde\n^-, \so(d))$.

We choose a local coset representative of the form
\begin{equation}\label{eq:coset-rep}
  g= \underbrace{e^{x^0 H} e^{x^a P_a} e^{u Z}}_{g_0} \underbrace{e^{v^a B_a}}_b.
\end{equation}
Here $g_0$ is the coset representative of the generalised
non-relativistic spacetime and $b$ is a Galilei boost parametrised by
$v^a$.

The role of the coordinate $u$ of the central charge $Z$ in
non-relativistic theories to construct a Wess--Zumino term was first
discussed in \cite{Gauntlett:1990nk}. The coordinates $x^0,x^a,u$
suggest an interpretation of these coordinates as relativistic
coordinates in a space of one higher dimension, here $d+2$.

We now calculate the pull-back of the Maurer--Cartan form along the
coset representative:
\begin{equation}
  \Omega = g^{-1} dg = b^{-1} \Omega_0 b + b^{-1} db,
\end{equation}
where $\Omega_0 = g_0^{-1} dg_0$.  These are easy to calculate and one
finds
\begin{equation}
  b^{-1}db = dv^a B_a
\end{equation}
and
\begin{equation}
  \Omega_0 = dx^0 H + dx^a P_a + \left( du - \frac{x^2}{2R^2}
  dx^0\right) Z - \frac{x^a}{R} dx^0 B_a.
\end{equation}
We then calculate $b^{-1} \Omega_0 b$ to arrive at the final
expression
\begin{equation}
  \Omega = \left( du - \frac{x^2}{2R^2} dx^0 - \tfrac12 v^2
    dx^0 - v_a dx^a \right) Z + dx^0 H + (dv^a - \tfrac1R x^a dx^0)
  B_a + (dx^a + v^a dx^0) P_a.
\end{equation}
The $\so(d)$-invariants in the adjoint representation are $H$ and $Z$
and hence the $\so(d)$-invariant lagrangian is built out of the $H$
and $Z$ components of $\Omega$.  The $H$-component is an exact form,
hence it does not contribute to the Euler--Lagrange equations.  We
will therefore concentrate on the $Z$-component.  Pulling it back to
the interval parametrising the worldline of the particle, we arrive at
the following lagrangian
\begin{equation}
  L = \dot u - \tfrac12 \left( \frac{x^2}{R^2} + v^2 \right) \dot x^0
  - v_a \dot x^a.
\end{equation}
The first term is again a total derivative, so it does not contribute
to the Euler--Lagrange equations. Its role is to make the lagrangian
invariant, since without it the lagrangian is only quasi-invariant.
In other words, it is a Wess--Zumino term.

 Solving for $v^a$ via its equation
$\frac{\partial L}{\partial v^a} = 0$, we find
\begin{equation}
  v^a = - \frac{\dot x^a}{\dot x^0}
\end{equation}
and re-introducing this into the lagrangian, we obtain
\begin{equation}\label{eq:BNHLag}
  L = \frac{\dot x^2}{2\dot x^0} - \frac{x^2}{2R^2}
  \dot x^0.
\end{equation}
If we choose the gauge $\dot x^0 = 1$, so we use $x^0$ as the
parameter along the worldline, we see that $L$ is indeed the
lagrangian for a $d$-dimensional harmonic oscillator with
characteristic frequency $\frac1R$.  Taking the limit $R \to \infty$
in the lagrangian, we arrive at the lagrangian for a non-relativistic
spinless particle of unit mass:
\begin{equation}
  L = \frac{\dot x^2}{2\dot x^0}.
\end{equation}

Had we considered instead the central extension of the other
Newton--Hooke algebra $\n^+ = \n^+_{\gamma=-1}$ in
Table~\ref{tab:KLAs}, we would have obtained the inverted harmonic
oscillator \cite{Gao:2001sr}.

\subsubsection{Massless Galilei  Particle}

Is there a massless particle associated to the (unextended) Galilei
algebra?  The answer is yes\footnote{We acknowledge
discussions with Axel Kleinschmidt on this point.}.  The model was
first introduced by Souriau \cite{Souriau}.  A massless relativistic
particle follows a direction on the lightcone.  In the
non-relativistic case, since the speed of light is infinite, the
particle follows a spatial longitudinal direction, say $x^d$.
In this case, the unbroken group is generated by $L_{ij}, B_i, P_d$;
that is, the infinitesimal rotations $L_{ij}$ on the hyperplane
spanned by $x^1,x^2,\dots,x^{d-1}$, the infinitesimal galilean boosts
$B_i$ in directions perpendicular to $x^d$ and the infinitesimal
longitudinal translations along $x^d$.  The evolution space has Klein
pair $(\g, \iso(d-1))$, where $\g$ is the Galilei algebra, as in
Table~\ref{tab:KLAs}, and the $\iso(d-1)$ subalgebra is spanned by
$L_{ij}, B_i$, for $i,j =1,\dots,d-1$.

A local coset representative  is
\begin{equation}\label{eq:gP-massless-gal}
  g = \underbrace{e^{t H + x^i P_i + x^d P_d}}_{g_0} \underbrace{e^{\theta^i R_i} e^{v B_d}}_b,
\end{equation}
with $i=1,\dots,d-1$, where $R_i:= L_{id}$ are the broken rotations
and $B_d$ is the broken longitudinal boost.

The pull-back of the Maurer--Cartan form is given as usual by
\begin{equation}
  \Omega = \Ad_{b^{-1}} \Omega_0 + b^{-1} db,
\end{equation}
where
\begin{equation}
  \Omega_0 = g_0^{-1} dg_0 = dt H + dx^i P_i + dx^d P_d.
\end{equation}
The $ISO(d-1)$-invariant subspace of $\g$ is spanned by $P_d$ and
$B_d$, hence we need to extract those components of $\Omega$ in order
to write down the lagrangian.  We notice that
\begin{equation}
  b^{-1}db = e^{-v B_d} e^{-\theta^i R_i} d (e^{\theta^i R_i} e^{v
    B_d}) = e^{-v B_d} \left(  e^{-\theta^i R_i} d e^{\theta^i R_i}
  \right) e^{v B_d} + dv B_d.
\end{equation}
Since $[R_i,R_j] = - L_{ij}$, the expression in parenthesis lives in
the span of $L_{ij}, R_i$ and hence the first of the above terms lives
in the span of $L_{ij}, R_i, B_i$.  Therefore the only term in
$b^{-1}db$ which contributes to the lagrangian is $dv B_d$.

We calculate $\Ad_{b^{-1}} \Omega_0$ paying particular attention to
the $P_d$ component:
\begin{equation}
  \begin{split}
    \Ad_{b^{-1}} \Omega_0 &= dt H - v dt P_d + \exp(\ad_{-\theta\cdot
      R}) dx^i P_i + \exp(\ad_{-\theta\cdot R}) dx^d P_d\\
    &= dt H + \left(dx^i + \frac{\cos\|\theta\| -1}{\|\theta\|^2}
      \theta^i\theta_j dx^j - \frac{\sin\|\theta\|}{\|\theta\|}
      \theta^i dx^d\right) P_i\\
    & \qquad {} + \left( \cos\|\theta\| dx^d +
      \frac{\sin\|\theta\|}{\|\theta\|} \theta_i dx^i - v dt \right) P_d,
  \end{split}
\end{equation}
where $\|\theta\|^2 =\delta_{ij} \theta^i \theta^j$.  In summary,
\begin{equation}
  \Omega = dv B_d + \left( \cos\|\theta\| dx^d +
      \frac{\sin\|\theta\|}{\|\theta\|} \theta_i dx^i - v dt \right)
    P_d + \cdots
\end{equation}
omitting terms which are not $ISO(d-1)$-invariant.  The $B_d$
component is exact, so that it does not contribute to the
Euler--Lagrange equations.  Therefore we concentrate on the $P_d$
component and introducing the ``colour'' $k$ \cite{Souriau}, we
may write the lagrangian as the pull-back to the interval of the
$P_d$-component:
\begin{equation}
  L = k \left(\cos\|\theta\| \dot x^d +
      \frac{\sin\|\theta\|}{\|\theta\|} \theta_i \dot x^i - v \dot t \right).
\end{equation}

We calculate the spatial canonical momentum $\vec p =
(p_1,p_2,\dots,p_d)$ to be
\begin{equation}
  p_d = \frac{\partial L}{\partial \dot x^d} = k \cos\|\theta\|
  \qquad\text{and}\qquad p_i = \frac{\partial L}{\partial \dot x^i} =
  k \frac{\sin\|\theta\|}{\|\theta\|} \theta_i,
\end{equation}
from where we see that $\vec p \cdot \vec p = k^2$.  Introducing the
associated unconstrained momentum, we implement the above constraint
via a Lagrange multiplier $e$ to arrive at the lagrangian
\begin{equation}
  L = p_d \dot x^d + p_i \dot x^i - v \dot t + \tfrac12 e \left( \vec
    p \cdot \vec p - k^2 \right).
\end{equation}
Notice that the $v$ equation of motion is $\frac{\partial L}{\partial v} = - k
\dot t = 0$, so that propagation is instantaneous.

The quantisation of the mass-shell constraint gives the Helmholtz equation
\begin{equation}\label{eq:masslessg}
  (\nabla^2 + k^2) \Phi(t,\vec x)=0,
\end{equation}
which agrees with the field equation of the galilean magnetic
Klein--Gordon field (see equation~\eqref{case1bflat}).  The approach
to quantum field theory in terms of particle variables is know as the
\emph{world approach} to field theory and it was first considered by
Feynman  \cite{Feynman:1950ir,Feynman:1951gn}, see also for example
\cite{Schubert:2001he,Casalbuoni:1974pj}.

\subsection{Carroll particle lagrangians}

In this section we will construct the action of a massive (timelike) particle in
Carroll spacetime.  The case of a massless Carroll particle can be
obtained from from the massive one by taking the mass to be zero.
We will also construct the lagrangian of a tachyonic particle.  The
presence of these three kinds of particles (timelike, lightlike and
tachyonic) is due to the causal structure of the Carroll geometry.
This is analogous to what happens in lorentzian geometry, but in
contrast with the galilean case, where the notion of mass is not
related to the causal structure.

The Carroll algebra is denoted $\car$ and given in
Table~\ref{tab:KLAs}.  Besides the Lie brackets in
equations~\eqref{eq:gen-kla-1}, \eqref{eq:gen-kla-2} and
\eqref{eq:gen-kla-3}, which are shared by all kinematical Lie
algebras, the only nonzero bracket in $\car$ is
\begin{equation}\label{eq:Calgebra-again}
  [B_a, P_b] = \delta_{ab} H.
\end{equation}
In contrast to the Galilei algebra, the Carroll algebra (for $d\geq3$)
does not allow nontrivial central extensions, although $H$ is a
central element.

\subsubsection{Massive Carroll particle}

We construct the timelike massive Carroll particle lagrangian \cite{Bergshoeff:2014jla}
\cite{Duval:2014uoa} using the method of nonlinear realisations.  The
Klein pair for the evolution space is $(\car, \so(d))$, where $\so(d)$
is the span of the rotations $L_{ab}$.  A coset representative for the
corresponding coset space is given by
\begin{equation}
  g = \underbrace{e^{t H + x^a P_a}}_{g_0} \underbrace{e^{v^a B_a}}_{b},
\end{equation}
where $t, x^a$ are the Goldstone bosons associated to spacetime
translations and $v^a$ are the Goldstone bosons associated to the
broken boosts.

The Maurer-Cartan form $\Omega$ is given by
\begin{equation}
  \Omega = g^{-1} dg = \Ad_{b^{-1}} \Omega_0 + b^{-1}db,
\end{equation}
where
\begin{equation}
  \Omega_0 = g_0^{-1} dg_0 = dt H + dx^a P_a \qquad\text{and}\qquad
  b^{-1}db = dv^a B_a.
\end{equation}
It follows that
\begin{equation}
  \Ad_{b^{-1}} \Omega_0 = dt H + \exp(-\ad_{v^aB_a}) dx^b P_b = dt H +
  dx^b (P_b - v_b H) = (dt - v_b dx^b) H + dx^b P_b.
\end{equation}
The lagrangian is the pull-back to the interval of the rotationally
invariant component of $\Omega$, which is the component along $H$:
\begin{equation}
  L = M (-\dot t + v_a \dot x^a)
\end{equation}
where we have introduced a mass $M$.  Notice that the ordinary massive
particle does not move: the momentum of the Carroll particle $p_a = M
v_a$ and there is no relation between the momentum and the velocity of
the particle.  The canonical lagrangian is obtained by introducing a
Lagrange multiplier $e$:
\begin{equation}
  L_{\text{can}} = -E \dot t + p_a \dot x^a - \tfrac12 e \left( E^2 -
    M^2 \right).
\end{equation}
Note that in this form we have introduced also negative energies that
are allowed in the Carroll case. Since $H$ is a Casimir, its
eigenvalues can take any real value: positive negative or zero.
Physically, a timelike or lightlike Carroll particle does not move.

The quantisation of the mass-shell constraint for a Carroll massive
particle gives the wave equation \cite{Bergshoeff:2014jla}
\begin{equation}\label{eq:carroll-wave}
  \left(\frac{\partial^2}{\partial t^2}+M^2\right)\Phi(t,\vec x)=0;
\end{equation}
that is, the equation of motion of the carrollian electric
Klein-Gordon field theory, see \eqref{eq:case3c-flat}.

\subsubsection{Tachyonic Carroll Particle}

Here we construct the tachyon Carroll particle lagrangian \cite{deBoer:2021jej}.
A relativistic tachyon has a spacelike momentum, but in the
ultra-local limit, the lightcone collapses to the timeline and
hence any momentum having a nonzero component along a spacelike
direction is tachyonic.  For example, we may take the momentum purely
along the $d$-direction: $\alpha = M \pi^d \in \car^*$ in the dual of
the Carroll algebra.  The resulting coadjoint orbit has Klein pair
$(\car,\h_\alpha)$, where $\h_\alpha$ is spanned by $L_{ij}, B_i, B_d,
P_d, P_0$, where now $i,j = 1,\dots,d-1$.  The subalgebra $\h_\alpha$
is isomorphic to the direct sum of the $\iso(d-1)$ algebra generated
by $L_{ij},B_i$ and the Heisenberg algebra generated by
$B_d,P_d,P_0$.  The evolution space is obtained by breaking the
translation symmetry in the $d$-direction.  Therefore the Klein pair
for the evolution space is $(\car,\h)$, where $\h$ is spanned by
$L_{ij}, B_i, B_d, P_0$ and isomorphic now to $\iso(d-1) \oplus
\RR^2$.  This Klein pair is not reductive, but we may choose a
complement $\m$ spanned by $R_i := L_{id}, P_i, P_d$.  It is not
reductive because $[B_d, P_d] = P_0 \not\in \m$.  Nevertheless the
image $\bar P_d$ of $P_d$ in $\car/\h$ is an invariant of the linear
isotropy representation of $\h$ on $\car/\h$.

Let us choose a coset representative
\begin{equation}\label{eq:gP-tachyon-carroll}
  g = \underbrace{e^{x^d P_d} e^{x^iP_i}}_{g_0} \underbrace{e^{\theta^i R_i}}_{b}
\end{equation}
and pull back the left-invariant Maurer--Cartan form.  This will take
values in the Carroll algebra, but we project to $\car/\h$ and keep the
invariant components, which here is only the component along $P_d$.
Equivalently, we calculate the dual pairing $\left<\alpha,
  g^{-1}dg\right>$.  A calculation shows that
\begin{equation}
  g^{-1}dg = \left( \cos\|\theta\| dx^d +
    \frac{\sin\|\theta\|}{\|\theta\|}\theta_i dx^i \right) P_d + \cdots
\end{equation}
where $\|\theta\|^2 = \delta_{ij}\theta^i\theta^j$, hence
\begin{equation}
  \left<\alpha, g^{-1}dg\right> = M \left<\pi^d, g^{-1}dg\right> =M
  \left( \cos\|\theta\| dx^d +
    \frac{\sin\|\theta\|}{\|\theta\|}\theta_i dx^i \right).
\end{equation}
The lagrangian is now obtained by pulling back this component to the
interval parametrising the worldline of the particle:
\begin{equation}
  L = M \left( \cos\|\theta\| \dot x^d +
    \frac{\sin\|\theta\|}{\|\theta\|}\theta_i \dot x^i \right).
\end{equation}
The canonical momenta are given by
\begin{equation}
  p_d := \frac{\partial L}{\partial \dot x^d} = M \cos\|\theta\|
  \qquad\text{and}\qquad p_i := \frac{\partial L}{\partial \dot x^i} =
  M \frac{\sin\|\theta\|}{\|\theta\|}\theta_i,
\end{equation}
from where it follows that they are constrained:
\begin{equation}
  \vec p \cdot \vec p := p_d^2 + \sum_i p_i^2 = M^2.
\end{equation}
We may implement this constraint via a Lagrange multiplier to arrive
at the lagrangian
\begin{equation}
  L = p_d \dot x^d + p_i \dot x^i + \tfrac12 \lambda (\vec p \cdot
  \vec p - M^2).
\end{equation}
Since $\dot x^0$ does not appear in the action, its momentum $p_0$ is
also zero. We may implement that constraint with a second Lagrange
multiplier and write the canonical lagrangian as
\begin{equation}
  L = p_A \dot x^A + \tfrac12 \lambda (\vec p \cdot \vec p - M^2) + \mu p_0,
\end{equation}
Notice that the mass-shell constraint $\vec p \cdot \vec p - M^2=0$
coincides with the one of the massless galilean particle and also that
the energy ($p_0$) of a tachyon particle is zero.  The associated
wave equation reduces to a Helmholtz equation:
\begin{equation}\label{eq:magcarroll}
  (\vec{\nabla}^2+M^2)\Phi(t,\vec x)=0 \qquad\text{and}\qquad
  \frac{\partial}{\partial t} \Phi(t,\vec x)=0,
\end{equation}
which is related to the equations of motion of magnetic Carroll field
theory as in equation~\eqref{case3bflat}.  The field equations
resulting from \eqref{case3bflat} are not exactly \eqref{eq:magcarroll},
but include a source term for the Helmholtz equation.  The relation
among galilean and Carroll particles has been studied in
\cite{Gomis:2022spp} based on the duality among Galilei and Carroll
algebras \cite{Barducci:2018wuj}, at the level of the associated wave
equations~\eqref{eq:masslessg} and \eqref{eq:carroll-wave}.

\subsection{One- and two-dimensional particle dynamics with
  $\SL(2,\RR)$ symmetry}
\label{sec:sl2R-invariant-dynamics}

In this section we will give several examples of one- and
two-dimensional particle dynamics with $\SL(2,\RR)$ symmetry.  There
are three two-dimensional homogeneous spaces of $\SL(2,\RR)$:
corresponding to the hyperbolic plane, the lightcone and (anti)~de
Sitter spacetime.  In addition, $\SL(2,\RR)$ acts on the real
projective line $\RR P^1$ (also known as one-dimensional conformal
space) via projective transformations.  Among the particle dynamics
discussed here, we will recover the conformal mechanics of
\cite{deAlfaro:1976vlx}, see, e.g.,
\cite{Ivanov:1988vw,Ivanov:1988it,deAzcarraga:1998ni,Fedoruk:2011ua}
and the Schwarzian particle action of
\cite{Kitaev:2017awl,Maldacena:2016hyu,Stanford:2017thb}.

As already discussed in Section~\ref{sec:homog-kinem-spac}, there are
three spatially isotropic homogeneous spaces associated to the Lorentz
group $\SO(d,1)$: namely, hyperbolic space $\eH_d$, de Sitter
spacetime $\dS_d$ and the future lightcone $\LL_d$.  The picture is
the familiar foliation of Minkowski spacetime into orbits of the
Lorentz group.  Whereas de Sitter spacetime is a maximally symmetric
lorentzian manifold and hyperbolic space is a maximally symmetric
riemannian manifold, the future lightcone is what we could term a
maximally symmetric carrollian manifold.  Each of these three spaces
is described infinitesimally by a Klein pair $(\g,\h)$ where
$\g = \so(d,1)$ and
\begin{equation}
  \h \cong
  \begin{cases}
    \so(d) & (\eH_d)\\
    \so(d-1,1) & (\dS_d)\\
    \iso(d-1) & (\LL_d)
  \end{cases}
\end{equation}
In this section we will concentrate in the case $d=2$ and we will use
the isomorphism $\so(2,1) \cong \sl(2,\RR)$.  Each of the above Klein
pairs can thus be realised geometrically as coset spaces
$\SL(2,\RR)/\Hgr$ for some one-dimensional connected closed Lie subgroup
$H \subset \SL(2,\RR)$. Up to conjugation in $\SL(2,\RR)$, there are three connected closed
one-dimensional Lie subgroups $H \subset \SL(2,\RR)$:
\begin{eqnarray}
    \text{(elliptic)} & \qquad H &= \left\{ \begin{pmatrix} \cos\theta & - \sin\theta \\ \sin\theta & \cos\theta  \end{pmatrix} ~\middle |~ \theta \in  \RR/2\pi\ZZ \right\}\\
    \text{(hyperbolic)} & \qquad H &= \left\{ \begin{pmatrix} \cosh\tau & \sinh\tau \\ \sinh\tau & \cosh\tau  \end{pmatrix} ~\middle |~ \tau\in \RR \right\}\\
  \text{(parabolic)} & \qquad H &= \left\{ \begin{pmatrix} 1 & \zeta \\ 0 & 1  \end{pmatrix} ~\middle |~ \zeta\in \RR \right\}.
\end{eqnarray}
They can be distinguished by the trace of the non-identity elements:
$<2$ in the elliptic case, $>2$ in the hyperbolic case and $=2$ in the
parabolic case.  They can also be distinguished by the causal nature
of the vectors they leave invariant in the three-dimensional vector
representation of $\so(2,1)$: timelike in the elliptic case, spacelike
in the hyperbolic case and lightlike in the parabolic case.

Since the vector representation of $\SL(2,\RR)$ is isomorphic to the
coadjoint representation, these homogeneous spaces can also be
interpreted as coadjoint orbits and hence, according to Souriau, as
the space of motions of elementary systems.  The evolution spaces can
in all cases be interpreted as the Lie group $\SL(2,\RR)$ itself.  It
is then a matter of interpretation how to project the trajectories on
the evolution space into particle trajectories in the spacetime.

The Lie algebras of these Lie subgroups are given by
\begin{eqnarray}
    \text{(elliptic)} & \qquad \h &= \left\{ \begin{pmatrix} 0 & - z  \\ z & 0  \end{pmatrix} ~\middle |~ z\in \RR \right\}\\
    \text{(hyperbolic)} & \qquad \h &= \left\{ \begin{pmatrix} 0 & z \\ z & 0  \end{pmatrix} ~\middle |~  z\in \RR \right\}\\
  \text{(parabolic)} & \qquad \h &= \left\{ \begin{pmatrix} 0 & z \\ 0 & 0  \end{pmatrix} ~\middle |~ z\in \RR \right\}.
\end{eqnarray}
We will write $\g = \h \oplus \m$ in each case with
\begin{eqnarray}
    \text{(elliptic)} & \qquad \m &= \left\{ \begin{pmatrix} x & y  \\ y & -x  \end{pmatrix} ~\middle |~ x,y\in \RR \right\}\\
    \text{(hyperbolic)} & \qquad \m &= \left\{ \begin{pmatrix} x & -y \\ y & x  \end{pmatrix} ~\middle |~  x,y\in \RR \right\}\\
  \text{(parabolic)} & \qquad \m &= \left\{ \begin{pmatrix} x & 0 \\ y & -x  \end{pmatrix} ~\middle |~ x,y\in \RR \right\}.
\end{eqnarray}
In the elliptic and hyperbolic cases, the split $\g = \h \oplus \m$ is
reductive, so that $[\h,\m] \subset \m$ in the obvious notation,
whereas in the parabolic case no reductive split exists.  Let us write
$\m = \left\{x P_1 + y P_2 ~\middle |~ x,y \in \RR\right\}$ and $\h =
\left\{z B ~\middle |~ z \in \RR\right\}$ in all cases, which defines
the matrices $B,P_1,P_2$:
\begin{eqnarray}
\text{(elliptic)} & \qquad B &= \begin{pmatrix}0 & -1 \\ 1 &
  0 \end{pmatrix} \qquad P_1 = \begin{pmatrix}1 & 0 \\ 0 & -1 \end{pmatrix} \qquad P_2 = \begin{pmatrix}0 & 1 \\ 1 & 0 \end{pmatrix} \\
\text{(hyperbolic)} & \qquad B &= \begin{pmatrix}0 & 1 \\ 1 &
  0 \end{pmatrix} \qquad P_1 = \begin{pmatrix}1 & 0 \\ 0 & -1 \end{pmatrix} \qquad P_2 = \begin{pmatrix}0 & -1 \\ 1 & 0 \end{pmatrix} \\
\text{(parabolic)} & \qquad B &= \begin{pmatrix}0 & 1 \\ 0 &
  0 \end{pmatrix} \qquad P_1 = \begin{pmatrix}1 & 0 \\ 0 & -1 \end{pmatrix} \qquad P_2 = \begin{pmatrix}0 & 0 \\ 1 & 0 \end{pmatrix}.
\end{eqnarray}
In the elliptic case, there is a positive-definite inner product on
$\m$ which is $\Hgr$-invariant: $\left<P_a,P_b\right> = \delta_{ab}$,
whereas in the hyperbolic case there is an $\Hgr$-invariant lorentzian
inner product on $\m$ given by $\left<P_a,P_b\right>= \eta_{ab}$, with
$\eta$ diagonal with $\eta_{11}= - \eta_{22} = 1$.  In the parabolic
case, $P_1$ is $\Hgr$-invariant (in $\g/\h$) and so is the degenerate
bilinear form $b$ whose only nonzero entry is $b(P_2,P_2)$.

We shall describe $\SL(2,\RR)$-invariant particle dynamics on each of
the coset manifolds $\SL(2,\RR)/\Hgr$, where $\Hgr$ is either an elliptic,
hyperbolic or parabolic subgroup.  To do so we will parametrise a
neighbourhood of the identity of $\SL(2,\RR)$ via $g : \RR^3 \to
\SL(2,\RR)$ where
\begin{equation}
  g(x,y,z) = \underbrace{e^{y P_2} e^{x P_1}}_{g_0} \underbrace{e^{z B}}_b,
\end{equation}
where $g_0$ is a coset representative for the spacetime and $b$
corresponds to the extra generator that is broken by the presence of
the particle.  Notice that $B,P_1,P_2$ are defined differently in each of
the three cases, as show above, consistent with this interpretation.

The left-invariant Maurer--Cartan one-form on $\SL(2,\RR)$ pulls back
to $g^{-1}dg \in \Omega^1(\RR^3,\g)$.  Choosing $\alpha \in \g^*$, we
have that
$L_\alpha := \left<\alpha,g^{-1}dg\right> \in \Omega^1(\RR^3)$, where
we have used $\left<-,-\right>$ to denote the dual pairing between
$\g$ and $\g^*$.  Let $I := [a,b] \subset \RR$ and let
$\gamma : I \to \RR^3$ be a regular curve.  We may pull back
$L_\alpha$ via $\gamma$ to produce a one-form
$\gamma^* L_\alpha \in \Omega^1(I)$ which we may integrate to arrive
at the following action functional:
\begin{equation}
  S_\alpha = \int_I \gamma^* L_\alpha.
\end{equation}
We will see that after partially solving the Euler--Lagrange
equations, $S_\alpha$ will induce an action for particle dynamics in
$\SL(2,\RR)/\Hgr$.

\subsubsection{Particle dynamics on the hyperbolic plane}
\label{sec:part-dynam-H}

Despite the name, the hyperbolic plane $\eH_2$ is the quotient of
$SL(2,\RR)$ by an elliptic subgroup.  Let us write
$ g^{-1}dg = \theta^1 P_1 + \theta^2 P_2 + \theta^3 B$, where
\begin{equation}
  \begin{split}
    \theta^1 &= \cosh(2x) \cos(2z) dy - \sin(2z) dx\\
    \theta^2 &= \cos(2z) dx + \cosh(2x) \sin(2z) dy\\
    \theta^3 &= dz + \sinh(2x) dy.
  \end{split}
\end{equation}
The invariant metric on $\SL(2,\RR)/\Hgr$ is given (up to homothety)
by
\begin{equation}
  ds^2 = (\theta^1)^2 + (\theta^2)^2 = dx^2 + \cosh(2x)^2 dy^2.
\end{equation}
The action is given by
\begin{equation}
  S_\alpha = \int_a^b \left(  \alpha_1 (\cosh(2x) \cos(2z) \dot y -
    \sin(2z) \dot x) + \alpha_2 (\cos(2z) \dot x + \cosh(2x)\sin(2z)
    \dot y)+ \alpha_3(\dot z + \sinh(2x) \dot y) \right) dt.
\end{equation}
The Euler--Lagrange equation for $z$ is simply $\frac{\d L}{\d z} =
0$, which is equivalent to
\begin{equation}
  \alpha_1 (\cosh(2x) \sin(2z) \dot y + \cos(2z) \dot x) = \alpha_2
  (\cosh(2x)\cos(2z)\dot y - \sin(2z)\dot x),
\end{equation}
from where we may solve (implicitly) for $z$ as follows:
\begin{equation}
  \tan (2z) = \frac{\alpha_2 \cosh(2x) \dot y - \alpha_1 \dot
    x}{\alpha_1 \cosh(2x)\dot y + \alpha_2 \dot x}.
\end{equation}
Reinserting into the action (and dropping total derivatives), we
arrive at
\begin{equation}
  S'_\alpha = \int_a^b \left(\sqrt{\alpha_1^2 + \alpha_2^2} \sqrt{\dot
    x^2 + \cosh(2x)^2 \dot y^2} + \alpha_3 \sinh(2x) \dot y \right) dt.
\end{equation}
We recognise the first term as the line element in $\eH_2$  with
hyperbolic metric
\begin{equation}\label{eq:metric-H}
  ds^2 = (\alpha_1^2 + \alpha_2^2) (dx^2 + \cosh(2x)^2 dy^2),
\end{equation}
whereas the second term is the coupling to a Maxwell field
\begin{equation}
  A = \alpha_3 \sinh(2x) dy,
\end{equation}
whose fieldstrength
\begin{equation}
  F = dA = 2 \alpha_3 \cosh(2x) dx \wedge dy =
  \frac{2\alpha_3}{\alpha_1^2 + \alpha_2^2} \dvol,
\end{equation}
where $\dvol$ is the hyperbolic area form of the metric in
equation~\eqref{eq:metric-H}.

\subsubsection{Particle dynamics on de~Sitter spacetime}
\label{sec:part-dynam-dS}

This case is very similar \emph{mutatis mutandis} to the previous
case, although now we quotient by a hyperbolic subgroup.  Again we
write $g^{-1}dg = \theta^1 P_1 + \theta^2 P_2 + \theta^3 B$, where
\begin{equation}
  \begin{split}
    \theta^1 &= \cosh(2x) \cosh(2z) dy + \sinh(2z) dx\\
    \theta^2 &= \cosh(2z) dx + \cosh(2x) \sinh(2z) dy\\
    \theta^3 &= dz - \sinh(2x) dy.
  \end{split}
\end{equation}
The invariant metric on $\SL(2,\RR)/\Hgr$ is now given (up to homothety)
by
\begin{equation}
  ds^2 = (\theta^1)^2 - (\theta^2)^2 = - dx^2 + \cosh(2x)^2 dy^2.
\end{equation}
As we see from the metric, $x$ is a time coordinate and $y$ is a space
coordinate.  This metric could also be re-interpreted as
$\mathsf{AdS}_2$, by reinterpreting $x$ as space and $y$ as time.

The action is now given by
\begin{equation}
  S_\alpha = \int_a^b \left(  \alpha_1 (\cosh(2x) \cosh(2z) \dot y +
    \sinh(2z) \dot x) + \alpha_2 (\cosh(2z) \dot x + \cosh(2x)\sinh(2z)
    \dot y)+ \alpha_3(\dot z - \sinh(2x) \dot y) \right) dt.
\end{equation}
The Euler--Lagrange equation for $z$ is again simply $\frac{\d L}{\d z} =
0$, which translates into
\begin{equation}
  \alpha_1 (\cosh(2x) \sinh(2z) \dot y + \cosh(2z) \dot x) + \alpha_2
  (\cosh(2x)\cosh(2z)\dot y + \sinh(2z)\dot x),
\end{equation}
and which allows us to solve for $z$ implicitly:
\begin{equation}
  \tanh (2z) = \frac{-(\alpha_2 \cosh(2x) \dot y + \alpha_1 \dot x)}{\alpha_1 \cosh(2x)\dot y + \alpha_2 \dot x}.
\end{equation}
Reinserting into the action (and dropping total derivatives), we
arrive at (see, also, \cite[eq.(5.10)]{Anabalon:2006ii})
\begin{equation}
  S'_\alpha = \int_a^b \left(\sqrt{\alpha_1^2 - \alpha_2^2} \sqrt{-\dot
    x^2 + \cosh(2x)^2 \dot y^2} - \alpha_3 \sinh(2x) \dot y \right) dt.
\end{equation}
We recognise the first term as the line element in $\dS_2$  with
metric
\begin{equation}\label{eq:metric-dS}
  ds^2 = (\alpha_1^2 - \alpha_2^2) (-dx^2 + \cosh(2x)^2 dy^2),
\end{equation}
whereas the second term is the coupling to a Maxwell field
\begin{equation}
  A = -\alpha_3 \sinh(2x) dy,
\end{equation}
whose fieldstrength
\begin{equation}
  F = dA = - 2 \alpha_3 \cosh(2x) dx \wedge dy =
  \frac{2\alpha_3}{\alpha_2^2 - \alpha_1^2} \dvol,
\end{equation}
where $\dvol$ is now the area form of the de Sitter metric in
equation~\eqref{eq:metric-dS}.

\subsubsection{Particle dynamics on the lightcone}
\label{sec:part-dynam-L}

Finally, we discuss the parabolic case.  Again we write $g^{-1}dg =
\theta^1 P_1 + \theta^2 P_2 + \theta^3 B$, where now
\begin{equation}
  \begin{split}
    \theta^1 &= e^{-2x}dy \\
    \theta^2 &= dx - z e^{-2x} dy\\
    \theta^3 &= dz - 2 z dx + z^2 e^{-2x} dy.
  \end{split}
\end{equation}
There is no $\SL(2,\RR)$-invariant metric here, but only a carrollian
structure $(\kappa,\eta)$, where the carrollian vector field is $\kappa =
z \d_x + e^{2x} \d_y$ and the carrollian degenerate metric is given by
$\eta = (dx - z e^{-2x}dy)^2$.

The action is now given by
\begin{equation}
  S_\alpha = \int_a^b \left(  \alpha_1 e^{-2x} \dot y + \alpha_2 (\dot
    x - z e^{-2x}\dot y) + \alpha_3 (\dot z - 2z \dot x + z^2 e^{-2x} \dot y) \right) dt.
\end{equation}
The Euler--Lagrange equation for $z$ is again simply $\frac{\d L}{\d z} =
0$, which is easily solved for $z$:
\begin{equation}
  z = \frac{\alpha_2}{2\alpha_3} + e^{2x} \frac{\dot x}{\dot y}.
\end{equation}
Reinserting into the action (and dropping total derivatives), we
arrive at
\begin{equation}
  S'_\alpha = \int_a^b \left(\left( \alpha_1 -
    \frac{\alpha_2^2}{4\alpha_3}\right) e^{-2x} \dot y^2 - \alpha_3 e^{2x}
  \dot x^2 \right)\frac{dt}{\dot y}.
\end{equation}
Choosing the ``static gauge'' where $\dot y  = 1$ and changing
variables to $u = e^x$, we arrive at the following action
\begin{equation}
  S''_\alpha = \int_a^b \left(\left( \alpha_1 -
    \frac{\alpha_2^2}{4\alpha_3}\right) u^{-2} - \alpha_3 \dot u^2 \right) dt,
\end{equation}
which we recognise as a version of the one-dimensional conformal
mechanics of \cite{deAlfaro:1976vlx}.

\subsubsection{One-dimensional Schwarzian particle}
\label{sec:schwarzian}

Here we will rederive the $\SL(2,\RR)$-invariant Schwarzian action of
\cite{Kitaev:2017awl,Maldacena:2016hyu,Stanford:2017thb} using the
method of nonlinear realisations and the inverse Higgs mechanism
applied to $\SL(2,\RR)$.  An alternative derivation using nonlinear
realisations for $\SL(2,\RR)\times R^+$ can be found in
\cite{Galajinsky:2019lak}.

Let $\RR P^1$ denote the real projective line: the space of straight
lines through the origin in the plane $\RR^2$.  The real projective
line is diffeomorphic to the circle.  Given a diffeomorphism $\varphi : \RR
P^1 \to \RR P^1$, we define its \emph{Schwarzian derivative} (or
simply its Schwarzian) by the formula
\begin{equation}\label{eq:schwarzian-derivative}
  \Sch(\varphi) := \frac{\varphi'''}{\varphi'} - \frac32
  \left(\frac{\varphi''}{\varphi'} \right)^2,
\end{equation}
where the primes represent derivatives with respect to the local
coordinate on $\RR P^1$.  The Schwarzian defines a quadratic
differential on $\RR P^1$ or, in physical terms, a quasiprimary field
with weight $2$ under diffeomorphisms of the circle and it plays an
important rôle in projective geometry (see, e.g., \cite{MR2177471}).
One of its most important properties is its invariance under
$\PSL(2,\RR)$ Möbius transformations:
\begin{equation}
  \varphi \mapsto \frac{a \varphi + b}{c \varphi + d}
  \qquad\text{with}\qquad  ad - bc = 1.
\end{equation}

We will re-use the basis for $\sl(2,\RR)$ in
Section~\ref{sec:part-dynam-L}, but with a change of notation to
reflect that $\SL(2,\RR)$ is the one-dimensional group of
conformal transformations.  Therefore we introduce the basis $H,K,D$
for $\sl(2,\RR)$ where
\begin{equation}
  K = \begin{pmatrix}0 & 1 \\ 0 &
  0 \end{pmatrix}, \qquad D = \begin{pmatrix}1 & 0 \\ 0 &
  -1 \end{pmatrix} \qquad\text{and}\qquad H = \begin{pmatrix}0 & 0 \\
  1 & 0 \end{pmatrix}.
\end{equation}
Here $H$ generates translations, $K$ generates special conformal
transformations and $D$ generates dilatations, which are all the
one-dimensional conformal transformations.  The ad-invariant inner
product on $\sl(2,\RR)$, which is a multiple of the Killing form, can
be normalised to $\left<D,D\right>=2$ and $\left<H,K\right>=1$ in this
basis.

We will choose a local chart $(\rho,y,u)$ for $\SL(2,\RR)$ different
than the one we introduced in Section~\ref{sec:part-dynam-L} to derive
the lagrangian for one-dimensional conformal mechanics.  We shall
parametrise group elements near the identity by
\begin{equation}
  g = \underbrace{e^{\rho H}}_{g_0} e^{yK} e^{uD},
\end{equation}
where $g_0$ is a local coset representative for the one-dimensional
conformal space thought of as the coset space $\SL(2,\RR)/\Hgr$, with
$\Hgr$ the two-dimensional non-abelian Lie group generated by $K$ and
$D$.  Explicitly, the above parametrisation is
\begin{equation}
  g =
  \begin{pmatrix}
    e^u & e^{-u} y\\ e^u \rho & e^{-u} (1 + \rho y)
  \end{pmatrix}.
\end{equation}
The Maurer--Cartan form is given by
\begin{equation}
  \begin{split}
    \Omega = g^{-1}dg &= \Omega_H H + \Omega_D D + \Omega_K K \\
    &= e^{2u} d\rho H + (du - y d\rho) D + e^{-2u} (dy - y^2 d\rho) K.
  \end{split}
\end{equation}
In contrast to what we did in Section~\ref{sec:part-dynam-L}, the
lagrangian here will not be linear in the components of the
Maurer--Cartan form, but rather quadratic, resulting from applying the
inverse Higgs mechanism to the lagrangian for geodesic motion on
$\SL(2,\RR)$ relative to the bi-invariant metric
\begin{equation}
  \left<\Omega,\Omega\right> = 2 \Omega_D^2 + 2 \Omega_H \Omega_K = 2 \left( du^2 - 2 y du d\rho + d\rho dy \right).
\end{equation}
The geodesic lagrangian is obtained by pulling back the metric to
the interval parametrising the world-line of the particle:
\begin{equation}\label{eq:geodesics-sl2R}
  L = \tfrac12 \left<g^{-1}g', g^{-1} g'\right> = u'^2 + (y' - 2 y u') \rho',
\end{equation}
where we are using primes to denote differentiation with respect to
the parameter of the world-line of the particle.  This lagrangian is
invariant under both left and right multiplication by
$\SL(2,\RR)$.  For example, under infinitesimal left multiplication
with parameter $\alpha H + \beta D + \gamma K$, we have
\begin{equation}\label{eq:leftglobalsymmSch}
  \delta u = \beta + \gamma \rho, \qquad \delta \rho = \alpha - 2
  \beta \rho + \gamma \rho^2 \qquad\text{and}\qquad \delta y = \gamma
  + 2 y \beta + 2 \rho y \gamma.
\end{equation}
We recognise in \eqref{eq:leftglobalsymmSch} the transformation of
the Goldstone field $\rho$ under an infinitesimal Möbius
transformation.

We can reduce the number of Goldstone fields in the action by imposing
some conditions on the Maurer--Cartan form, a procedure also known as
the inverse Higgs mechanism \cite{Ivanov:1975zq}.  In the present
context, the conditions are familiar from Drinfel'd--Sokolov reduction
\cite{Drinfeld:1984qv,Polyakov:1989dm} and are given by
\begin{equation}\label{eq:constraintSL2R}
  \Omega_H = 1 \qquad\text{and}\qquad \Omega_D = 0.
\end{equation}
This breaks the symmetry of the lagrangian under right multiplication,
leaving only the global symmetry described infinitesimally in
equation~\eqref{eq:leftglobalsymmSch}.

We can solve the constraints~\eqref{eq:constraintSL2R} explicitly for
$u,y$ in terms of $\rho$:
\begin{equation}\label{eq:constraintSL2R-solved}
  y = \frac{u'}{\rho'} \qquad\text{and}\qquad u = \tfrac12 \log\left( \frac1{\rho'} \right).
\end{equation}
It follows that
\begin{equation}
  u'= -\tfrac12 \frac{\rho''}{\rho'} \qquad\text{and}\qquad   y' =
  \frac{\rho''^2}{\rho'^3} - \frac12 \frac{\rho'''}{\rho'^2}.
\end{equation}
Substituting this in the lagrangian \eqref{eq:geodesics-sl2R}, we
obtain
\begin{equation}
  L = -\frac12 \left( \frac{\rho'''}{\rho'} - \frac32
    \left(\frac{\rho''}{\rho'} \right)^2 \right) = -\tfrac12 \Sch(\rho).
\end{equation}

In summary, it is possible to obtain the Schwarzian action using the
inverse Higgs mechanism.  It is also possible to obtain the Schwarzian
action by integrating out the gauge transformations of the particle
model with variables $x^\mu, \lambda$ and lagrangian $L=\frac 12 \dot
x^2-\frac 12 \lambda x^2$, as in \cite{Siegel:1988ru,Gomis:1993pp}.

\subsection{Non-relativistic limit of relativistic particle actions}
\label{sec:nrlimits-particles}

In this section we obtain some of the non-lorenztian particle
dynamics studied in the previous sections as non-relativistic limits
of relativistic particles.

\subsubsection{Non-relativistic limit of the $\mathsf{AdS}_{d+1}$ particle  action}

We consider the action of a massive particle propagating in
$\mathsf{AdS}_{d+1}$, with metric
\begin{equation}\label{eq:AdS-metric}
  ds^2 = - \cosh^2{\frac{r}{R}}(dX^0)^2+\left(
\frac{\sinh{\frac{r}{R}}}{{\frac{r}{R}}}\right)^2 (dX^a)^2-
\left(\left( \frac{\sinh{\frac{r}{R}}}{{\frac{r}{R}}}\right)^2-1\right)
(dr)^2,
\end{equation}
where $r=\sqrt{X_a X^a}$.  The particle lagrangian is that describing
geodesic motion on this geometry:
\begin{equation}\label{eq:AdS-lag}
  L = -m\sqrt{\cosh^2{\frac{r}{R}}(\dot X^0)^2-\left(
\frac{\sinh{\frac{r}{R}}}{{\frac{r}{R}}}\right)^2 (\dot X^a)^2+
\left(\left(\frac{\sinh{\frac{r}{R}}}{{\frac{r}{R}}}\right)^2-1\right)
(\dot r)^2}.
\end{equation}
In order to take the non-relativistic limit, we introduce an
invertible change of variables with a dimensionless parameter $\omega$:
\begin{equation}
  X^0=\omega x^0, \qquad  m=\omega M \qquad\text{and}\qquad  R= \omega \tilde R.
\end{equation}
After this change of variable, the lagrangian becomes
\begin{equation}
L = -M\omega^2 \dot x^0 + \frac{M(\dot x^a)^2}{2\dot x^0}- \dot
x^0\frac{Mr^2}{2R^2} + O(\omega^{-2}).
\end{equation}
The omitted terms $O(\omega^{-2})$ will not contribute in the limit
$\omega \to \infty$, but this limit is problematic due to the presence
of a quadratically divergent term.  This term may be cancelled if we
introduce at the relativistic level a coupling to a constant
electromagnetic field \cite{Gomis:2000bd,Gomis:2005pg} $A$ (with
$F=dA=0$) in such a way we preserve the same physical degrees of
freedom:
\begin{equation}
  L_{\text{em}}= A_\mu e^\mu, \quad A_\mu=(M\omega, \vec 0),
\end{equation}
where $e^\mu$ are the components of the vielbein of the
metric~\eqref{eq:AdS-metric}.

Doing so and taking the limit $\omega \to \infty$, the lagrangian
becomes
\begin{equation}
  L=\frac{M(\dot x^a)^2}{2\dot x^0}-\dot x^0\frac{M r^2}{2R^2},
\end{equation}
which takes the expected form of the reparametrization invariant
$\n^-$-particle lagrangian in equation~\eqref{eq:BNHLag}.

\subsubsection{Massless galilean particle and the non-relativistic limit of a tachyon}

Now we will consider the non-relativistic limit of a tachyon.  We
start with the relativistic canonical action of a tachyon of mass m
\begin{equation}
  S= \int d\tau \left( p_A\dot x^A- \frac{e}{2} \left( {p}^{\, 2} -m^2c^2\right)\right).
\end{equation}
The non-relativistic limit is defined by taking $c \to \infty$ in
\begin{equation}
  x^0=ct, \qquad\text{and}\quad  p_0=-\frac{E}{c}
\end{equation}
while keeping the \emph{colour} $k=mc$ finite. The action
becomes~\cite{Batlle:2017cfa}
\begin{equation}\label{eq:massG}
  S= \int d\tau \left( -E \dot{t} + \vec{p}\cdot \dot{\vec{x}}- \frac{e}{2} \left(  \vec{p}^{\, 2} -k^2 \right)\right).
\end{equation}
If we eliminate the momenta we have
\begin{equation}\label{eq:massG1}
  S= \int d\tau \left( -E \dot{t} + \frac{k}{2} \sqrt{\dot{\vec{{x}}}^2}\right).
\end{equation}
In this form the action can be interpreted a relativistic tachyonic
particle with an instantaneous interaction \cite{Barducci:2018wuj}.  The
field theory associated to this particle model is the galilean
magnetic Klein--Gordon field theory as in
equation~\eqref{case1bflat}.

The non-relativistic limit of the one-dimensional conformal mechanics
and of the Schwarzian particle has been studied in
\cite{Grumiller:2020elf,Gomis:2020wxp}.

\subsection{Carrollian limits of particle actions}

In this section we obtain some of the non-lorenztian particle
dynamics studied in the previous sections as carrollian limits of
relativistic particles.

\subsubsection{Carrollian limit of a massive particle in $\mathsf{AdS}_{d+1}$ background}

We consider the canonical action of a massive particle in
AdS$_{d+1}$ background
\begin{equation}
  S= \int d\tau \left( p_\mu\dot x^\mu- \frac{e}{2} \left(
      g^{\mu\nu}p_\mu p_\nu -m^2\right)\right),
\end{equation}
where $g^{\mu\nu}$ is the inverse metric of \eqref{eq:AdS-metric}.

The carrollian limit is defined by taking $\omega \to \infty$ in
\begin{equation}\label{eq:carrolllimit}
  x^0=\frac{t}{\omega}, \qquad p_0=-\omega E,, \qquad\text{and}\qquad m=M\omega,
\end{equation}
and keeping $R$ fixed.  It is understood that, before taking the
limit,  we rescale the Einbein variable like
\begin{equation}
  e\to\frac{-e}{\omega^2}.
\end{equation}
The carrollian action is given by
\begin{equation}
  \label{eq:carrollian-action}
  S_{C} =  \int d\tau\, \left( - E \dot{t} + \dot{\vec{x}}\cdot\vec{p}
    -\frac{e}{2} \left(\frac{E^2}{\cosh^2\frac{r}{R}} - M^2\right)  \right).
\end{equation}
A particle in AdS Carroll does not move. The Carroll particle in flat
space time is obtained by sending $R\to\infty$ and it can be written as
\begin{equation}
  S=\int d\tau(-M\sqrt{\dot t^2}+M\vec p\dot{\vec x}),
\end{equation}
which can be interpreted\footnote{We acknowledge discussions with
  Roberto Casalbuoni on this point.} as a timelike relativistic
particle which is at rest in a given point in space: $\dot{\vec x}=0$
\cite{Barducci:2018wuj}.  The field theory associated to this particle
model is the Carroll electric Klein--Gordon field theory as in
equation~\eqref{eq:case3c-flat} \cite{Bergshoeff:2014jla}.

The massless Carroll particle is obtained by putting $M=0$ in
equation~\eqref{eq:carrollian-action}.

\subsubsection{Carrollian limit of relativistic tachyon}

We consider the action of a relativistic tachyon in configuration
space the action is given by
\begin{equation}
   S=-mc\int d\tau \sqrt{({\dot {\vec x}})^2-({\dot {x}^0})^2}.
\end{equation}
In order to take the carrollian limit, it is useful to introduce the
carrollian time and mass $M$ given by
\begin{equation}
  s=C, \qquad x^0=C ct \qquad\text{and}\qquad mc=MC.
\end{equation}
Substituting in the action, we obtain
\begin{equation}
  S = M C \int d\tau \sqrt{ \dot{\vec{x}}^{2} - \frac{\dot{s}^2}{C^2}}.
\end{equation}
The Carroll limit in these variables is given by taking
\begin{equation}
  s\to\infty \qquad\text{and}\qquad MC\to \tilde M.
\end{equation}
The Carroll action of a tachyon is given by \cite{deBoer:2021jej}
\begin{equation}
  L=-\tilde M\sqrt{({\dot {\vec X}})^2}.
\end{equation}
The canonical action is given by
\begin{equation}\label{eq:Carrolltach}
  S= \int d\tau \left( -E \dot{t} + \vec{p}\cdot \dot{\vec{x}}-
    \frac{e}{2} \left(  \vec{p}^{\, 2} -\tilde M^2 \right)-\mu
    E\right).
\end{equation}
The quantisation of the constraints gives the Helmholtz equation
\begin{equation}
  (\nabla^2+M^2)\Phi(t,\vec x)=0 \qquad\text{and}\qquad
  \frac{\partial}{\partial t}\Phi(t,\vec x)=0.
\end{equation}
Note that the Helmholtz equation as also appearing the quantisation of
massless galilean particles. The relation among Galilei and Carroll
particles in its $v/c$ coorrections is analysed in \cite{Gomis:2022spp}.

This ends our discussion of the dynamics of non-lorentzian particles.

\section{Gravity}
\label{sec:gravity}

This section contains three subsections. In the first two subsections
we will describe gravity from a kinematical point of view by a gauging
procedure that uses the Lie algebra of symmetries that underlies the
theory as a starting point. We call it a gauging procedure because
there are additional steps involved as compared to gauging a Lie
algebra of internal symmetries leading to Yang-Mills which makes the
relation between the final result and the original Lie algebra less
direct (see, e.g., \cite{Chamseddine:1976bf}). In the first subsection
we explain this gauging procedure for the relativistic case while in
the second subsection we will focus on three non-lorentzian algebras:
the Bargmann algebra underlying Newton-Cartan gravity, the Galilei
algebra and the Carroll algebra. In the third subsection we will
describe Newton-Cartan gravity from a dynamical point of view by
defining a suitable non-relativistic limit of the Einstein equations
of motion. Next, we will discuss the non-lorentzian gravity theories
underlying the Galilei and Carroll algebras, called Galilei gravity
and Carroll gravity, respectively.

\subsection{Gauging the Poincaré algebra}
\label{subsec:gauging}

We first consider the relativistic case. Our starting point is the Poincaré algebra
\begin{multicols}{2}
\begin{subequations}\label{eq:PoincareAlgebraCommutators}
\setlength{\abovedisplayskip}{-15pt}
\allowdisplaybreaks
\begin{align}
[P_{{A}},P_{{B}}]& = 0\, ,\\
    [M_{{A}{B}},P_{{C}}]
    & = 2\eta_{{C}[{B}}P_{{A}]}\, , \\
    [M_{{A}{B}},M_{{C}{D}}]
    &= 4\eta_{[{A}[{C}}M_{{D}]{B}]}\, ,
\end{align}
\end{subequations}
\end{multicols}
\noindent
where $P_{{A}}$ and $M_{{A}{B}}$ are the generators of
spacetime translations and Lorentz transformations, respectively.  The
capital indices run over ${A}=0,...,d$ and we have chosen the
Minkowski metric to have mostly plus signature. In this subsection we will apply a gauging procedure to this Poincaré algebra keeping the relativistic symmetries intact.

As a first step in the gauging procedure, we associate to the translation and Lorentz rotation generators the independent gauge fields $E_\mu{}^{ A}$ and $\Omega_\mu{}^{A B}$ which we call the Vielbein and Lorentz spin-connection, respectively:
\begin{equation}
A_\mu^I T_I  = E_\mu{}^{ A} P_{ A} + \frac{1}{2} \Omega_\mu{}^{ A B}M_{ A B}\,.
\end{equation}
These gauge fields transform as covariant vectors under  a general coordinate transformation with parameter $\xi^\mu$ while  their P-transformations corresponding to the translation generators $P_{ A}$, with parameters $\eta^{ A}$, and their Lorentz transformation rules corresponding to the Lorentz generators $M_{ A B}$, with parameters $\Lambda^{ A B}$, follow from the structure constants $f^I{}_{JK}$ of the Poincaré algebra:
\begin{equation}\label{transfgf}
\delta A_\mu^I =   \xi^{\lambda}\partial_{\lambda} A_\mu^I  +\partial_{\mu}\xi^{\lambda} A_\lambda^I + \partial_\mu\Lambda^I - f^I{}_{JK}\Lambda^JA_\mu^K
\end{equation}
or
\begin{subequations}
\begin{eqnarray}\label{gctL}
  \delta E_{\mu}{}^{{A}}
  & = &
        \xi^{\lambda}\partial_{\lambda} E_{\mu}{}^{{A}}
        +\partial_{\mu}\xi^{\lambda} E_{\lambda} {}^{{A}} +  \partial_{\mu} \eta^{{A}} -\Omega_{\mu}{}^{{A}}{}_{{B}}\eta^{{B}}
        +\Lambda^{{A}}{}_{{B}}E_{\mu}{}^{{B}}\, ,
  \\[4pt]
  \delta \Omega_{\mu}{}^{{A}{B}}
  & = &
        \xi^{\lambda}\partial_{\lambda}  \Omega_{\mu}{}^{{A}{B}}
        +\partial_{\mu}\xi^{\lambda}  \Omega_{\lambda}{}^{{A}{B}}
        +\partial_{\mu} \Lambda^{{A}{B}}
        +2\Lambda^{{C} [{A}}\Omega_{\mu}{}^{{B}]}{}_{{C}}\,.\label{second}
\end{eqnarray}
\end{subequations}

We now wish to argue that in the context of general relativity, the general coordinate transformations, Lorentz rotations and P-transformations do not define three independent symmetries of the Einstein-Hilbert action. To write down such an Einstein-Hilbert action we first define the curvature tensors associated to each gauge field as follows:
\begin{equation}\label{curvatures}
R_{\mu\nu}{}^I(T) = 2\partial_{[\mu} A_{\nu]}^I + \frac{1}{2}f^I{}_{JK}A_\mu^J A_\nu^K
\end{equation}
or
\begin{eqnarray}
R_{\mu\nu}{}^{ A}(P) &=& 2\partial_{[\mu}E_{\nu]}{}^{ A} - 2 \Omega_{[\mu}{}^{ A}{}_{ B}E_{\nu]}{}^{ B}\,\\[.1truecm]
R_{\mu\nu}{}^{ A B}(M) &=& 2\partial_{[\mu}\Omega_{\nu]}{}^{ A B} +2 \Omega_{[\mu}{}^{ A C}\Omega_{\nu]}{}^{ B}{}_{ C}\,.
\end{eqnarray}
The Ricci tensor and Ricci scalar are  defined by
\begin{eqnarray}
&&R_\mu{}^{ A}(M) = - E^\nu{}_{ C} R_{\mu\nu}{}^{ C A}(M)\,,\hskip 1.5truecm R(M) = E^\mu{}_{ A} R_\mu{}^{ A}(M),
\end{eqnarray}
where we have used the inverse Vielbein field $E^\mu{}_{ A}$ defined by
\begin{equation}
E^\mu{}_{ A} E_\mu{}^{ B} = \delta_{ A}{}^{ B}\,,\hskip 2truecm E^\mu{}_{ A} E_\nu{}^{ A} = \delta_\nu{}^\mu\,.
\end{equation}

We now consider the  Einstein-Hilbert action (without cosmological constant)
\begin{equation}\label{standard}
S_{\rm EH} = \frac{1}{16 \pi {\rm G}_N} \int d^{d+1} x\,  E R(M)\,,
\end{equation}
where $E$ is the determinant of the Vielbein field $E_\mu{}^{ A}$ and ${\rm G}_N$ is Newton's constant. By construction this action is invariant under general coordinate transformations and Lorentz rotations. However, except for $d=2$, it is not manifestly invariant under the P-transformations given in equation~\eqref{gctL} of the Poincaré algebra. This can for instance be seen by writing the Einstein-Hilbert action \eqref{standard} in the equivalent form\,\footnote{The equivalence between the expressions \eqref{standard} and \eqref{equivalent} can be seen by writing out the definition of the determinant $E$ in equation~\eqref{standard}:
\begin{equation}
E = \frac{1}{(d+1)!} \epsilon^{\mu_0\cdots \mu_{d}}\epsilon_{{ A}_0\cdots { A}_{d}}E_{\mu_0}{}^{{ A}_0}\cdots E_{\mu_{d}}{}^{{ A}_{d}}\,.
\end{equation}}
\begin{equation}\label{equivalent}
S_{\rm EH} = \frac{1}{16 \pi {\rm G}_N} \int d^{d+1} x\,
\epsilon^{\mu_0\cdots \mu_{d}}\epsilon_{{ A}_0\cdots { A}_{d}}E_{\mu_0}{}^{{ A}_0}\cdots E_{\mu_{d-2}}{}^{{ A}_{d-2}}
R_{\mu_{d-1}\mu_{d}}{}^{{ A}_{d-1}{ A}_{d}}(M)\,.
\end{equation}
The special thing about $d=2$ is that the Einstein-Hilbert action as given in \eqref{equivalent} reduces to the Chern-Simons form
\begin{equation}\label{CSform}
S_{\rm EH} = \frac{1}{16 \pi {\rm G}_N} \int d^{3} x\,
\epsilon^{\mu\nu\rho}\epsilon_{ A B C}E_{\mu}{}^{{ A}}
R_{\nu\rho}{}^{ B C}(M)\,,
\end{equation}
which is manifestly invariant under all the gauge symmetries of the three-dimensional Poincaré algebra.  In $d=3$, one could consider, besides the term
\begin{equation}
\epsilon^{\mu\nu\rho\sigma}\epsilon_{ A B C D}E_\mu{}^{ A}E_\nu{}^{ B} R_{\rho\sigma}{}^{ C D}(M)
\end{equation}
given in \eqref{equivalent} the so-called Holst term \cite{Holst:1995pc}
\begin{equation}
\alpha \epsilon^{\mu\nu\rho\sigma}E_\mu{}^{ A}E_\nu{}^{ B} R_{\rho\sigma}{}_{ A B}(M)\,,
\end{equation}
where $\alpha$ is a real parameter. The two terms together  give rise to the usual  Einstein equations. This can be seen by first noting that varying the action with respect to the spin-connection gives the same equation of motion as without the Holst term. This follows from the following identity:
\begin{equation}
X^{ A B} +\alpha \epsilon^{ A B C D}X_{ C D} =0\hskip .5truecm \rightarrow\hskip .5truecm X^{ AB}=0\,,
\end{equation}
where  $X^{ AB}$ is a three-form given by
\begin{equation}
X_{\mu\nu\rho}^{ A B} = R_{[\mu\nu}{}^{[ A}(P)E_{\rho]}{}^{ B]}\,.
\end{equation}
The field equation $X^{ A B}=0$ implies $R_{\mu\nu}{}^{ A}(P)=0$ which is a curvature constraint that can be used  to solve for the spin-connection as will be explained below, see the solution given in equation~\eqref{solution}. Next, varying the action with respect to the Vielbein, the Holst term does not contribute to the equations of motion for a dependent spin-connection due to the Riemann tensor identity $R_{[ A B C] D}(M)=0$.

Although, for $d>2$, the Einstein-Hilbert action is not invariant under P-transformations, it does  transform into terms that vanish upon using  the
equation of motion of the spin connection field which is given by
\begin{equation}\label{eomO}
R_{\mu\nu}{}^{{A}}(P)=0\,.
\end{equation}
Such a variation can always be
cancelled by adding terms to the $P$-transformation rule of the spin
connection. After a long calculation one finds the result given in equation~\eqref{Ptransf}. Note that, except for the first term, all terms in the transformation rule \eqref{Ptransf} are proportional to the Ricci tensor and Ricci scalar and therefore vanish upon using the equations of motion corresponding to the inverse Vielbein field $E^\mu{}_{ A}$, i.e.~the Einstein equations:
\begin{equation}
R_\mu{}^{ A}(M) - 2E_\mu{}^{ A}R(M) =0\,.
\end{equation}
 One thus ends up with a set of P-transformations that do not straightforwardly follow from the Poincaré algebra. Instead of doing the long calculation mentioned above to obtain the transformation rule \eqref{Ptransf}, there is an easier way to  derive the P-transformations of the spin-connection field by making use of the fact that  these P-transformations are not new but, instead, related to the general coordinate transformations and the Lorentz transformations of the Poincaré algebra. To show this relation  we need to make use of a special symmetry which in the literature is called a `trivial'  or `equation of motion' symmetry (see,  e.g., \cite{Freedman:2012zz}. These symmetries, which are easier to derive than the P-transformation of the spin-connection field,  are called `trivial' because they have the  distinguishing
feature that all terms in the transformation rules vanish upon using the equations of motion. They therefore  correspond to vanishing Noether charges.
A most simple example of a trivial symmetry is provided by the following action describing two real Klein-Gordon scalars $A$ and $B$:
\begin{equation}
S =   \int d^{d+1} x\, \frac{1}{2}\big( A\Box A + B\Box B\big)\,.
\end{equation}
 This action is invariant under the trivial symmetries with parameter $\lambda$
 \begin{equation}
 \delta A = \lambda \Box B\,,\hskip 2truecm \delta B = -\lambda \Box A\,.
 \end{equation}
 We can see this by writing
 \begin{equation}\label{seen1}
 \delta S = \lambda \frac{\delta S}{\delta\phi^i}\Omega^{ij}\frac{\delta S}{\delta\phi^j} =0
 \end{equation}
for  $\phi^i = (A,B)$ and using the fact that $\Omega^{ij} = \epsilon^{ij}$ is anti-symmetric.

 Similarly, the Einstein-Hilbert action is invariant under the following trivial symmetries with  parameters $\sigma^{ A}$:
\begin{subequations}
\begin{eqnarray}\label{eom}
  \delta E_{\mu}{}^{{A}}
  & = &
        R_{\mu\nu}{}^{{A}}(P) \sigma^{\nu}\,,
  \\[4pt]
  \delta \Omega_{\mu}{}^{{A}{B}}
  & = &
        -R_{\mu}{}^{[{A}}(M) \sigma^{{B}]}
        -\frac{1}{2}  E_{\mu}{}^{[{A}}R_{{C}}{}^{{B}]}(M) \sigma^{{C}}
        +\frac{3}{4}  E_{\mu}{}^{[{A}} R(M) \sigma^{{B}]}\,,
\end{eqnarray}
\end{subequations}
with $\sigma^{\nu} \equiv \sigma^{{B}} E^{\nu}{}_{{B}}$. Like in the example of the two scalar fields the Vielbein field transforms to the equation of the spin-connection field while the spin-connection field transforms to the equation  of motion of the Vielbein field leading to a zero variation of the action as follows:
\begin{equation}
\delta S \sim
\begin{pmatrix}
  \frac{\delta S}{\delta E_\mu{}^{ A}} &\frac{\delta S}{\delta \Omega_\rho{}^{ C D}}
\end{pmatrix}
\begin{pmatrix}
  0&E^\mu{}_{ C}E^\rho{}_{ A}\\
  -E^\mu{}_{ C}E^\rho{}_{ A} &0
\end{pmatrix}
\begin{pmatrix}
  \frac{\delta S}{\delta E_\mu{}^{ A}}\\
    \frac{\delta S}{\delta \Omega_\rho{}^{ C D}}
\end{pmatrix}
\sigma_{ D} =0\,.
\end{equation}

Using these trivial symmetries, we can write the P-transformation given in
equation~\eqref{gctL} of the Vielbein field as the sum of a special general
coordinate transformation, Lorentz transformation and trivial symmetry
transformation with parameters given by
\begin{equation}
  \xi^{\mu}
  = \eta^{\mu}\,,
  \hskip 1truecm
  \Lambda^{{A}{B}}=\eta^{\lambda}\Omega_{\lambda}{}^{{A}{B}}\,,
  \hskip 1truecm
  \sigma^{{A}} =  \eta^{{A}}\,,
\end{equation}
with $\eta^{\mu} \equiv \eta^{{B}} E^{\mu}{}_{{B}}$.
Since the same decomposition rule must apply to the spin-connection field, it follows that the P-transformation of this spin-connection field is given by
\begin{eqnarray}\label{Ptransf}
  \delta_\eta \Omega_{\mu}{}^{{A}{B}}
  & = &
        \eta^{\lambda} R_{\lambda\mu}{}^{{A}{B}}(M)
        +R_{\mu}{}^{[{A}}(M) \eta^{{B}]}
        +E_{\mu}{}^{[{A}}R_{{C}}{}^{{B}]}(M) \eta^{{C}}
        +E_{\mu}{}^{[{A}} R(M) \eta^{{B}]}\,,
\end{eqnarray}
which is  the same expression that one obtains by requiring that the Einstein-Hilbert term is invariant under P-transformations.

Summarizing, the P-transformations given in eqs.~\eqref{gctL}, \eqref{second}, \eqref{Ptransf} and the general coordinate transformations given in the same equations~\eqref{gctL} and \eqref{second} do not define two independent symmetries of the first-order Einstein-Hilbert action \eqref{standard}. In fact, if they would be independent symmetries, the theory would
  have no propagating degrees of freedom left. Both symmetries have their
advantages. On the one hand,  the general
coordinate transformations have a nice geometrical interpretation, but, on the other hand, the
$P$-transformations are more directly related to the underlying Poincaré algebra.

When taking the non-relativistic limit of general relativity  in  subsection \ref{subsec:limi} , we prefer to work with the second-order formulation of general relativity. The reason for this is that for matter-coupled gravity theories, such as supergravity, it is more convenient to work in such a second-order formulation.
In that case, it is understood that the equation of motion \eqref{eomO} has been used to solve for the spin-connection field $\Omega$ in terms of the Vielbeine $E_\mu{}^{ A}$ and their inverses $E^\mu{}_{ A}$. To solve this constraint it is convenient to introduce the notation
\begin{equation}
E_{\mu\nu}{}^{ A} \equiv \partial_{[\mu}E_{\nu]}{}^{ A}
\end{equation}
and to write the constraint \eqref{eomO} in terms of flat indices as
\begin{equation}\label{flatcurvature}
2E_{ A B C} -\Omega_{ A C B} + \Omega_{ B C A}=0\,.
\end{equation}
Following the solution of the Christoffel symbol in general relativity, we add three times this equation with the flat indices cyclic interchanged and multiply one of the three equations with a minus sign. Adding up the three resulting equations leads to the solution
\begin{equation}\label{solution}
\Omega_{ A B C} = 2E_{ A[ B C]} +E_{ B C  A}\hskip .3truecm \textrm{or}\hskip .3truecm
\Omega_\mu{}^{ A B} = -2 E_\mu{}^{[ A B]} + E^{ A B}{}_\mu\,.
\end{equation}

The independent fields  are then given by the Vielbein fields $E_\mu{}^{ A}$ only. They   transform under general coordinate transformations and local lorentz rotations as follows:
\begin{eqnarray}
\delta E_\mu{}^{ A} &=& \xi^\lambda\partial_\lambda E_\mu{}^{ A} + \partial_\mu \xi^\lambda E_\lambda{}^{ A}
+\Lambda^{ A}{}_{ B} E_\mu{}^{ B}\,.
\end{eqnarray}
The general coordinate transformations are not affected by the NR limit we consider in   subsection \ref{subsec:limi}, they are the same before and after taking the limit.

\subsection{Gauging non-lorentzian algebras}
\label{subsec:gauging2}

We next consider the non-lorentzian case. There are several non-lorentzian algebras we could consider. As specific examples we will consider the Galilei algebra, its central extension called the Bargmann algebra and the so-called Carroll algebra.
\vskip .3truecm

\noindent {\bf The Galilei algebra.}\ \ Before discussing the Bargmann algebra that underlies the symmetries of NC gravity, we will first, as a warming up exercise,  shortly discuss
the special case of the Bargmann algebra with {\it zero} central extension, i.e.~the Galilei algebra. In the next Chapter, we will show how the symmetries corresponding to the Galilei algebra  arise if one takes the so-called Galilei limit of a real Klein-Gordon scalar field. Here, we will show how the Galilei algebra can be obtained as a particular contraction of the Poincaré algebra and how the structure constants of this Galilei algebra fix the transformation rules of the gauge fields under the Galilei symmetries.

To show how the  Galilei algebra is obtained by a contraction of the Poincaré algebra,  we first  decompose the relativistic flat Lorentz index ${ A}$ into ${ A}=\{0,a\} $ with $a=(1,\dots,d)$, and redefine, using a contraction parameter $\omega$,  the Poincaré generators according to
\begin{eqnarray}
P_0 &=& \omega^{-1} H \,, \label{eq: GalP0 redef} \\
J_{0a} &=& \omega G_a \,, \label{eq: GalJ0a redef}
\end{eqnarray}
where $H$ and $G_a$ are  the generators of time translations and boosts, respectively. The generators $P_{a}$ of space translations and $J_{ab}$ of spatial rotations are not redefined. Next, taking the limit $\omega \rightarrow \infty$ we obtain the following Galilei algebra:
\begin{eqnarray}\label{Galgebra}
&&[J_{ab}, P_c] = 2\delta_{c[a}P_{b]}\,,\hskip 1.5truecm [J_{ab}, G_c] = 2\delta_{c[a}G_{b]}\,,\nonumber\\[.2truecm]
&&[J_{ab}, J_{cd}] = 4
\delta_{[a[d}\,J_{c]b]}\,,\hskip 1.3truecm  [H,G_{a}] = P_{a}\,.
\end{eqnarray}

Following \cite{Andringa:2010it}, we next associate to each generator of the Galilei algebra a gauge field as follows:
\begin{equation}
A_\mu^I T_I  = \tau_\mu H + e_\mu{}^{a} P_{a} + \frac{1}{2} \omega_\mu{}^{ ab}J_{ab}+ \omega_\mu{}^{a}G_{a}\,.
\end{equation}
Using the general formula \eqref{transfgf}, the gauge fields transform as  1-forms under general coordinate transformations while  under spatial rotations with parameters $\lambda^{a}{}_{b}$ and galilean boosts with parameters $\lambda^{a}$   they transform as follows:
\begin{eqnarray} \label{Gtransf1}
\delta \tau_\mu &=& 0\,,  \\[.1truecm]
\delta e_\mu{}^{a}  &=& \lambda^{a}\tau_\mu + \lambda^a{}_b e_\mu{}^{b}  \,, \\[.1truecm]
\delta \omega_\mu{}^{ab}  &=& (D_\mu \lambda)^{a b}, \label{eq: Gal trans omegaab}\\[.1truecm]
\delta \omega_\mu{}^{a}  &=& (D_\mu \lambda)^{a} + \lambda^{a}{}_{b}\omega_\mu{}^{b}  \,.\label{Gtransf4}
\end{eqnarray}
Here $D_\mu$ is the covariant derivative with respect to spatial rotations, e.g., $(D_\mu\lambda)^{a}= \partial_\mu\lambda^{a} -\omega_\mu{}^a{}_b\lambda^{b}$.
\vskip .3truecm

\noindent {\bf The Bargmann algebra.}\ \ Our starting point is now the centrally extended Galilei algebra which is called the Bargmann algebra. The reason that we need to add one more generator to the Galilei algebra, which has the same number of generators as the Poincaré algebra,  is that in the relativistic case energy is equivalent to mass but in the non-relativistic case energy and mass  are two separately conserved quantities. The corresponding Noether symmetries lead to two generators in the Bargmann algebra: the time translation generator corresponding to the conservation of energy and the central charge or mass generator corresponding to the conservation of mass.

The Bargmann algebra can be obtained by performing a special Wigner-In\"on\"u contraction of the direct product of the Poincaré algebra given in equation~\eqref{eq:PoincareAlgebraCommutators} with a U(1) algebra with generator  $\mathcal{Z}$.
As a first step we make the following invertable redefinition of the relativistic generators
\begin{align}\begin{split}\label{contraction}
 P_0 &= \frac{1}{2\omega}\,H +\omega\,Z \,,
 \hskip1.5cm M_{ab} = J_{ab} \,,
 \hskip1.5cm M_{a0}=\omega\, G_{a} \,,\\[.1truecm]
 P_{a} &= P_{a}\,, \hskip3.5cm \mathcal{Z}= \frac{1}{2\omega}\, H -\omega\, Z\,,
\end{split}\end{align}
where $\omega$ is a (dimensionless) contraction parameter and where we have decomposed  the flat space-time index $ A$ into a time-like $0$-index and
spatial $a$-indices as $A = (0,a)$. Note the off-diagonal nature of the redefinitions. Would we have restricted to rescaling each generator separately, we would only be able to obtain the Galilei algebra times U(1). Using the redefinition \eqref{contraction}, we find that the redefined generators, after taking the limit that $\omega$ goes to infinity,  generate the following Bargmann algebra:
\begin{align}\begin{split}\label{Bargmannalg}
 \big[P_{a},J_{bc}\big] &= 2\,\delta_{a[b}\,P_{c]} \,, \hskip1cm \big[J_{ab},J_{cd}\big]=4\,\delta_{[a[c}\,J_{d]b]} \,,\\[.1truecm]
 \big[G_{a},J_{bc}\big] &= 2\,\delta_{a[b}\,G_{c]} \,, \hskip1.8cm \big[H,G_{a}\big] = P_{a} \,,
 \hskip1.5cm \big[P_{a},G_{b}\big] = \delta_{ab}\,Z \,,
\end{split}\end{align}
where the generator $Z$ has taken the role of the central charge generator.

We next associate to each generator of the Bargmann algebra a gauge field as follows:
\begin{equation}
A_\mu^I T_I  = \tau_\mu H + e_\mu{}^{a} P_{a} + \frac{1}{2} \omega_\mu{}^{ ab}J_{ab}+ \omega_\mu{}^{a}G_{a}+ m_\mu Z\,.
\end{equation}
Using the general formula \eqref{transfgf}, the gauge fields transform as  1-forms under general coordinate transformations while  under spatial rotations with parameters $\lambda^{a}{}_{b}$, galilean boosts with parameters $\lambda^{a}$ and central charge transformations with parameter $\sigma$  they transform as follows:
\begin{align}\begin{split} \label{NCtraforules}
 \delta \tau_\mu &=0 \,, \\[.1truecm]
 \delta e_\mu{}^{a} &=\lambda^{a}{}_{b} \, e_\mu{}^{b} +\lambda^{a}\tau_\mu \,, \\[.1truecm]
  \delta \omega_\mu{}^{ab} &= \partial_\mu\lambda^{ab} +2\,\lambda^{[a}{}_{c}\,\omega_\mu{}^{cb]} \,, \\[.1truecm]
 \delta \omega_\mu{}^{a} &= \partial_\mu \lambda^{a} +\lambda^{a}{}_{b}\,\omega_\mu{}^{b}
                         -\omega_\mu{}^{a}{}_{c}\,\lambda^{c}\,,\\[.1truecm]
 \delta m_\mu &=\partial_\mu\sigma +\lambda_{a} \,e_\mu{}^{a} \,.
\end{split}\end{align}
Using the general formula \eqref{curvatures} the curvatures corresponding to these gauge fields that transform covariantly under these symmetries are given by
\begin{eqnarray}
R_{\mu\nu}(H) &=& 2\partial_{[\mu}\tau_{\nu]}\,,\\[.1truecm]
R_{\mu\nu}{}^{a}(P) &=& 2\partial_{[\mu}e_{\nu]}{}^{a} -2 \omega_{[\mu}{}^{ab}e_{\nu]b} -2\omega_{[\mu}{}^{a}\tau_{\nu]}\,,\\[.1truecm]
R_{\mu\nu}{}^{ a b}(J)  &=& 2\partial_{[\mu}\omega_{\nu]}{}^{ a b} +2 \omega_{[\mu}{}^{ a c}\omega_{\nu]}{}^{ b}{}_{c}\,,\\[.1truecm]
R_{\mu\nu}{}^{a}(G) &=& 2\partial_{[\mu}\omega_{\nu]}{}^{ a } -2 \omega_{[\mu}{}^{a}{}_{b} \omega_{\nu]}{}^{b}\,,\\[.1truecm]
R_{\mu\nu}(Z) &=& 2\partial_{[\mu}m_{\nu]} -2\omega_{[\mu}{}^{a}e_{\nu]a}\,.
\end{eqnarray}
Note that the  curvature $R_{\mu\nu}(H)$ corresponding to $\tau_\mu$ does not contain any of the other gauge fields. It therefore can describe an {\it intrinsic torsion} \cite{Figueroa-OFarrill:2020gpr}. Imposing a constraint on this curvature leads to a purely geometric constraint.\,\footnote{The following discussion on the intrinsic torsion also applies  when we gauge the Galilei algebra.} This is quite different from the conventional curvature constraints, to be discussed below, that will be used to solve some gauge fields in terms of the others. Instead of $R_{\mu\nu}(H)$  we will sometimes use a notation in terms of the torsion tensor
\begin{equation}
T_{\mu\nu} = \partial_{[\mu}\tau_{\nu]}\,.
\end{equation}
 One may  distinguish between the following three different cases:\footnote{One cannot impose $T_{0a}=0$ since such a constraint is not invariant under galilean boost transformations.}
\begin{eqnarray}
T_{\mu\nu} =0&:&\ \ \textrm{zero torsion}\,,\label{zerotorsion}\\[.1truecm]
T_{ab}=0&:&\ \ \textrm{twistless torsional}\,,\\[.1truecm]
T_{\mu\nu} \ne 0&:&\ \ \textrm{general torsion}\,.
\end{eqnarray}
We have used here the projective inverse NR Vielbeine $\tau^\mu$ and $e^\mu{}_{a}$ defined by
\begin{equation}\label{inverseVb}
\tau_\mu\tau^\mu =1\,,\hskip .5truecm \tau_\mu e^\mu{}_{a} = \tau^\mu e_{\mu}{}^{a} = 0\,,\hskip .5truecm e_\mu{}^{a} e^\nu{}_{a} + \tau_\mu\tau^\nu = \delta_\mu{}^\nu\,.
\end{equation}
to convert curved indices into flat indices. For instance,
\begin{equation}
T_{ab} = e^\mu{}_{a} e^\nu{}_{b}T_{\mu\nu}\,.
\end{equation}
The zero torsion case defines a Newtonian spacetime with a co-dimension 1 foliation or, equivalently, a preferred time direction $t$ given by $\tau_\mu = \partial_\mu t$. Any observer traveling along a curve $\mathcal{C}$ from a time slice $\Sigma_{t_A}$ at $t=t_A$ to a time slice $\Sigma_{t_B}$ at $t=t_B$ will measure a time difference $\Delta T$ given by
\begin{equation}
\Delta T = \int_{t_A}^{t_B} dx^\mu \tau_\mu = t_B - t_A
\end{equation}
independent of the curve $\mathcal{C}$. The twistless torsional case leads to a spacetime with a hypersurface orthogonality condition of the clock fucction $\tau_\mu$. Such spacetimes are encountered in Lifschitz holography \cite{Christensen:2013lma}.

Using the projective inverses of the  timelike and spatial Vielbein fields $\tau_\mu$ and $e_\mu{}^{a}$, the
so-called `conventional constraint' equations\,\footnote{For the use of conventional constraints in gravity and supergravity, see, e.g., \cite{Freedman:2012zz,Ortin:2015hya}.}
\begin{equation}\label{conventional}
R_{\mu\nu}{}^{a}(P) = R_{\mu\nu}(Z) =0
\end{equation}
provide precisely sufficient equations to solve  the spin-connection fields for spatial rotations and galilean boosts in terms of the other independent gauge fields. For the zero torsion case these gauge fields are solved, by doing a similar calculation as in the relativistic case (see after
equation~\eqref{flatcurvature}),  as follows:
\begin{subequations}\label{standardomega}
\begin{align}
&{\omega}_{\mu}{}^{ab} (\tau,e,m)= -2  e_{\mu}{}^{[ab]} + e_{\mu c}e^{abc} - \tau_\mu m^{ab}\,,\\[.1truecm]
&{\omega}_{\mu }{}^{a}(\tau,e,m) = e_{\mu 0}{}^{a}  - e_{\mu c}e_0{}^{ac} + m_\mu{}^{a}  - \tau_\mu m^{a0}\,.
\end{align}
\end{subequations}
Here we have defined
\begin{equation}
e_{\mu\nu}{}^{a} = \partial_{[\mu}e_{\nu]}{}^{a}\,,\hskip 1truecm m_{\mu\nu} = \partial_{[\mu}m_{\nu]}\,.
\end{equation}
Furthermore, we have again used the inverse NR Vielbeine
to convert curved indices into flat indices. For instance
\begin{equation}
 e_{\mu 0}{}^{a} = \tau^\nu e_{\mu\nu}{}^{a}\,.
\end{equation}
We note that the transformation of the dependent spin-connection fields  is identical to the transformations of the independent spin-connection fields as given in eq~\eqref{NCtraforules}, i.e.
\begin{equation}
\delta {\omega}_{\mu}{}^{ab} (\tau,e,m) = \delta{\omega}_{\mu}{}^{ab}\,,\hskip 2truecm \delta {\omega}_{\mu }{}^{a}(\tau,e,m) = \delta{\omega}_{\mu }{}^{a}\,.
\end{equation}
This is due to the fact that the curvatures in the conventional constraint equations \eqref{conventional} do not transform to any of the other curvatures under spatial rotations, galilean boosts and central charge transformations. From now on we will assume that the spin-connections are dependent fields but we will not indicate their dependence anymore. Finally, we note that, by solving the conventional constraints \eqref{conventional}, we work by definition in a second-order formulation.

Sofar, we did not yet discuss the $P_{ A} = (P_0,P_{a})$-transformations with parameters $(\eta, \eta^{a})$ of the gauge fields. According to the Bargmann algebra they are given by
\begin{eqnarray}\label{P2}
\delta \tau_\mu &=& \partial_\mu \eta\,,\\[.1truecm]
\delta e_\mu{}^{a} &=& \partial_\mu \eta^{a} -\omega_\mu{}^{ab}\eta_{b}\,,\\[.1truecm]
\delta m_\mu &=& - \omega_{\mu a}\eta^{a} \,.
\end{eqnarray}

To show how these P-transformations are related to general coordinate transformations, we consider the following general identity valid for any  Lie algebra
with structure constants $f^I{}_{JK}$:
\begin{equation}
   0 = \delta_{gct}(\xi^{\lambda})A_{\mu}{}^I +
          \xi^\lambda R_{\mu\lambda}{}^I(T)
   - \sum_{\substack{\{J\}}} \delta(\xi^{\lambda}A_{\lambda}{}^{J})A_{\mu}{}^I\,,
\label{veryimportantequation}
\end{equation}
where the index $I$ labels the gauge fields $A_\mu{}^I$ and corresponding curvatures $R_{\mu\nu}{}^I(T)$ of the gauge algebra. The sum in the last term is over all gauge fields. To see how this identity works, let us set, for instance,
 $I=a$ for the $P_{a}$-transformations and consider the parameters
\begin{equation}
\xi^\lambda=\tau^\lambda \eta  + e_{a}{}^\lambda\eta^{a}\hskip .5truecm \textrm{or}\hskip .5truecm \eta = \xi^\lambda\tau_\lambda\,,\ \eta^{a}  = \xi^\lambda e_\lambda{}^{a}\,.
\end{equation}
We can then bring the contribution of $e_\mu{}^{a}$ to the sum in the last term of (\ref{veryimportantequation}) to the left-hand
side of the equation to obtain the following relation between a $P_{a}$-transformation with parameter $\eta^{a}$ and a general coordinate transformation with parameter $\xi^\lambda = e_{a}{}^\lambda\eta^{a}$:
\begin{equation}
\delta_P(\eta^{b}) e_{\mu}^{a}
   =  \delta_{gct}(\xi^{\lambda})e_{\mu}^{a} + \xi^{\lambda}R_{\mu\lambda}{}^{a}(P)
    - \delta_M(\xi^{\lambda}\omega_{\lambda}^{ab})e_{\mu}^{a}\,. \label{Poincarepexchange}
\end{equation}
The same kind of identity holds for each gauge field that transform under a P-transformation, i.e., in our case $\tau_\mu\,, e_\mu{}^{a}$ and $m_\mu$, see equation~\eqref{P2}: one can relate the $P$-transformation of these gauge fields to a general coordinate transformation plus other symmetries of the Bargmann algebra by setting the curvature of  these gauge fields to zero. Following this rule we precisely obtain the zero torsion constraint \eqref{zerotorsion} and the two conventional constraints \eqref{conventional}. Remarkably, these constraints allow us to solve for the remaining gauge fields, i.e. the two spin-connection fields, and hence, as dependent gauge fields, they automatically have a P-transformation that is related to a general coordinate transformation since this was already proven for all the independent gauge fields.

We note  that for non-zero torsion, the conventional constraints \eqref{conventional} do not transform to each other anymore under all the symmetries of the theory. To achieve this, one needs to add to these conventional constraints additional (independent) torsion tensors with the correct transformation properties. This leads to a notion of {\it torsional} NC geometry that is discussed in \cite{Bergshoeff:2022fzb}.
\vskip .3truecm

\noindent {\bf The Carroll algebra.}\ \ Carroll symmetries emerge if one considers an {\it ultra-local} limit of general relativity which  is the opposite of taking a NR limit. At first sight this seems a strange thing to do. However, Carroll symmetries have shown up in several recent investigations in different connections such as strong coupling limits of gravity \cite{Henneaux:1979vn,Henneaux:1981su}, flat space holography \cite{Duval:2014uva}, black hole horizons \cite{Donnay:2019jiz}, de Sitter cosmology and dark matter \cite{deBoer:2021jej} and even fractons  \cite{Casalbuoni:2021fel,Bidussi:2021nmp}.  Here, for completeness, we shortly discuss the gauging of the Carrol algebra and point out some differences with the Galilei algebra.

To define the contraction of the Poincaré algebra that gives rise to the Carroll algebra, we decompose the  $A$-index into ${ A}=\{0,a\} $ with $a=(1,\dots,d)$, and redefine the Poincaré generators according to
\begin{eqnarray}
P_0 &=& \omega H \,, \label{eq: P0 redef} \\
J_{0a} &=& \omega G_{a} \,, \label{eq: J0a redef}
\end{eqnarray}
where $H$ and $G_{a}$ are  the generators of time translations and boosts, respectively. The generators $P_{a}$ of space translations and $J_{ab}$ of spatial rotations are not redefined. Next, taking the limit $\omega \rightarrow \infty$ we obtain the following Carroll algebra:
\begin{eqnarray}\label{eq:Calgebra}
&&[J_{ab}, P_{c}] = 2\delta_{c[a}P_{b]}\,,\hskip 1.5truecm [J_{ab}, G_{c}] = 2\delta_{c[a}G_{b]}\,,\nonumber\\[.2truecm]
&&[J_{ab}, J_{cd}] = 4
\delta_{[a[d}\,J_{c]b]}\,,\hskip 1.3truecm  [P_{a},G_{b}] = \delta_{ab}H\,.
\end{eqnarray}

We next associate to each generator of the Carroll algebra a gauge field as follows:
\begin{equation}
A_\mu^I T_I  = \tau_\mu H + e_\mu{}^{a} P_{a} + \frac{1}{2} \omega_\mu{}^{ ab}J_{ab}+ \omega_\mu{}^{a}G_{a}\,.
\end{equation}
Using the general formula \eqref{transfgf}, the gauge fields transform as  1-forms under general coordinate transformations while  under spatial rotations with parameters $\lambda^{a}{}_{b}$ and  Carroll boosts with parameters $\lambda^{a}$   they transform as follows:
\begin{eqnarray} \label{bossymm1}
\delta \tau_\mu &=&   e_\mu{}^{a}\lambda_{a}\,, \nonumber \\  [.1truecm]
\delta e_\mu{}^{a} &=&   \lambda^{a}{}_{b} e_\mu{}^{b}\,, \nonumber \\[.1truecm]
\delta \omega_\mu{}^{ab} &=&  (D_\mu\lambda)^{ab}\,,  \\[.1truecm]
\delta \omega_\mu{}^{a} &=& (D_\mu\lambda)^{a} +  \lambda^{a}{}_{b} \omega_{\mu }{}^{b}
\nonumber \,,
\end{eqnarray}
where $D_\mu$ is the covariant derivative with respect to spatial rotations.
Note that, in contrast to the Galilei algebra, $\tau_\mu$ transforms under a boost transformation while $e_\mu{}^{a}$ is invariant. Another important difference with the Galilei algebra is that the Carroll algebra does not allow for a central extension.

Unlike  the Galilei or Bargmann algebra, all Carroll curvatures contain a spin-connection field. A priori such curvatures are part of conventional constraints, needed to solve for the spin-connection fields, and, therefore cannot describe an intrinsic torsion like the tensor $T_{\mu\nu}$ in the Galilei and Bargmann case. However, given the $R_{\mu\nu}{}^{a}(P)$  curvature
\begin{equation}
R_{\mu\nu}{}^{a}(P) = e_{\mu\nu}{}^{a} - \omega_{[\mu}{}^{ab}e_{\nu]b}\,,\hskip 1.5truecm e_{\mu\nu}{}^{a} = \partial_\mu e_\nu{}^{a} - \partial_\nu e_\mu{}^{a}\,,
\end{equation}
 it turns out that the following boost-invariant projection $K^{ab} = K^{ba}$ does not contain any spin-connection field:
\begin{equation}
K^{ab} = \tau^\mu e^{\nu(a}R_{\mu\nu}{}^{b)}(P) = \tau^\mu e^{\nu(a}e_{\mu\nu}{}^{b)}(P)\,.
\end{equation}
Using that $\tau^\mu e_\mu{}^{a}=0$ one can show that $K^{ab}$ is nothing else as than the spatial components of the extrinsic curvature:
\begin{equation} \label{defK}
K_{ab} = e_{a}{}^\mu e_{b}{}^\nu K_{\mu\nu}\,,\hskip 1truecm K_{\mu\nu} = \tau^\lambda \partial_\lambda h_{\mu\nu} + \partial_\mu \tau^\lambda  h_{\lambda \nu} + \partial_\nu \tau^\lambda  h_{\lambda \mu}
\end{equation}
with $h_{\mu\nu} = e_\mu{}^{a} e_\nu{}^{b}\delta_{ab}$.

\subsection{Taking limits}
\label{subsec:limi}

The aim of this section is to define the limits of general relativity that correspond to the non-lorentzian algebras we defined in the previous section as Wigner-In\"on\"u contractions of the Poincaré algebra. Our main target is the Bargmann algebra but, for completeness, we will also shortly discuss the  limits corresponding to the Galilei and Carroll algebra leading to Galilei and Carroll gravity, respectively.

Generically, to define a limit in all three cases, we will perform the following two steps:

\begin{itemize}
\item we make an invertible field redefinition writing all relativistic fields in terms of the would-be  fields of the limiting theory and a dimensionless contraction parameter $\omega$. The invertibility implies that the number of fields before and after taking the limit remains the same. The would-be limiting fields only become the true limiting fields after taking the NR limit in the second step. Before this step we are just rewriting the  general relativity theory.
\item Either in the action or equations  of motion we take the  limit that $\omega$ goes to $\infty$. We do not allow divergent terms in the action. A noteworthy feature of several  of the limits that we will be taking is that they are based upon a cancellation of the leading divergence by different contributions.   The limiting action  is given by all terms of order $\omega^0$. Taking the limit of the equations of motion the resulting  equations of motion are given by the terms of leading order in $\omega$. Independent of this we will also take the  limit of the transformation rules.
\end{itemize}
\vskip .05truecm
 One should distinguish between taking limits from making expansions. In an expansion each field is expanded as an infinite sum of terms of increasing powers of $\omega^{-1}$. The leading terms in such an expansion do not necessarily correspond a redefinition defining a limit. For instance, in a post-Newtonian expansion of general relativity one does not introduce the additional field $M_\mu$. Instead, $m_\mu$ occurs as the sub-leading term in an expansion of $E_\mu{}^0$. Some results about limits can, however, be read off from making an expansion. For instance, the first leading term in the expansion in $\omega$ of a relativistic lagrangian is always invariant under the corresponding  non-lorentzian symmetry \cite{Batlle:2016iel,Bergshoeff:2017btm}.
\vskip .2truecm

\noindent {\bf Galilei gravity.}\ \ We first consider the case of Galilei gravity.
Using a first-order formulation, an invariant action for Galilei  gravity  can be obtained by taking a specific  NR limit of the Einstein-Hilbert action.  To define this limit, we redefine the gauge fields and symmetry parameters with a dimensionless parameter parameter $\omega$ as follows \cite{Bergshoeff:2017btm}:
\begin{eqnarray}
E_\mu^0 &=& \omega \tau_\mu \,, \;\;\quad \Omega^{0a}_\mu \;=\; \omega^{-1} \omega^{a}_\mu \,, \label{eq: Galrescal1}\\
E_\mu^{a} &=& e^{a}_\mu \,, \quad \;\;\;\, \; \Omega^{ab}_\mu \;=\; \omega^{ab}_\mu \,, \label{eq: Galrescal2}\\
\Lambda^{0a} &=& \omega^{-1} \lambda^{a}  \,,  \;\;  \Lambda^{ab} \;=\; \lambda^{ab} \,.\label{eq: Galrescal4}
\end{eqnarray}

Substituting the above field redefinitions into the Einstein-Hilbert action,
redefining Newton's constant $G_N = \omega G_G$ and taking the $\omega\rightarrow\infty$ limit  we end up with the following Galilei  action
\begin{equation}
S_{\text{Gal}}= - \frac{1}{2\kappa} \int e  R_{\mu\nu}{}^{ab}(J) e^\mu_{a} e^\nu_{b}\,,\label{eq: ActionGal}
\end{equation}
where $\kappa = 8\pi G_G$ and $e={\rm det}\, (\tau_\mu,e_\mu{}^{a})$ is the non-relativistic determinant.  The projective inverses $\tau^\mu$ and $e^\mu{}_{a}$ transform under the Galilei boosts and spatial rotations as follows:
\begin{eqnarray}
\delta \tau^\mu &=& - \lambda^{a} e_{a}^\mu \,, \hskip 2truecm
\delta e^\mu_{a}  =  \lambda^{ab}e_{b}^\mu \,.
\end{eqnarray}
One may verify that the Galilei action \eqref{eq: ActionGal} is not only Galilei invariant but it also has an emergent local scaling symmetry given by
\begin{eqnarray}
\tau_\mu & \rightarrow & \lambda(x)^{-(d-2)} \tau_\mu  \,,\hskip 1truecm
e^{a}_\mu \rightarrow  \lambda(x) e^{a}_\mu \,,\label{eq: rescale}
\end{eqnarray}
where $\lambda(x)$ is an arbitrary function. This emergent local scaling symmetry implies that there is a so-called `missing' equation of motion that
does not follow from the variation of the  Galilei action \eqref{eq: ActionGal}. This missing equation of motion can be obtained by taking the limit of the relativistic equations of motion. We will encounter a similar situation when discussing NC gravity below.

For any $d >2$ the equations of motion that follow from the variation of the Galilei action \eqref{eq: ActionGal} lead
 to the following  constraint on the geometry
\begin{equation}
T_{ab} = e^\mu_{a} e^\nu_{b} T_{\mu\nu} =  0\,. \label{eq: RHab is zero}
\end{equation}
This constraint  means that this geometry has twistless torsion.

For $d>2$ the equations of motion  can be used to solve for the spatial rotation spin connection $\omega_\mu{}^{ab}$ as
\begin{equation}
\omega_\mu^{ab} = \tau_{\mu}A^{a b} + e_{\mu c} \left(e^{\rho [a} e^{b]\nu}\partial_{\rho}{e^{c}_{\nu}} + e^{\rho[a} e^{c]\nu}\partial_{\rho}{e^{b}_{\nu}} - e^{\rho[b} e^{c]\nu}\partial_{\rho}{e^{a}_{\nu}}\right)
+ \frac{4}{d-2}  e^{[a}_{\mu} \tau^{b]0}\,, \label{eq: Gal solomegaab}
\end{equation}
except for $A^{ab}$ which is an undetermined anti-symmetric tensor component of $\omega_\mu{}^{ab}$.
 In the second order formulation the constraint \eqref{eq: RHab is zero} arises from the variation with respect to $A^{ab}$. Hence, we can interpret $A^{ab}$ as a Lagrange multiplier. Indeed, in the case $d>2$, plugging expression \eqref{eq: Gal solomegaab} into the action \eqref{eq: ActionGal} to obtain it in a second order formulation leads to
\begin{equation}
S_{\text{Gal}}= - \frac{1}{2\kappa} \int e \left( \left. R_{\mu\nu}{}^{ab}(J) e^\mu_{a} e^\nu_{b} \right|_{A^{ab}=0} + A^{ab}T_{ab} \right) \,.\label{eq: ActionGal sec order}
\end{equation}
This makes manifest the fact that the variation with respect to $A^{ab}$ of the second order action in equation \eqref{eq: ActionGal sec order} reproduces the constraint \eqref{eq: RHab is zero}.

The case $d=2$ is special. In that case we may write $\omega_\mu{}^{ab} =\epsilon^{ab}\omega_\mu$ and it turns out that this $\omega_\mu$ cannot be determined from the field equations, i.e.~there is no second-order formulation. Also, in contrast to the $d>2$ case, the equations of motion imply a stronger geometrical constraint, namely the zero torsion constraint
\begin{equation}
T_{\mu\nu} = 0 \,. \label{eq: RH0a is zero}
\end{equation}
Using the identity $e\epsilon^{ab}e_a^\mu e_b^\nu = 2\epsilon^{\mu\nu\rho}\tau_\rho$, which is valid for $d=2$, the galilean action \eqref{eq: ActionGal} can be rewritten as
\begin{equation}
S_{\text{Gal 3D}}= - \frac{1}{2\kappa} \int \epsilon^{\mu\nu\rho} \tau_\mu\partial_\nu\omega_\rho\,.\label{eq: ActionGal 3D}
\end{equation}
This form of the action makes manifest that its variation with respect to $\omega_\mu$ precisely reproduces the zero torsion constraint
\eqref{eq: RH0a is zero}. We note that the Galilei algebra in $d=2$ only allows for a degenerate invariant bilinear form. The above action corresponds to the Chern-Simons action for the Galilei algebra with this degenerate bilinear form. The degeneracy of the form explains why not all fields occur in the action.

\vskip .3truecm

\noindent {\bf NC gravity}\ \ We will now derive the equations of motion describing pure Newton-Cartan (NC) gravity in $d+1$ dimensions by taking a specific non-relativistic (NR) limit of general relativity.

In the second-order formulation that we are using here, we need to express the relativistic Vielbein field $E_\mu{}^{ A}$ into the would-be non-relativistic fields of NC gravity, that we described in the previous subsection,
in an invertible way using a contraction parameter. Inspired by the standard  Wigner-In\"onu contraction   of the Poincaré algebra we first write
\begin{equation}\label{limitG}
E_\mu{}^{ 0} = \omega\tau_\mu\,,\hskip 2truecm E_\mu{}^{{ a}} = e_\mu{}^{a}\,,
\end{equation}
where we have decomposed $ A = (0,a), \omega$ is a dimensionless parameter,  $\tau_\mu$ is the clock function and $e_\mu{}^{a}$ are the rulers.
It is clear that this limit  cannot  give rise to NC gravity
because in the NR case energy is not the same as mass and hence we need two gauge fields, one for energy and one for mass, that in the previous subsection we called $\tau_\mu$ and $m_\mu$, respectively. Indeed, the NR limit defined by equation~\eqref{limitG}  gives rise to the Galilei gravity we discussed above. The additional mass operator gives rise to a central extension of the Galilei algebra called the Bargmann algebra. We saw in the previous subsection that, in order to obtain this Bargmann algebra from the Wigner-In\"on\"u contraction of a relativistic algebra, we must extend the Poncar\'e algebra with an additional U(1) generator. In terms of gauge fields this implies that we should extend  general relativity with an additional gauge field $M_\mu$ before taking the limit.\,\footnote{ We note that in a Post-Newtonian approximation of general relativity there is no need to add this extra gauge field $M_\mu$ since the lowest order terms in such an approximation do not need to constitute an invertible field redefinition.} In order not to extend general relativity with extra degrees of freedom we impose by hand that the field equation of  $M_\mu$ is given by the following zero flux condition\,\footnote{Here and in the following we will indicate the equation of motion of a field with square brackets.}
\begin{equation}\label{additional}
[M]_{\mu\nu} = \partial_\mu M_\nu - \partial_\nu M_\mu =0\,.
\end{equation}
Note that, without extending general relativity any further,  this field equation does not follow from a relativistic action and therefore the specific  limit we are considering can only be taken at the level of the equations of motion, i.e.~the Einstein equations.

Given the extended general relativity theory, we consider the following redefinitions \cite{Bergshoeff:2015sic}:
\begin{equation}
E_\mu{}^{ 0} = \omega\tau_\mu + \frac{\alpha+1}{\omega} m_\mu\,,\hskip 1truecm  E_\mu{}^{{ a}} = e_\mu{}^{a}\,,\hskip 1truecm M_\mu = \omega\tau_\mu + \frac{\alpha}{\omega} m_\mu \,,
\end{equation}
where  $\alpha$ is a real parameter related to the following field redefinition:
\begin{equation}
\tau_\mu  \rightarrow \tau_\mu +\frac{\alpha}{\omega^2}m_\mu\,.
\end{equation}
From now on, we will take $\alpha=0$:
\begin{equation}\label{a=0}
E_\mu{}^{ 0} = \omega\tau_\mu + \frac{1}{\omega} m_\mu\,,\hskip 1truecm  E_\mu{}^{{ a}} = e_\mu{}^{a}\,,\hskip 1truecm M_\mu = \omega\tau_\mu\,.
\end{equation}
Note that the relativistic inverse Vielbeine are redefined as follows
\begin{equation}\label{inversea=0}
E^\mu_{0} = \frac{1}{\omega}\tau^\mu + \cdots \,,\hskip 1.5truecm E^\mu{}_{a} = e^\mu{}_{a} + \cdots \,,
\end{equation}
where the NR inverse Vielbeine  $\tau^\mu$ and $e^\mu{}_{a}$ were  defined in equation~\eqref{inverseVb}.
We have only given here the leading order redefinitions. The lower order dotted terms in \eqref{inversea=0} do not contribute to the final answer when taking the NR limit.

As a simple example of how the limit works we consider the following lagrangian describing a relativistic particle of mass $M$:
\begin{equation}
S = -M\int d\tau\,\sqrt{-E_\mu{}^A{\dot X}^\mu E_\nu{}^B {\dot X}^\nu \eta_{AB}} - M_\mu{\dot X}^\mu\,.
\end{equation}
The last term represents a coupling of the gauge field $M_\mu$ to the particle via a Wess-Zumino term. Substituting the field redefinitions \eqref{a=0} into this action and redefining the mass $M$ with $M=\omega m$, we obtain, after taking the limit $\omega \to \infty$ and expanding the square root, the following lagrangian describing the coupling of a non-relativistic particle of mass $m$ with embedding coordinates $X^\mu(\tau)$ to a NC background \cite{DePietri:1994je}:
\begin{equation}
S = \frac{m}{2} \int d\tau\,\bigg\{ \frac{e_\mu{}^{a}{\dot X}^\mu e_\nu{}^{b} {\dot X}^\nu\delta_{ab}}{\tau_\rho {\dot X}^\rho} - 2 m_\mu{\dot X}^\mu \bigg\}\,.
\end{equation}
One can show that this action, due to the second term, is invariant under galilean boost transformations. For a flat spacetime with $\tau_\mu  = \delta_{\mu,0}\,, e_\mu{}^{a} = \delta_\mu{}^{a}=\delta_\mu{}^i$ and $m_\mu = \partial_\mu c$ the action reads
\begin{equation}
S_{\rm flat\ spacetime} = \frac{m}{2} \int d\tau\,\bigg\{ \frac{{\dot X}^i  {\dot X}^j \delta_{ij}}{{\dot t}} - 2 {\dot c} \bigg\}\,.
\end{equation}
which describes the coupling of a massive particle to a Newton
potential $\Phi$, which is in accordance with
equation~\eqref{eq:BNHLag} for $R\to \infty$.

We now consider the relativistic Einstein equations
\begin{equation}
\delta S = - \frac{1}{8 \pi {\rm G}_N} \int d^{d+1} x\,  E \,\delta E_\mu{}^{ A} E^{\mu  B} [G]_{ A B}\,,
\end{equation}
with
\begin{equation}\label{Einsteineqs}
[G]_{ A B}  = R_{ A B}(\Omega) - \frac{1}{2}\eta_{ A B} R(\Omega) =0\,.
\end{equation}
This field equation is symmetric in the $\ A$ and $ B$ since  the Ricci tensor is symmetric\,\footnote{This follows from inserting into the Bianchi identity for the curvature corresponding to the spacetime translation generator  the (conventional) constraint that this curvature is zero.}
\begin{equation}\label{symm}
R_{ A B}(\Omega) = R_{ B A}(\Omega)\,.
\end{equation}

Performing the field redefinitions \eqref{a=0} and \eqref{inversea=0} we find
\begin{subequations}
\label{eq:screscale}
\begin{align}
\Omega_{\mu}{}^{ab}&=\omega^{2}\, \accentset{(2)}{\omega}_{\mu}{}^{ab}+
\accentset{(0)}{\omega}_{\mu}{}^{ab} + \cdots\,,\\[.1truecm]
\Omega_{\mu}{}^{0a}&=\omega\ \accentset{(1)}{\omega}_{\mu}{}^{a}+\omega^{-1}\,\accentset{(-1)}{\omega}_{\mu}{}^{a} + \cdots\,,
\end{align}
\end{subequations}
where the $\omega'$s denote expansion coefficients of the relativistic spin-connection fields $\Omega$. The special expansion coefficients  $\accentset{(0)}{\omega}_{\mu}{}^{ab}$ and $\accentset{(-1)}{\omega}_{\mu}{}^{a}$ will serve as spin-connection fields in the  non-relativistic case and will be denoted by
\begin{equation}
\accentset{(2)}{\omega}_{\mu}{}^{ab} = \omega_\mu{}^{ab}\,,\hskip 2truecm  \accentset{(-1)}{\omega}_{\mu}{}^{a} = \omega_\mu{}^{a}\,.
\end{equation}
We find that the different expansion coefficients are given by
\begin{subequations}\label{eq:leading}
\begin{align}
&\accentset{(2)}{\omega}_\mu{}^{ab}= -\tau_\mu T^{ab}\,,\\[.1truecm]
&{\omega}_{\mu}{}^{ab} = -2  e_{\mu}{}^{ab} + e_{\mu c}e^{abc} - \tau_\mu m^{ab}\,,\\[.1truecm]
&\accentset{(1)}{\omega}_\mu{}^{a}=  e_{\mu b}T^{ba}-2\,\tau_{\mu }T^{a0}\,, \\[.1truecm]
&{\omega}_{\mu }{}^{a} = e_{\mu 0}{}^{a}  - e_{\mu c}e_0{}^{ac} + m_\mu{}^{a}  - \tau_\mu m^{a0}\,.
\end{align}
\end{subequations}
Here we have defined
\begin{equation}
 e_{\mu\nu}{}^{a} = \partial_{[\mu}e_{\nu]}{}^{a}\,,\hskip 1truecm m_{\mu\nu} = \partial_{[\mu}m_{\nu]}\,.
\end{equation}
Like before, we have used the inverse NR Vielbeine
to convert curved indices into flat indices.

We now substitute the expansions \eqref{eq:screscale} of the relativistic spin-connection fields into the Einstein equations \eqref{Einsteineqs} and the expansion \eqref{a=0} of the relativistic vector field $M_\mu$ into the additional equation of motion \eqref{additional}.
The leading  terms in the expressions for the relativistic  spin-connection fields are all proportional to the torsion tensor $T_{\mu\nu}$. Upon inserting these terms into the Einstein equations  will lead to leading order and sub-leading order terms that are also proportional to $T_{\mu\nu}$. On the other hand, looking at the leading order term of the additional equation of motion \eqref{additional} we already conclude that the torsion is zero:
$T_{\mu\nu} = 0$\,. Substituting this zero torsion constraint  into the expanded Einstein equations, we find that the leading order terms of these equations are not anymore given by the terms proportional to the vanishing torsion but instead by terms that involve the NR fields $\omega_\mu{}^{ab}$ and $\omega_\mu{}^{a}$. We thus find that  the different components of the relativistic Enstein tensor $[G]_{ A B}$ give rise to the following NC equations of motion:
\begin{eqnarray}
&&[G]_{00}\,:\hskip .2 truecm R_{0a}{}^{a}(G)=0\,,\label{NC1}\\[.1truecm]
&&[G]_{0a}\,:\hskip .22 truecm  R_{0ca}{}^{c}(J) = 0\,,\label{NC2}\\[.1truecm]
&&[G]_{ab}\,:\hskip  .23truecm  R_{acb}{}^{c}(J) = 0\label{NC3}\,,
\end{eqnarray}
where the curvatures for the galilean boosts and spatial rotations have been defined in the previous subsection.
To derive these equations of motion, we have made use  of the identity
\begin{equation}
R_{ab}{}^{b}(G) = R_{0b}{}^{b}{}_{a}(J)\,,
\end{equation}
which follows from  taking the NR limit of the relativistic identity \eqref{symm}.

In a flat Newtonian spacetime we have
\begin{equation}\label{restrictions}
\tau_\mu = \delta_\mu{}^0\,,\hskip 1truecm e_\mu{}^{a} = \delta_\mu{}^{a}\,,\hskip 1truecm  m_\mu = \tau_\mu\, \Phi\,,
\end{equation}
where $\Phi$ is the (time-independent) Newton potential.  The only non-trivial spin-connection field for this special case is given by $\omega_0{}^{a} = \partial^{a}\Phi$ and the only non-trivial equation of motion reads
\begin{equation}\label{Poisson}
R_{0a}{}^{a}(G) = \partial_{a}\omega_0{}^{a} = \partial_{a}\partial^{a}\Phi =0\,,
\end{equation}
thus recovering the well-known sourceless Laplace's equation for the Newton potential. The restrictions \eqref{restrictions} can be seen as gauge-fixing conditions for the diffeomorphisms restricting to frames with constant acceleration only. The NC equations \eqref{NC1}-\eqref{NC3} can then be viewed as the extension of the Laplace equation \eqref{Poisson} to arbitrary frames.

\vskip .3truecm

\noindent {\bf Carroll gravity}\ \ Finally, we consider the case of Carroll gravity.\,\footnote{For other recent work on Caroll gravity, see
\cite{Henneaux:2021yzg,Hansen:2021fxi,Perez:2021abf,Perez:2022jpr}.}
We will  derive an invariant action for Carroll gravity by taking the ultra-local limit of the Einstein-Hilbert action. To define this limit, we redefine the gauge fields and symmetry parameters with a dimensionless  parameter $\omega$ as follows \cite{Bergshoeff:2017btm}:\,\footnote{For a different approach to Carroll gravity, see \cite{Hartong:2015xda}.}
\begin{eqnarray}
E_\mu^0 &=& \omega^{-1} \tau_\mu \,, \quad \Omega^{0a}_\mu \;=\; \omega^{-1} \omega^{a}_\mu \,, \label{eq: rescal1}\\
E_\mu^{a} &=& e^{a}_\mu \,, \quad \quad \;\;\, \Omega^{ab}_\mu \;=\; \omega^{ab}_\mu \,, \label{eq: rescal2}\\
\Lambda^{0a} &=& \omega^{-1} \lambda^{a} \,, \quad \;\;\, \Lambda^{ab} \;=\; \lambda^{ab} \,.\label{eq: rescal4}
\end{eqnarray}
Substituting  the field redefinitions  (\ref{eq: rescal1}) and (\ref{eq: rescal2}) into  the Einstein-Hilbert action, redefining Newton's constant as $G_N =\omega^{-1} G_C$ and taking the $\omega\rightarrow\infty$ limit,  we end up with the following Carroll action\footnote{This limit shows similarities with the strong coupling limit considered in \cite{Henneaux:1979vn}, \cite{Henneaux:1981su}. Note that both limits lead to a theory with a Carroll-invariant vacuum solution. This suggests that, although looking different at first sight, the result of the two limits might be the same up to field redefinitions.}
\begin{equation}
S_{\text{Car}}= - \frac{1}{16\pi G_C} \int e \left(2\tau^\mu e^\nu_{a} R(G)_{\mu\nu}{}^{a}+e^\mu_{a} e^\nu_{b} R(J)_{\mu\nu}{}^{ab}\right)\,. \label{eq: ActionCar}
\end{equation}
Here $e={\rm det}\, (\tau_\mu,e_\mu{}^{a})$. The projective inverses $\tau^\mu$ and $e^\mu{}_{a}$ transform under boosts and spatial rotations as follows:
\begin{eqnarray}
\delta \tau^\mu =0\,,\hskip 1.5truecm
\delta e^\mu_{a}  =  - \lambda^{a}\tau^\mu + \lambda^{ab}e_{b}^\mu \,. \label{eq: varinvviel Car}
\end{eqnarray}

The field equations corresponding to the first-order Carroll action \eqref{eq: ActionCar}
can be used to solve for the spin connections
\begin{eqnarray}
\omega_{\mu}{}^{a} &=& \tau_\mu \tau^\nu e^{\rho a} \partial_{[\nu}\tau_{\rho]} + e^{\nu a}\partial_{[\mu}\tau_{\nu]} + S^{ab} e^{b}_\mu \,, \label{eq: solomegaa}\\[.1truecm]
\omega_\mu{}^{ab} &=& - 2 e^{\rho [a}\partial_{[\mu} e_{\rho]}^{b]} + e_{\mu c} e^{\rho a} e^{\nu b}\partial_{[\rho} e_{\nu]}^{c}  \,,\label{eq: solomegaab}
\end{eqnarray}
except for a symmetric component $S^{ab}=S^{(ab)} = e^{\mu(a} \omega_\mu^{b)}$ of the boost spin connection $\omega_\mu{}^{a}$ which remains undetermined.

Plugging the dependent expressions for the spin connections \eqref{eq:
  solomegaa} and \eqref{eq: solomegaab} into the Carroll action
\eqref{eq: ActionCar} we obtain
\begin{equation}\label{eq: Car act S lag multi}
  S_{\text{Car}}= - \frac{1}{16\pi G_C} \int e \left(
    \left.2\tau^\mu e^\nu_{a}
      R(G)^{a}{}_{\mu\nu}\right|_{S^{ab}=0}+e^\mu_{a} e^\nu_{b}
    R(J)^{ab}{}_{\mu\nu} +
    2K_{ab}S^{ab}-2\delta^{ab}\delta_{cd}K_{ab}S^{cd}\right)\,.
\end{equation}
From this expression of the action it follows that the equation of
motion for $S^{ab}$ implies that $K_{ab}=0$. In other words, we
conclude that $S^{ab}$ is actually a Lagrange multiplier that
enforces the intrinsic torsion constraint $K_{ab}=0$ with $K_{ab}$
defined in equation~\eqref{defK}. This corresponds to the totally
geodesic Carroll structure mentioned in section 4.3.4.

Finally, we note that in $d=2$ the Carroll algebra can be equipped
with a non-degenerate, invariant bilinear form and as a consequence it
is possible to write down a Chern-Simons action for the Carroll
algebra. This Chern-Simons action is precisely the same as the action
given above for $d=2$.

\section{Field Theories}

In this section we will discuss the non-lorentzian (NL) field theories
for a complex and real massive spin-0 particle, a massive spin-1/2
particle and a massless spin-1 particle.\,\footnote{There is a huge
  literature on field theories with galilean and carrollian
  symmetries, see e.g.~\cite{Bagchi:2014ysa,Bagchi:2016bcd} for some
  early references.}  There are two approaches here. Either one takes
the NL limit of the relativistic field theory in a flat Minkowski
background and after taking the limit one couples the theory to NL
gravity or one first couples the model under consideration to general
relativity and next takes the NL limit of the matter coupled to
gravity system using the NL limits we derived in the previous section.
We will opt for this second option. In particular, for spin-0, we will
discuss the Galilei, Bargmann and Carroll limits while for spin-1/2
and spin-1 we will only discuss the Bargmann limit.

\subsection{Real Massive Spin-0}

We first  discuss the Galilei and Carroll limits of a real massive  scalar field. This leads to the following four cases:
\vskip .3truecm

\noindent{\bf spin-0 Galilei.}\ \  We consider the following lagrangian for a real scalar field:
\begin{align}\label{case1}
 E^{-1} \,\mathcal{L}_{\rm rel} = +\frac12\, E^{\mu 0} E^{\nu 0} \partial_\mu \Phi \partial_\nu \Phi  - \frac12\, E^{\mu a} E^{\nu b}\delta_{ab} \partial_\mu \Phi \partial_\nu \Phi -\frac12 \epsilon M^2 \Phi^2\,,
\end{align}
where $\epsilon = +1$ and $\epsilon = -1$ corresponds to a massive particle and a tachyon, respectively. Performing the Galilei redefinitions \eqref{eq: Galrescal1} and \eqref{eq: Galrescal2},  we obtain
\begin{align}
 e^{-1} \,\mathcal{L}_{\rm rel} = +\frac{1}{2\omega^2} \tau^\mu\tau^\nu \partial_\mu \Phi \partial_\nu \Phi  -\frac12\, e^{\mu a} e^{\nu b}\delta_{ab} \partial_\mu \Phi \partial_\nu \Phi -\frac12 \epsilon M^2 \Phi^2\,,
\end{align}
where $e = \det (\tau_\mu,e_\mu{}^{a})$ and where we have ignored an overall power of $\omega$.\,\footnote{Such an overall power can be cancelled by a further redefinition of the fields.}
There are now two ways to proceed. First, by  choosing  $\epsilon=-1$ and taking the limit $\omega \to \infty$, we obtain the following `magnetic' Galilei lagrangian:
\begin{align}
 e^{-1} \,\mathcal{L}_{\rm magnetic\ Galilei} =  - \frac12\, e^{\mu a} e^{\nu b}\delta_{ab} \partial_\mu \Phi \partial_\nu \Phi +\frac12 M^2 \Phi^2\,.
\end{align}
The flat spacetime lagrangian and the corresponding transformation rules are given by
\begin{align}\label{case1bflat}
 \mathcal{L}_{\rm magnetic\ Galilei}(flat\ spacetime)  =  - \frac12\,  (\partial_{i}\Phi)^2   +\frac12 M^2 \Phi^2\,
\end{align}
and
\begin{equation}
\delta \Phi =  \Big(\zeta\,\partial_t +\xi^{i}\partial_{i} -\lambda^{i}\,t\,\partial_{i} -x^{j}\lambda^{i}{}_{j}\,\partial_{i}
              \Big)\Phi \,.
\end{equation}
This limit was considered in the context of taking the limit of a tachyonic particle lagrangian \cite{Batlle:2017cfa} where it leads to the massless Galilei particle of Souriau with `colour' $M$ \cite{Souriau}.

A second option is to first redefine $\Phi = \omega\phi, M = \omega^{-1}m$ and obtain the following lagrangian:
\begin{align}\label{case1c}
 e^{-1} \,\mathcal{L}_{\rm rel} = +\frac12\, \tau^\mu\tau^\nu \partial_\mu \phi \partial_\nu \phi  -\frac12\, \omega^2\, e^{\mu a} e^{\nu b}\delta_{ab} \partial_\mu \phi \partial_\nu \phi -\frac12\epsilon m^2 \phi^2\,,
\end{align}
To deal with the quadratic divergence in the second term, we use a result of  \cite{Gomis:2005pg} and rewrite  the lagrangian, introducing auxiliary fields  $ \chi^{a}$, as follows:\,\footnote{The general expression is that for each $X$, the quadratic divergence $\omega^2 X^2$ can be rewritten, introducing an auxiliary field $\chi$, as $-\frac{1}{\omega^2} \chi^2 -2 \chi X$.}
\begin{align}\label{case1d}
 e^{-1} \,\mathcal{L}_{\rm rel} = +\frac12\, \tau^\mu\tau^\nu
  \partial_\mu \phi \partial_\nu \phi + \frac1{2\omega^2}\chi^{a}\chi_{a} + \chi^{a}e^\mu{}_{a}\partial_\mu\phi -\frac12\epsilon  m^2 \phi^2\,.
\end{align}
Next, choosing $\epsilon = +1$ and taking  the limit $\omega\to \infty$ the auxiliary fields $\chi^{a}$ become Lagrange multipliers and we obtain the following lagrangian:
\begin{align}
 e^{-1} \,\mathcal{L}_{\rm electric\ Galilei} = +\frac12\, \tau^\mu\tau^\nu \partial_\mu \phi \partial_\nu \phi  + \chi^{a}e^\mu{}_{a}\partial_\mu\phi -\frac12 m^2 \phi^2
 \end{align}
The flat spacetime lagrangian and the corresponding transformation rules are given by
\begin{align}
 \mathcal{L}_{\rm elctric\ Galilei}(flat\ spacetime) = +\frac12\,  (\partial_t \phi)^2  + \chi^{i}\partial_{i}\phi -\frac12 m^2 \phi^2
 \end{align}
and
\begin{equation}
\delta \phi =  \Big(\zeta\,\partial_t +\xi^{i}\partial_{i} -\lambda^{i}\,t\,\partial_{i} -x^{j}\lambda^{i}{}_{j}\,\partial_{i}
              \Big)\phi \,,\hskip 1truecm 
              \delta \chi^i =  \Big(\zeta\,\partial_t +\xi^{j}\partial_{j} -\lambda^j\, t\,\partial_j
              -x^{j}\lambda^{k}{}_{j}\,\partial_{k}
              \Big)\chi^i
               +\lambda^i(\partial_t\phi)\,.
\end{equation}

\vskip .5truecm

\noindent{\bf spin-0 Carroll.}\ \  This case was recently considered in \cite{deBoer:2021jej} in connection with dark matter and inflation and in
\cite{Henneaux:2021yzg} using field theory in an Hamiltonian formulation. The two types of Carroll limits considered here have also been considered in the context of $p$-brane sigma models  using a lagrangian formulation \cite{Bergshoeff:2020xhv}. The fact that there are two types of Carroll limits also follows from the duality between the Galilei and Carrol symmetries considered in  \cite{Barducci:2018wuj}.

We consider the same lagrangian for a real scalar field as in the Galilei case:
\begin{align}
 E^{-1} \,\mathcal{L}_{\rm rel} = +\frac12\, E^{\mu 0} E^{\nu 0} \partial_\mu \Phi \partial_\nu \Phi  - \frac12\, E^{\mu a} E^{\nu b}\delta_{ab} \partial_\mu \Phi \partial_\nu \Phi -\frac12 \epsilon M^2 \Phi^2\,.
\end{align}
but now perform   the Carroll redefinitions \eqref{eq: rescal1} and \eqref{eq: rescal2}. In this way we obtain
\begin{align}
 e^{-1} \,\mathcal{L}_{\rm rel} = +\frac12\, \omega^2 \tau^\mu\tau^\nu \partial_\mu \Phi \partial_\nu \Phi  -\frac12\, e^{\mu a} e^{\nu b}\delta_{ab} \partial_\mu \Phi \partial_\nu \Phi -\frac12\epsilon  M^2 \Phi^2\,,
\end{align}
where $e = \det (\tau_\mu,e_\mu{}^{a})$ and where we have ignored an overall power of $\omega$.

Like in the Galilei case, there are now two options to proceed. First,
to deal with the  quadratic divergence in the first term, we rewrite the lagrangian introducing an auxiliary field $\chi$ as follows:
\begin{align}\label{case3a}
 e^{-1} \,\mathcal{L}_{\rm rel} = -\frac12\, \frac1{\omega^2}\chi^2 - \chi \tau^\mu\partial_\mu\Phi  - \frac12\, e^{\mu a} e^{\nu b}\delta_{ab} \partial_\mu \Phi \partial_\nu \Phi -\frac12\epsilon  M^2 \Phi^2\,.
\end{align}
Next, choosing $\epsilon=-1$ and taking the limit $\omega \to \infty$, we see that $\chi$ has become a Lagrange multiplier and  we obtain the following magnetic Carroll lagrangian:
\begin{align}\label{case3b}
 e^{-1} \,\mathcal{L}_{\rm magnetic \ Carroll} = - \chi \tau^\mu\partial_\mu\Phi  -  \frac12\, e^{\mu a} e^{\nu b}\delta_{ab} \partial_\mu \Phi \partial_\nu \Phi +\frac12 M^2 \Phi^2\,.
\end{align}
The flat spacetime lagrangian and the corresponding transformation rules are given by
\begin{align}\label{case3bflat}
 \mathcal{L}_{\rm magnetic \ Carroll}(flat\ spacetime) = - \chi (\partial_t\Phi)  -  \frac12\,  (\partial_i \Phi)^2  +\frac12 M^2 \Phi^2\,.
\end{align}
and
\begin{equation}
\delta \Phi =  \Big(\zeta\,\partial_t +\xi^{i}\partial_{i} - \lambda^i x_i\partial_t -x^{j}\lambda^{i}{}_{j}\,\partial_{i}
              \Big)\Phi \,,\hskip 1truecm 
              \delta \chi = \Big(\zeta\,\partial_t +\xi^{i}\partial_{i} - \lambda^i x_i\partial_t -x^{j}\lambda^{i}{}_{j}\,\partial_{i}
              \Big)\chi
                + \lambda^i(\partial_i\Phi)\,.
\end{equation}

A second option is to first redefine $\Phi =\frac{1}{\omega}\phi, M=\omega m$. We  then choose $\epsilon=+1$ and take  the limit $\omega \to \infty$ after which we obtain the following lagrangian \cite{Bergshoeff:2014jla}:
\begin{align}\label{eq:case3c}
 e^{-1} \,\mathcal{L}_{\rm electric\ Carroll } = +\frac12\,  \tau^\mu\tau^\nu \partial_\mu \phi \partial_\nu \phi -\frac12 m^2 \phi^2\,.
\end{align}
The flat spacetime lagrangian and the corresponding transformation rules are given by \cite{Bergshoeff:2014jla}
\begin{align}\label{eq:case3c-flat}
 e^{-1} \,\mathcal{L}_{\rm electric\ Carroll }(flat\ spacetime)  = +\frac12\,   (\partial_t \phi)^2 -\frac12 m^2 \phi^2\,.
\end{align}
and
\begin{equation}
\delta \phi =   \Big(\zeta\,\partial_t +\xi^{i}\partial_{i} -\lambda^i x_i\partial_t  -x^{j}\lambda^{i}{}_{j}\,\partial_{i}
              \Big)\phi\,.
\end{equation}

This concludes our discussion of the four limits of a real massive spin-0 particle.

\subsection{Complex Massive Spin-0}

Following \cite{Bergshoeff:2015sic}, we now discuss the standard Bargmann limit of a complex Klein-Gordon scalar field in a curved background.
In contrast to the real scalar discussed above the introduction of an extra vector gauge field has the effect that  the quadratic divergences  cancel and there is only one  way to take the limit.

Our starting point is a lagrangian for a relativistic massive complex scalar $\Phi$, with mass $M$, minimally coupled to an arbitrary gravitational background and the extra zero-flux U(1) gauge field $M_\mu$ that we introduced in the previous section:
\begin{align}\label{relscalar}
 E^{-1} \,\mathcal{L}_{\rm rel} = -\frac12\,g^{\mu\nu}\,D_\mu\Phi^*D_\nu\Phi -\frac{M^2}{2}\,|\Phi|^2 \,.
\end{align}
Here the covariant derivative is given by
\begin{align}
 D_\mu\Phi = \partial_\mu\Phi - {\rm i}\,M\,M_\mu\,\Phi \,.
\end{align}
Apart from invariance under diffeomorphisms, the above lagrangian is also invariant under a local U(1) symmetry given by the transformation rule
\begin{equation}
  \label{eq:PhiU1}
  \delta \Phi = {\rm{i}}\, M\, \Lambda\, \Phi \,.
\end{equation}
The  conserved current associated to this local U(1) symmetry, which is given by
\begin{equation}
  j^\mu_{\rm rel}  = \frac{M}{2{\rm{i}}} \, \Big( \Phi^* D^\mu \Phi - \Phi D^\mu \Phi^*\Big) \,,
\end{equation}
 expresses conservation of the number of particles minus the number of antiparticles.

Using the redefinitions of the previous section and redefining the mass parameter $M$ as
\begin{equation}
  \label{eq:massresc}
  M = \omega m \,,
\end{equation}
one finds that the $O(\omega^2)$ contribution to the lagrangian cancels with one contribution coming from the mass term and another one from the term that is quadratic in the U(1) gauge field. Therefore, the $\omega\rightarrow \infty$ limit is well-defined and leads to the following lagrangian for a Schr\"odinger field coupled to an arbitrary Newton-Cartan background:\,\footnote{We have ignored an overall factor of $\omega$ coming from the redefinition of $E = \omega \, e + \mathcal{O}(\omega^{-1})$. This factor is irrelevant as it amounts to an overall rescaling of the lagrangian that could be compensated by a redefinition of the scalar field. } \footnote{We have turned curved indices into flat indices using the inverse Newton-Cartan Vielbeine. Thus $\tilde{D}_0$, $\tilde{D}_{a}$ are shorthand for $\tau^\mu \tilde{D}_\mu$, $e^\mu{}_{a} \tilde{D}_\mu$. Spatial flat indices are raised and lowered with a Kronecker delta.}
\begin{align}\begin{split}\label{nrscalar}
e^{-1} \mathcal{L}_{\rm non-rel} = m\,\Big[\,\frac{{\rm{i}}}2\,\Big(\Phi^*\tilde D_0\Phi -\Phi\tilde D_0\Phi^*\Big)
             -\frac1{2m}\,\big|\tilde D_{a}\Phi\big|^2 \,\Big]  \,,
\end{split}\end{align}
where we have defined
\begin{align}
 \tilde D_\mu\Phi = \partial_\mu\Phi +{\rm{i}}\,m\,m_\mu\,\Phi \,.
\end{align}
The lagrangian \eqref{nrscalar}  is invariant under diffeomorphisms (with parameter $\xi^\mu$) and the local U(1) central charge transformation of the Bargmann algebra (with parameter $\sigma$), under which $\Phi$ transforms as
\begin{equation} \label{eq:nonrelU1Phi}
  \delta \Phi = \xi^\mu \partial_\mu \Phi  -{\rm{i}}\, m \, \sigma \, \Phi \,.
\end{equation}
One can then define the current associated to the central charge transformation by
\begin{equation}
  \label{eq:partconscurrent}
  j^\mu_{\rm non-rel} =  \tau^\mu \,|\Phi|^2 + e^\mu{}_{a}\, \frac{1}{2 m {\rm{i}}} \left( \Phi^*\tilde{D}^{a} \Phi - \Phi\tilde{D}^{a} \Phi^*  \right)\,.
\end{equation}
When choosing a flat background
\begin{align}
  \label{eq:flatbackground}
  \tau_\mu = \delta_\mu^t \,, \qquad\quad e_t{}^{a} = 0,\quad e_i{}^{a} = \delta_i^{a} \,, \qquad\quad m_\mu = 0 \,,
\end{align}
this current corresponds to the usual current of particle number or mass conservation. We thus explicitly see that, as expected for a non-relativistic limit, our NR limit procedure has suppressed antiparticles.

It is  instructive to look at the action on $\Phi$ of the symmetries that are left when the  flat background (\ref{eq:flatbackground}) is chosen.
The transformation rules (\ref{eq:nonrelU1Phi}) then reduce to those that leave these flat background fields invariant. They are determined by the following NR Killing equations
\begin{align}\begin{aligned}
  \label{eq:killeqsnr}
   \partial_\mu \xi^t &= 0 \,, \qquad &\partial_t \xi^i + \lambda^i &= 0 \,, \\
   \partial_i \xi^j + \lambda^j{}_i &= 0 \,, \qquad& \partial_t \sigma &= 0 \,, \qquad& \partial_i \sigma + \lambda_i &= 0 \,.
\end{aligned}\end{align}
The solution to these equations is given by
\begin{align}
  \label{eq:killvecsnr}
  \xi^t (x^\mu) = \zeta\,, \qquad \xi^i(x^\mu) = \xi^i - \lambda^i\, t - \lambda^i{}_j\, x^j\,, \qquad \sigma(x^\mu) = \sigma - \lambda^i \, x^i\,,
\end{align}
where the parameters $\zeta$, $\xi^i$, $\lambda^i$, $\lambda^{ij}$, $\sigma$ are now constants. These correspond to the usual time translation, spatial translations, galilean boosts, spatial rotations and central charge transformation of the rigid Bargmann algebra. One thus finds that
 $\Phi$ transforms as follows:
\begin{align} \label{eq:flatsymmscalnr}
 \delta\Phi = \Big(\zeta\,\partial_t +\xi^i\partial_i -\lambda^i\,t\,\partial_i -x^j\lambda^i{}_j\,\partial_i
              -{\rm{i}}\,m\,\sigma + {\rm{i}}\,m\,\lambda^i x^i\Big)\Phi \,.
\end{align}
The last term in this transformation rule corresponds to the phase
factor acquired by a Schr\"odinger field under rigid galilean boosts,
that is necessary to show Galilei invariance of the flat space
Schr\"odinger lagrangian. We note that this same Schr\"odinger
lagrangian is also invariant under an extra dilatation and special
conformal transformation, that extend the symmetries of the Bargmann
algebra denoted in (\ref{eq:flatsymmscalnr}) to the ones of the
Schr\"odinger algebra \cite{Niederer:1972zz}. So, even though we
started from a relativistic theory with no conformal invariance, we
end up with a NR theory that is invariant  under non-relativistic
conformal Schr\"odinger symmetries.

\subsection{Massive Spin-1/2}

Following \cite{Bergshoeff:2015sic}, our starting point is the Dirac lagrangian for a $d=3$ massive spin 1/2 particle described by a 4-component spinor $\Psi$ coupled to an arbitrary gravitational background and the U(1) gauge field $M_\mu$:
\begin{align}\label{relspin}
E^{-1} \mathcal{L}_{\rm rel} = \bar\Psi\slashed{D}\Psi -M\,\bar\Psi\Psi +{\rm h.c.}\,,
\end{align}
where  the covariant derivative is given by
\begin{align}
 D_\mu\Psi = \partial_\mu\Psi - \frac14\,\Omega_\mu{}^{ A B}\gamma_{ A B}\Psi + {\rm{i}}\,M\,M_\mu\,\Psi \,.
\end{align}
Before taking the limit, it is convenient to define  projected spinors  in terms of the original spinor as follows \cite{Gomis:2004pw}:
\begin{align} \label{eq:spinproj}
 \Psi_\pm  = \frac12\,\Big(\unity \pm {\rm{i}}\,\gamma^0\Big)\Psi\,, \hskip2cm \bar\Psi_\pm = \bar\Psi \,\frac12\,\Big(\unity \pm {\rm{i}}\,\gamma^0\Big) \,.
\end{align}
Besides redefining the bosonic gravitational fields we also redefine these  projected spinors as follows \cite{Gomis:2004pw}:
\begin{align} \label{eq:spinprojresc}
 \Psi_+ = \sqrt{\omega}\,\psi_+ \,, \hskip2cm \Psi_- = \frac1{\sqrt \omega}\,\psi_- \,.
\end{align}
Using all these redefinitions and taking $M = \omega m$, one finds that the action \eqref{relspin} upon taking the $\omega\rightarrow\infty$ limit reduces to
\begin{align}\begin{split}\label{4dNRdirac}
e^{-1} \mathcal{L}_{\rm non-rel} &= \bar\psi_+\gamma^0\tilde D_0\psi_+ +\bar\psi_+\gamma^{a}\tilde  D_{a}\psi_- +\bar\psi_-\gamma^{a}\tilde D_{a}\psi_+
       -2\,m\,\bar\psi_-\psi_- +{\rm h.c.}\,,
\end{split}\end{align}
where we have used the covariant derivatives
\begin{align}\begin{split}\label{4dcovD}
 \tilde D_\mu\psi_+ &= \partial_\mu\psi_+ -\frac14\,\omega_\mu{}^{ab}\gamma_{ab}\psi_+ -{\rm{i}}\,m\,m_\mu\,\psi_+ \,, \\
 \tilde D_\mu\psi_- &= \partial_\mu\psi_- -\frac14\,\omega_\mu{}^{ab}\gamma_{ab}\psi_- +\frac12\,\omega_\mu{}^{a}\gamma_{a0}\psi_+
                     -{\rm{i}}\,m\,m_\mu\,\psi_- \,.
\end{split}\end{align}
Note that all divergences have canceled.
The  invariance of the lagrangian \eqref{4dNRdirac} under galilean boosts is not manifest but can be checked by using the transformation rules
\begin{align}\begin{split}\label{nice}
 \delta \psi_+ &= \frac14\,\lambda^{ab}\gamma_{ab}\psi_+ + {\rm{i}} \, m \, \sigma \psi_+ \,, \\
 \delta \psi_- &= \frac14\,\lambda^{ab}\gamma_{ab}\psi_- -\frac12\,\lambda^{a}\gamma_{a0}\psi_+ + {\rm{i}} \, m \, \sigma \,\psi_- \,,
\end{split}\end{align}
that are easily found by applying all field redefinitions in the relativistic transformation rules and taking the limit $\omega \to \infty$.

The equations of motion corresponding to the  non-relativistic lagrangian  \eqref{4dNRdirac} are given by the L\'evy--Leblond equations
\begin{align}\begin{split}
  \gamma^0\tilde D_0 \psi_+ +\gamma^{a}\tilde D_{a} \psi_- &=0\,, \\[.1truecm]
  \gamma^{a}\tilde D_{a} \psi_+ -2\,m\,\psi_- &=0\,.
\end{split}\end{align}
The second equation can be used to solve for the auxiliary spinor $\psi_-$ and eliminate it from the lagrangian \eqref{4dNRdirac}. Substituting the solution for $\psi_-$ back into the first equation
we obtain the curved space generalization of the so-called
Schr\"odinger--Pauli equation:
\begin{align} \label{modLL}
 \Big[\gamma^0\tilde D_0  +\frac1{2m}\tilde D^{a} \tilde  D_{a}
        \Big]\psi_+ =0 \,.
\end{align}

\subsection{NR massless spin 1}

We now consider the massless spin 1 case. Our starting point is  the lagrangian of a real, massless, relativistic vector field coupled to gravity:
\begin{equation}
  \label{eq:realvecrel}
  E^{-1} \mathcal{L}_{\rm rel} = - \frac14\, E^\mu_{ A} E^{\rho  A}  E^\nu_{ B} E^{\sigma B} F_{\mu\nu} F_{\rho \sigma} \,,
\end{equation}
where $F_{\mu\nu}$ is the usual Maxwell field strength. Like for the spin 0 case, one can take an electric or magnetic galilean limit of electrodynamics or even an electric and magnetic carrollian limit. We will only discuss here the magnetic galilean limit since it shows the additional feature of an emergent symmetry, something that does not occur in the spin-0 case.
Redefining the gravitational background fields like in the previous section  leads to the following non-relativistic lagrangian
\begin{equation}\label{nrexpansion}
  e^{-1} \mathcal{L}_{\rm rel} = -\frac{1}{2\omega^2}\tau^\mu\tau^\nu F_{\mu a}F_{\nu }{}^a - \frac14\, F_{ab} F^{ab} \,.
\end{equation}
Taking the limit $\omega \to \infty$, we  obtain for a flat spacetime
\begin{equation}
  \label{eq:realvecnonrel}
   \mathcal{L}_{\rm non-rel} = - \frac14\, F_{ij} F^{ij} \,.
\end{equation}
Due to the absence of the field $A_0$ this lagrangian has an emergent
Stueckelberg symmetry  $\delta A_0(x) = \rho(x)$ while the
corresponding field equation of $A_0$ does not follow directly from
the non-relativistic lagrangian \eqref{eq:realvecnonrel}. The
situation is very similar to what happens when taking the limit of
Neveu-Schwarz gravity where the Poisson equation of the Newton
potential is missing \cite{Bergshoeff:2021bmc}. The missing equation
of motion can be obtained by taking the limit of the relativistic
equations of motion. The complete set of non-relativistic equations of
motion form a reducible but indecomposable representation under
galilean boosts which means that the equation of motion corresponding
to $A_0$ transforms to the equations of motion corresponding to $A_i$
but not the other way around. This shows the following connection
between the equation of motion corresponding to $A_0$ and the
lagrangian \eqref{eq:realvecnonrel}: the missing equation of motion
corresponding to $A_0$ is not invariant under galilean boosts by
itself but, instead transforms into the equations of motion that
follow from the non-relativistic lagrangian \eqref{eq:realvecnonrel}.

\subsection{Massless Spin 1 with an additional scalar field}
Allowing the option to add extra fields to the lagrangian, there is  yet another way to obtain a non-relativistic lagrangian from a relativistic one.
To be precise, extending the relativistic lagrangian with  a massless scalar $\rho$ we consider the following lagrangian  \cite{Bergshoeff:2015sic}:
\begin{equation}
  \label{eq:realvecscalrel}
E^{-1} \mathcal{L}_{\rm rel} = - \frac14\,  E^\mu_{ A} E^{\rho  A}  E^\nu_{ B} E^{\sigma B} F_{\mu\nu} F_{\rho \sigma} - \frac12\,  E^\mu_{ A} E^{\nu  A} \partial_\mu \rho\, \partial_\nu \rho \,.
\end{equation}
Defining  two  fields $A$ and $B$ as follows:
\begin{equation} \label{eq:AB}
  A = E^\mu{}_0 A_\mu - \rho \,, \qquad B = E^\mu{}_0 A_\mu + \rho \,,
\end{equation}
one can redefine the bosonic background fields like in the previous section  supplemented with the redefinitions
\begin{equation} \label{eq:ABtilde}
  A = \frac{1}{\omega} \tilde{A} \,, \qquad B = \omega \tilde{B} \,,
\end{equation}
to obtain the following non-relativistic lagrangian in the $\omega \rightarrow \infty$ limit\,\footnote{For a flat spacetime, the same lagrangian can be obtained by a null reduction \cite{Festuccia:2016caf}. A `T-dual way' to obtain the same Lagrangean is to take a so-called string limit of Maxwell in one dimension higher and to reduce over the spatial direction longitudinal to the string \cite{Bergshoeff:2021tfn}.}
\begin{equation}
  \label{eq:realvecscalnonrel}
  e^{-1} \mathcal{L}_{\rm non-rel} = \frac18\,\partial_0\tilde{B}\,\partial_0\tilde{B}
              +\frac12\,\tilde{D}_{a}\tilde{A}\,\partial^{a}\tilde{B} -\frac14\,F_{ab} F^{ab}
              -\frac12\,\tilde{D}^{a} A_{a}\, \partial_0 \tilde{B} \,,
\end{equation}
where the following derivatives were used
\begin{align} \label{eq:halfcovders}
  \tilde{D}_\mu \tilde{A} &= \partial_\mu \tilde{A} + \omega_\mu{}^{a} A_{a} \,, \nonumber \\
  \tilde{D}_\mu A_{a} &= \partial_\mu A_{a} - \omega_{\mu\, a}{}^{b} A_{b} + \frac12\, \omega_\mu{}_{a} \tilde{B} \,.
\end{align}
Note that the basic variables are a spatial vector $A_{a}$ with spatial flat indices and two extra fields $\tilde{A}$, $\tilde{B}$. These fields transform non-trivially under local spatial rotations and galilean boosts as follows:
\begin{align}
  \delta \tilde{A} &= -\lambda^{a} A_{a} \,, \qquad \qquad \qquad \delta \tilde{B} = 0\,, \nonumber \\
  \delta A_{a} &= \lambda_{a}{}^{b} A_{b}- \frac12\, \lambda^{a} \tilde{B} \,,
\end{align}
while they transform as scalars under general coordinate transformations.
It is with respect to these transformations that the above derivatives (\ref{eq:halfcovders}) are defined.
The  lagrangian (\ref{eq:realvecscalnonrel}) is also invariant under the U(1) gauge transformation
\begin{equation}
  \label{eq:nonrelU1}
  \delta \tilde{A} = \tau^\mu \partial_\mu \Lambda \,, \qquad\quad \delta A_{a} = e^\mu{}_{a} \partial_\mu \Lambda \,,
\end{equation}
although this invariance is not manifest.

To get a better physical understanding of the lagrangian
(\ref{eq:realvecscalnonrel}), we consider the equations of motion when
restricted to the flat background \eqref{eq:flatbackground} (such that
$i = a$):
\begin{equation}
  \begin{split}
  \partial^i \partial_i \tilde{B} &= 0 \,, \\
  \partial_i \partial_t \tilde{B} + \partial^j F_{ji} &= 0 \,, \\
  \partial_t \partial_t \tilde{B} - 2\, \partial^i \partial_i \tilde{A} + 2\, \partial_t \partial^i A_i &= 0 \,.
\end{split}
\end{equation}

One can consistently set $\tilde{B}$ to zero in these equations since
this constraint is invariant under all the symmetries of the theory.
The remaining equations for $\tilde{A}$ and $A_i$ then coincide with
the equations of galilean electromagnetism in the magnetic limit,
where $\tilde{A}$ plays the role of the electric potential. This
theory is not only invariant under the Galilei group, but also under
the galilean conformal group \cite{Niederer:1972zz, Hagen:1972pd,
  Duval:1990hj, Henkel:1993sg}. The latter is the conformal extension
of the Galilei group that is obtained by performing an
In\"on\"u-Wigner contraction of the relativistic conformal
group. Since the relativistic lagrangian we started from is
conformally invariant when restricted to flat space, it is not
surprising to see that the non-relativistic limit is invariant under
galilean conformal transformations.

\section{Conclusion}
\label{sec:conclusion}

In this review we summarised the basic properties of a number of
non-lorentzian theories. We first discussed the kinematical spaces and
corresponding symmetry algebras of these non-lorentzian theories. We
next constructed a number of actions describing the dynamics of
particles moving in these kinematical spaces. For this, we applied the
method of nonlinear realisations and explained the relation between
this method and the coadjoint orbit method.  We have also analysed
the non-lorentzian particles as a suitable non-relativistic limit of
relativistic particles.  We also discussed three types of
non-lorentzian gravity theories: Galilei gravity, Newton-Cartan
gravity and Carroll gravity. We not only showed how these gravity
theories can be obtained by applying a gauging procedure to an
underlying non-relativistic Lie algebra but also by taking a special
non-relativistic limit of general relativity. Introducing matter, we
discuss the coupling of gravity to field theories describing particles
of different spin.  We achieved this by starting from the relativistic
field theories coupled to general relativity and taking a
non-relativistic limit.

There are several ways to extend the results presented in this review.
As mentioned in the introduction, one could extend the degenerate
geometries we considered here to geometries that are characterised by
a foliation of a higher codimension. In particular, the geometries
with a codimension-2 foliation play an important role in describing
non-relativistic string theory, as in the article by Oling and Yan
\cite{Oling:2022fft}.

\section*{Acknowledgments}

We acknowledge many enlightening conversations on non-lorentzian
topics with the following people: Roberto Casalbuoni, Can Görmez, Ross
Grassie, Jelle Hartong, Emil Have, Yannick Herfray,
Axel Kleinschmidt, Johannes Landsteiner, Stefan Prohazka, Luca Romano,
Jan Rosseel, Jakob Salzer, Dieter Van den Bleeken and Kevin van
Helden.

The work of JG has been supported in part by MINECO
FPA2016-76005-C2-1-P and PID2019-105614GB-C21 and from the State
Agency for Research of the Spanish Ministry of Science and Innovation
through the Unit of Excellence Maria de Maeztu 2020-203 award to the
Institute of Cosmos Sciences (CEX2019-000918-M).

\providecommand{\href}[2]{#2}\begingroup\raggedright\endgroup

\end{document}